\documentclass{egpubl}
\usepackage{eurovis2021}

\SpecialIssuePaper         % uncomment for final version of Computer Graphics Forum, special issue
\usepackage[T1]{fontenc}
\usepackage{dfadobe}  

\usepackage{cite}  % comment out for biblatex with backend=biber
\BibtexOrBiblatex
\electronicVersion
\PrintedOrElectronic
\ifpdf \usepackage[pdftex]{graphicx} \pdfcompresslevel=9
\else \usepackage[dvips]{graphicx} \fi

\usepackage{egweblnk}
\usepackage{mathptmx}
\usepackage{url}

\usepackage{amsfonts}
\usepackage{amsmath}
\usepackage{amssymb}

\usepackage{color}
\usepackage{xcolor}

\definecolor{gray}{gray}{0.5}

\graphicspath{{Figures/}{./}}

\newcommand{\SE}{{\mathcal{H}}} % Shannon entropy
\newcommand{\CE}{{\mathcal{H}_{\text{CE}}}} % cross entropy
\newcommand{\MI}{{\mathcal{I}}} % mutual information
\newcommand{\DKL}{{\mathcal{D}_{\text{KL}}}} % KL-divergence
\newcommand{\DJS}{{\mathcal{D}_{\text{JS}}}} % JS-divergence
\newcommand{\Dnew}{{\mathcal{D}^k_{\text{new}}}} % the new divergence by Chen and Sbert
\newcommand{\Dncm}{{\mathcal{D}^k_{\text{ncm}}}} % the new asymmetric divergence by Chen and Sbert
\newcommand{\DM}{{D^k_{\text{M}}}} %

\newcommand{\DnewA}{{\mathcal{D}^{k=1}_{\text{new}}}}
\newcommand{\DnewB}{{\mathcal{D}^{k=2}_{\text{new}}}}
\newcommand{\DncmA}{{\mathcal{D}^{k=1}_{\text{ncm}}}}
\newcommand{\DncmB}{{\mathcal{D}^{k=2}_{\text{ncm}}}}

\usepackage[condensed]{tgheros}
\newenvironment{narrowfont}{\fontfamily{qhvc}\selectfont}{\par}
    
\title[A Bounded Measure for Estimating the Benefit of Visualization: Theoretical Discourse and Conceptual Evaluation]%
      {A Bounded Measure for Estimating the Benefit of Visualization: Theoretical Discourse and Conceptual Evaluation}

\author[M. Chen and M. Sbert]
{\parbox{\textwidth}{\centering Min Chen$^1$\orcid{0000-0001-5320-5729}
    \quad and \quad
    Mateu Sbert$^2$
    }
    \\
{\parbox{\textwidth}{\centering $^1$University of Oxford, UK
    \quad and \quad
    $^2$ University of Girona, Spain
    } 
}
}

\begin{document}

\teaser{
  \centering
  \begin{tabular}{@{}c@{\hspace{4mm}}c@{}}
    \includegraphics[height=40mm]{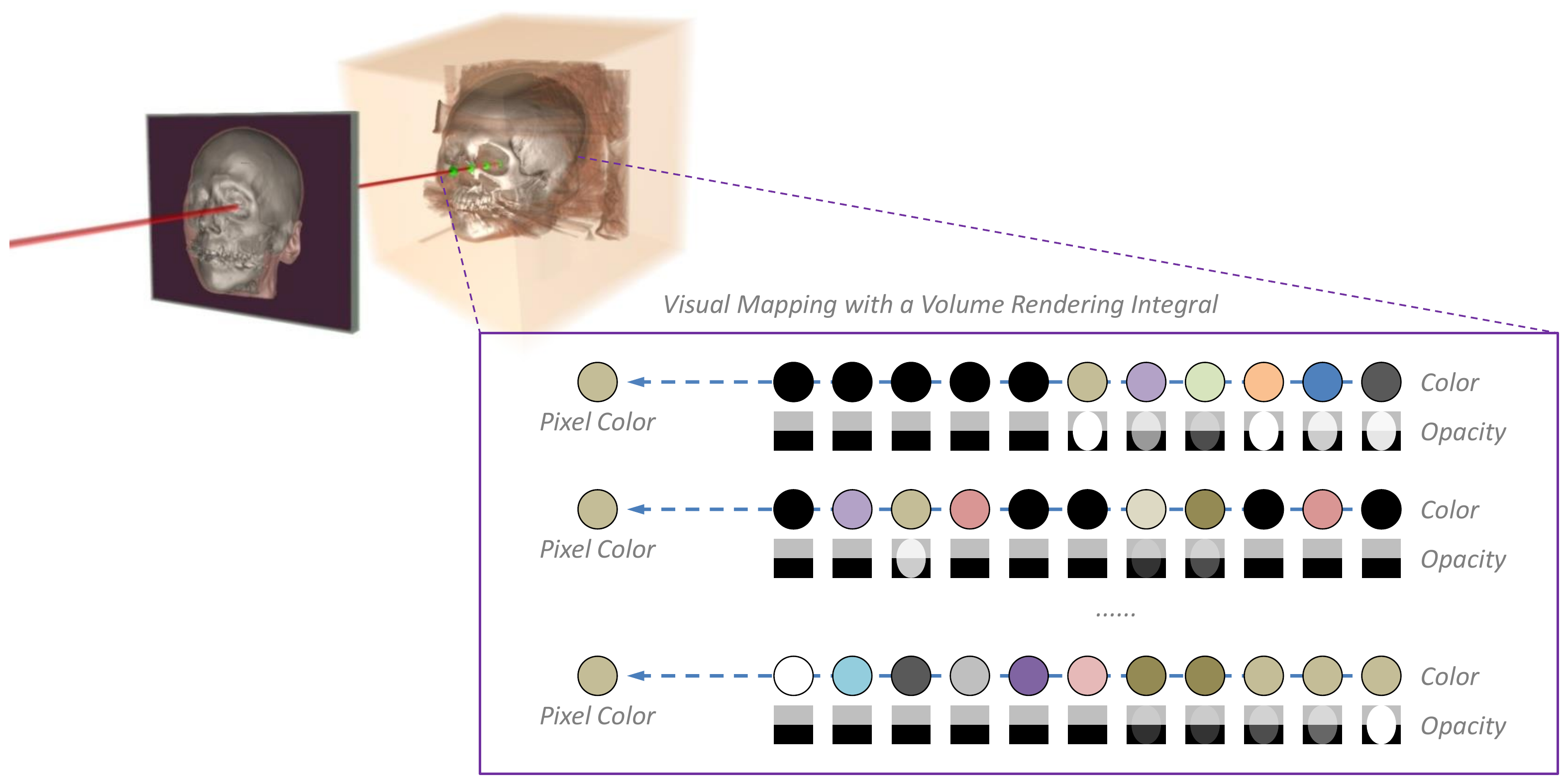} &
    \includegraphics[height=40mm]{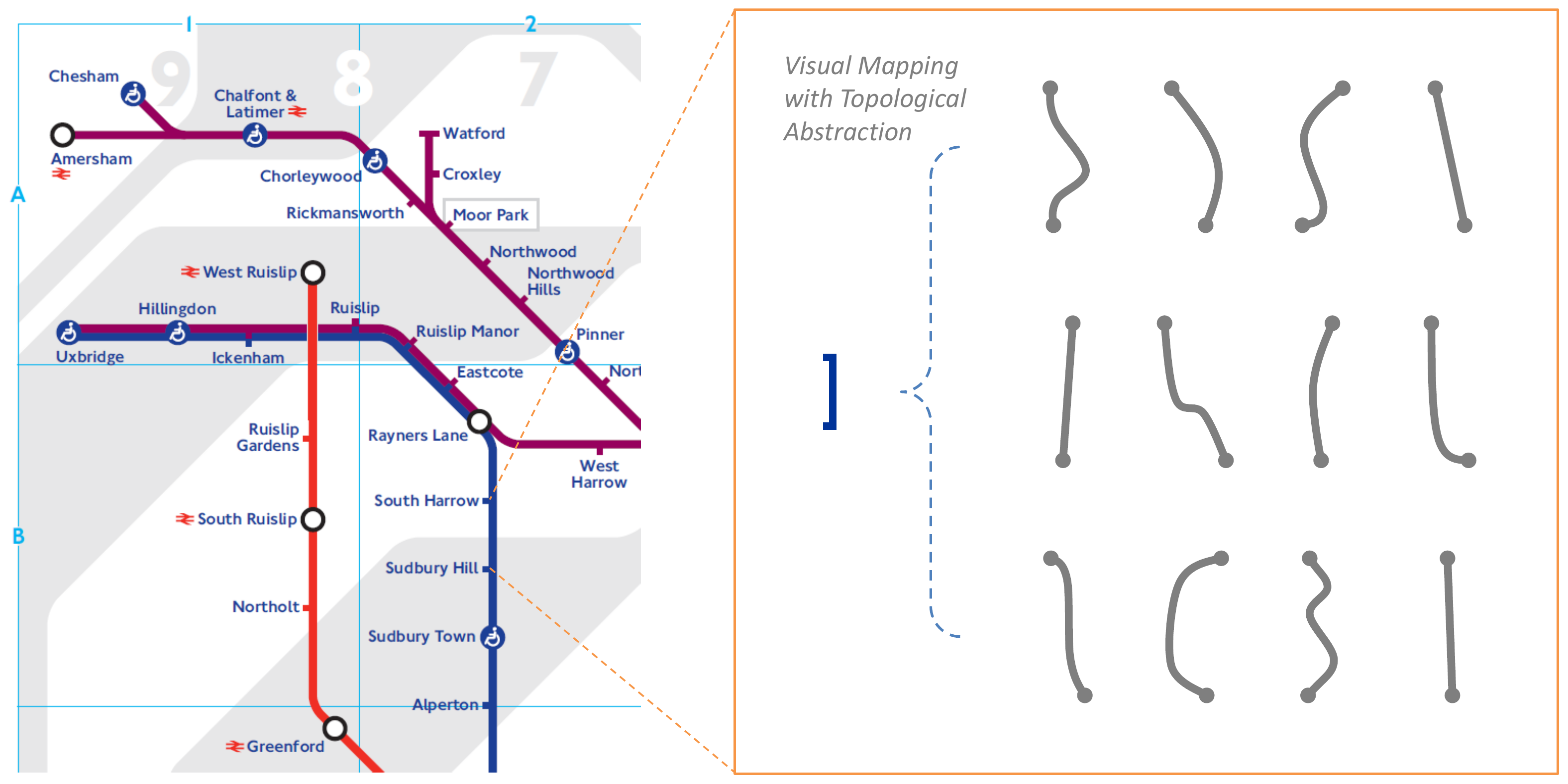}\\
    \small{(a) mapping from different sets of voxel values to the same pixel color} &
    \small{(b) mapping from different geographical paths to the same line segment}
  \end{tabular}
  \caption{Visual encoding typically features many-to-one mapping from data to visual representations, hence information loss. The significant amount of information loss in volume visualization and metro maps suggests that viewers not only can abide the information loss but also benefit from it. Measuring such benefits can lead to new advancements of visualization, in theory and practice.}
  \label{fig:InfoLoss}
}

\maketitle
%-------------------------------------------------------------------------
\begin{abstract}%
Information theory can be used to analyze the cost-benefit of visualization processes. However, the current measure of benefit contains an unbounded term that is neither easy to estimate nor intuitive to interpret. In this work, we propose to revise the existing cost-benefit measure by replacing the unbounded term with a bounded one. We examine a number of bounded measures that include the Jenson-Shannon divergence and a new divergence measure formulated as part of this work. We describe the rationale for proposing a new divergence measure. As the first part of comparative evaluation, we use visual analysis to support the multi-criteria comparison, narrowing the search down to several options with better mathematical properties. The theoretical discourse and conceptual evaluation in this paper provide the basis for further comparative evaluation through synthetic and experimental case studies, which are to be reported in a separate paper.
%-------------------------------------------------------------------------
%  ACM CCS 1998
%  (see https://www.acm.org/publications/computing-classification-system/1998)
% \begin{classification} % according to https://www.acm.org/publications/computing-classification-system/1998
% \CCScat{Computer Graphics}{I.3.3}{Picture/Image Generation}{Line and curve generation}
% \end{classification}
%-------------------------------------------------------------------------
%  ACM CCS 2012
%   (see https://www.acm.org/publications/class-2012)
%The tool at \url{http://dl.acm.org/ccs.cfm} can be used to generate
% CCS codes.
%Example:
% <concept_id>10010147.10010371.10010352.10010381</concept_id>
% <concept_desc>Computing methodologies~Collision detection</concept_desc>
% <concept_significance>300</concept_significance>
% <concept_id>10010583.10010588.10010559</concept_id>
% <concept_desc>Hardware~Sensors and actuators</concept_desc>
% <concept_significance>300</concept_significance>
% <concept_id>10010583.10010584.10010587</concept_id>
% <concept_desc>Hardware~PCB design and layout</concept_desc>
% <concept_significance>100</concept_significance>

% \ccsdesc[300]{Computing methodologies~Collision detection}
% \ccsdesc[300]{Hardware~Sensors and actuators}
% \ccsdesc[100]{Hardware~PCB design and layout}

\printccsdesc   
\end{abstract}


\section{Introduction}
\label{sec:Introduction}

To most of us, it seems rather intuitive that visualization should be accurate, different data values should be visually encoded differently, and visual distortion should be disallowed.
However, when we closely examine most (if not all) visualization images, we can notice that inaccuracy is ubiquitous.
The two examples in Figure \ref{fig:InfoLoss} evidence the presence of such inaccuracy.
In volume visualization, when a pixel is used to depict a set of voxels along a ray, many different sets of voxel values may result in the same pixel color.
In a metro map, a variety of complex geographical paths may be distorted and depicted as a straight line.
Since there is little doubt that volume visualization and metro maps are useful, some ``inaccurate'' visualization must be beneficial.

% It is now widely understood among visualization researchers and practitioners that the effectiveness of a visualization process depends on \emph{data}, \emph{user}, and \emph{task}.
% One important aspect of \emph{user} is a user's knowledge, which plays a critical role in reconstructing the information lost during visualization processes (e.g., data transformation and visual mapping).
% One major challenge in appreciating the significance of such knowledge is the difficulty to measure or estimate the knowledge used by a user during visualization.

In terms of information theory, the types of inaccuracy featured in Figure \ref{fig:InfoLoss} are different forms of information loss (or many-to-one mapping).
Chen and Golan proposed an information-theoretic measure \cite{Chen:2016:TVCG} for analyzing the cost-benefit of data intelligence workflows.
It enables us to consider the positive impact of information loss (e.g., reducing the cost of storing, processing, displaying, perceiving, and reasoning about the information) as well as its negative impact (e.g., being mislead by the information).
The measure provides a concise explanation about the benefit of visualization because visualization and other data intelligence processes (e.g., statistics and algorithms) all typically cause information loss and visualization allows human users to reduce the negative impact of information loss effectively using their knowledge. 

The mathematical formula of the measure features a term based on the Kullback-Leibler (KL) divergence \cite{Kullback:1951:AMS} for measuring the potential distortion of a user or a group of users in reconstructing the information that may have been lost or distorted during a visualization process.
% The cost-benefit ratio instigates that a user with more knowledge about the source data and its visual representation is likely to suffer less distortion.
While using the KL-divergence is mathematically intrinsic for measuring the potential distortion, its unboundedness property has some undesirable consequences.
The simplest phenomenon of making a false representation (i.e., always displaying 1 when a binary value is 0 or always 0 when it is 1) happens to be a singularity condition of the KL-divergence.
The amount of distortion measured by the KL-divergence often has much more bits than the entropy of the information space itself.
This is not intuitive to interpret and hinders practical applications.%

% Kijmongkolchai et al. applied the formula of Chen and Golan to the results of an empirical study for estimating users' knowledge used in visualization processes, and used a bounded approximation of the KL-divergence in their estimation \cite{Kijmongkolchai:2017:CGF}.

In this paper, we propose to replace the KL-divergence with a bounded term.
We first confirm the boundedness is a necessary property.
We then conduct multi-criteria decision analysis (MCDA) \cite{Ishizaka:2013:book} to compare a number of bounded measures, which include the Jensen–Shannon (JS) divergence \cite{Lin:1991:TIT}, and a new divergence measure $\Dnew$ (including its variations) formulated as part of this work.
We use visual analysis to aid the observation of the mathematical properties of these candidate measures, narrowing down from eight options to five.
In a separate but related paper (included in the supplementary materials), we use synthetic and experimental case studies to to instantiate values that may be returned by the five options. It also explores the relationship between measuring the benefit of visualization and measuring the viewers' knowledge used during visualization.
The search for the best way to measure the benefit of visualization will likely entail a long journey. The main contribution of this work is to initiate this endeavor.

% ====================
\section{Related Work}
\label{sec:RelatedWork}

Claude Shannon's landmark article in 1948 \cite{Shannon:1948:BSTJ} signifies the birth of information theory.
It has been underpinning the fields of data communication, compression, and encryption since.
As a mathematical framework, information theory provides a collection of useful measures, many of which, such as Shannon entropy \cite{Shannon:1948:BSTJ}, cross entropy \cite{Cover:2006:book}, mutual information \cite{Cover:2006:book}, and Kullback-Leibler divergence \cite{Kullback:1951:AMS} are widely used in applications of
physics, biology, neurology, psychology, and computer science
(e.g., visualization, computer graphics, computer vision, data mining, machine learning), and so on.
In this work, we also consider Jensen-Shannon divergence \cite{Lin:1991:TIT} in detail.

Information theory has been used extensively in visualization \cite{Chen:2016:book}.
It has enabled many applications in visualization, including
scene and shape complexity analysis by Feixas et al. \cite{Feixas:2001:CGF} and Rigau et al. \cite{Rigau:2005:SMA},
light source placement by Gumhold \cite{Gumhold:2002:Vis},
view selection in mesh rendering by V\'{a}zquez et al. \cite{Vazquez:2004:CGF} and Feixas et al. \cite{Feixas:2009:AP},
attribute selection by Ng and Martin \cite{Ng:2004:IV},
view selection in volume rendering by Bordoloi and Shen \cite{Bordoloi:2005:Vis}, and Takahashi and Takeshima \cite{Takahashi:2005:Vis},
multi-resolution volume visualization by Wang and Shen \cite{Wang:2006:TVCG},
focus of attention in volume rendering by Viola et al. \cite{Viola:2006:TVCG},
feature highlighting by J\"anicke and Scheuermann \cite{Jaenicke:2007:TVCG,Jaenicke:2010:CGA},
    and Wang et al. \cite{Wang:2008:TVCG},
transfer function design by Bruckner and M\"{o}ller \cite{Bruckner:2010:CGF},
	and Ruiz et al. \cite{Ruiz:2011:TVCG,Bramon:2013:JBHI},
multi-modal data fusion by Bramon et al. \cite{Bramon:2012:TVCG},
isosurface evaluation by Wei et al. \cite{Wei:2013:CGF},
measuring observation capacity by Bramon et al. \cite{Bramon:2013:CGF},
measuring information content by Biswas et al. \cite{Biswas:2013:TVCG},
proving the correctness of ``overview first, zoom, details-on-demand'' by Chen and J\"anicke \cite{Chen:2010:TVCG} and Chen et al. \cite{Chen:2016:book}, and
confirming visual multiplexing by Chen et al. \cite{Chen:2014:CGF}.

Ward first suggested that information theory might be an underpinning theory for visualization \cite{Purchase:2008:LNCS}.
Chen and J\"anicke \cite{Chen:2010:TVCG} outlined an information-theoretic framework for visualization, and it was further enriched by Xu et al. \cite{Xu:2010:TVCG} and Wang and Shen \cite{Wang:2011:E} in the context of scientific visualization.
Chen and Golan proposed an information-theoretic measure for analyzing the cost-benefit of visualization processes and visual analytics workflows \cite{Chen:2016:TVCG}.
It was used to frame an observation study showing that human developers usually entered a huge amount of knowledge into a machine learning model \cite{Tam:2017:TVCG}.
It motivated an empirical study confirming that knowledge could be detected and measured quantitatively via controlled experiments \cite{Kijmongkolchai:2017:CGF}. 
It was used to analyze the cost-benefit of different virtual reality applications
\cite{Chen:2019:TVCG}.
It formed the basis of a systematic methodology for improving the cost-benefit of visual analytics workflows \cite{Chen:2019:CGF}.
It survived qualitative falsification by using arguments in visualization \cite{Streeb:2019:TVCG}.
It offered a theoretical explanation of ``visual abstraction'' \cite{Viola:2019:book}.
The work reported in this paper continues the path of theoretical developments in visualization \cite{Chen:2017:CGA}, and is intended to improve the original cost-benefit formula \cite{Chen:2016:TVCG}, in order to make it a more intuitive and usable measurement in practical visualization applications.

\begin{figure}[t]
  \centering
  \includegraphics[width=80mm]{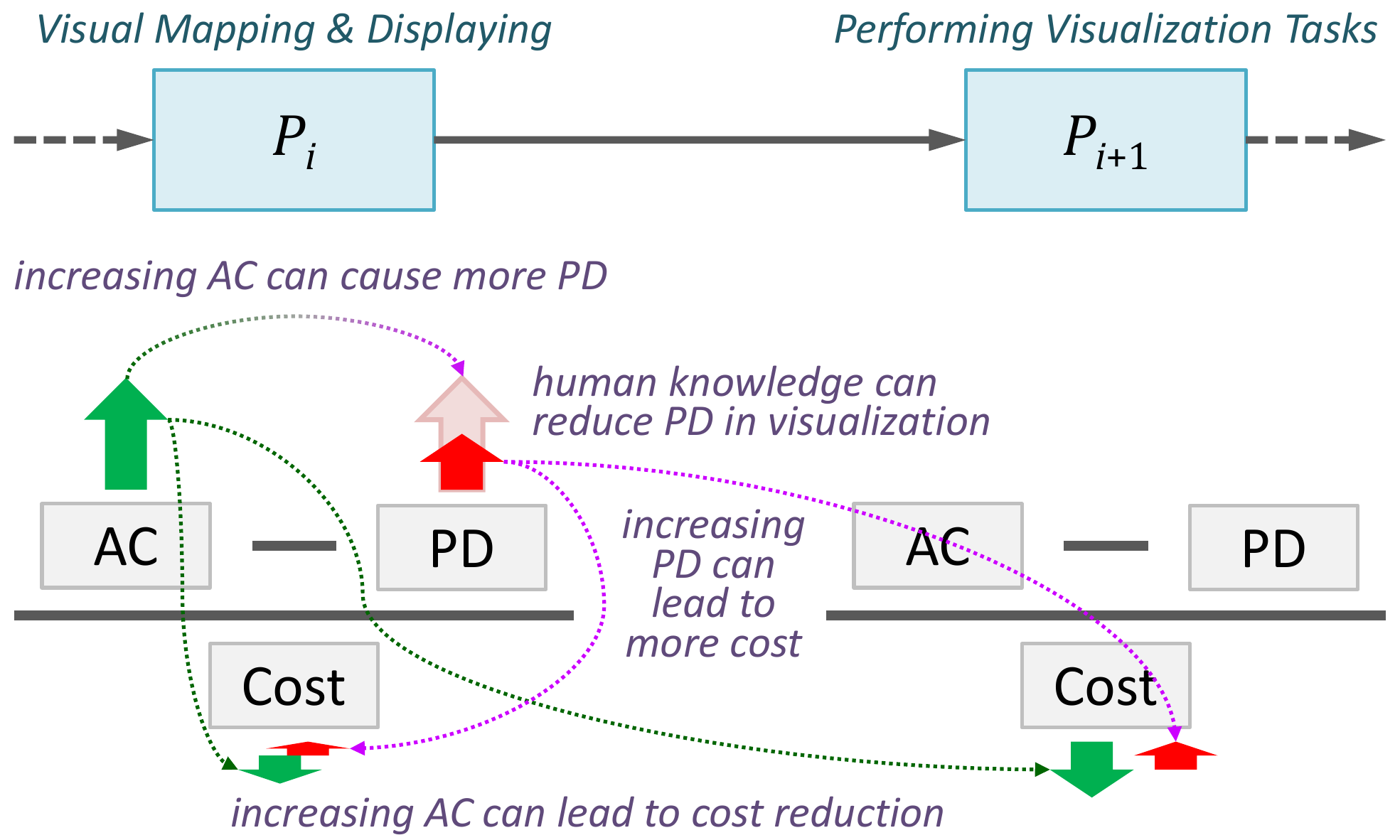}
  \caption{Alphabet compression may reduce the cost of $P_i$ and $P_{i+1}$, especially when human knowledge can reduce potential distortion.}
  \label{fig:Benefit}
  \vspace{-4mm}
\end{figure}

% ====================
\section{Overview and Motivation}
\label{sec:Motivation}
Visualization is useful in most data intelligence workflows, but the usefulness is not universally true because the effectiveness of visualization is usually data-, user-, and task-dependent.
The cost-benefit ratio proposed by Chen and Golan \cite{Chen:2016:TVCG} captures the essence of such dependency.
Below is the qualitative expression of the measure:
\begin{equation} \label{eq:CBM-1}
    \frac{\text{Benefit}}{\text{Cost}} = \frac{\text{Alphabet Compression} - \text{Potential Distortion}}{\text{Cost}}
\end{equation}

Consider the scenario of viewing some data through a particular visual representation.
The term \emph{Alphabet Compression} (AC) measures the amount of information loss due to visual abstraction \cite{Viola:2019:book}.
Since the visual representation is fixed in the scenario, AC is thus largely data-dependent.
AC is a positive measure reflecting the fact that visual abstraction must be useful in many cases though it may result in information loss.
This apparently counter-intuitive term is essential for asserting why visualization is useful. (Note that the term also helps assert the usefulness of statistics, algorithms, and interaction since they all usually cause information loss \cite{Chen:2019:CGF}. (See also Appendix \ref{app:OriginalTheory}.)

The positive implication of the term AC is counterbalanced by the term \emph{Potential Distortion}, while both being moderated by the term \emph{Cost}.
The term \emph{Cost} encompasses all costs of the visualization process, including computational costs (e.g., visual mapping and rendering), cognitive costs (e.g., cognitive load), and consequential costs (e.g., impact of errors).
As illustrated in Figure \ref{fig:Benefit}, increasing AC typically enables the reduction of cost (e.g., in terms of energy, or its approximation such as time or money).
% The measure of cost (e.g., in terms of energy, time, or money) is thus data-, user-, and task-dependent.

The term \emph{Potential Distortion} (PD) measures the informative divergence between viewing the data through visualization with information loss and reading the data without any information loss. The latter might be ideal but is usually at an unattainable cost except for values in a very small data space (i.e., in a small alphabet as discussed in \cite{Chen:2016:TVCG}).
As shown in Figure \ref{fig:Benefit}, increasing AC typically causes more PD.
PD is data-dependent or user-dependent.
Given the same data visualization with the same amount of information loss, one can postulate that a user with more knowledge about the data or visual representation usually suffers less distortion.
This postulation is a focus of this paper.

\begin{figure}[t]
\centering
\includegraphics[width=\linewidth]{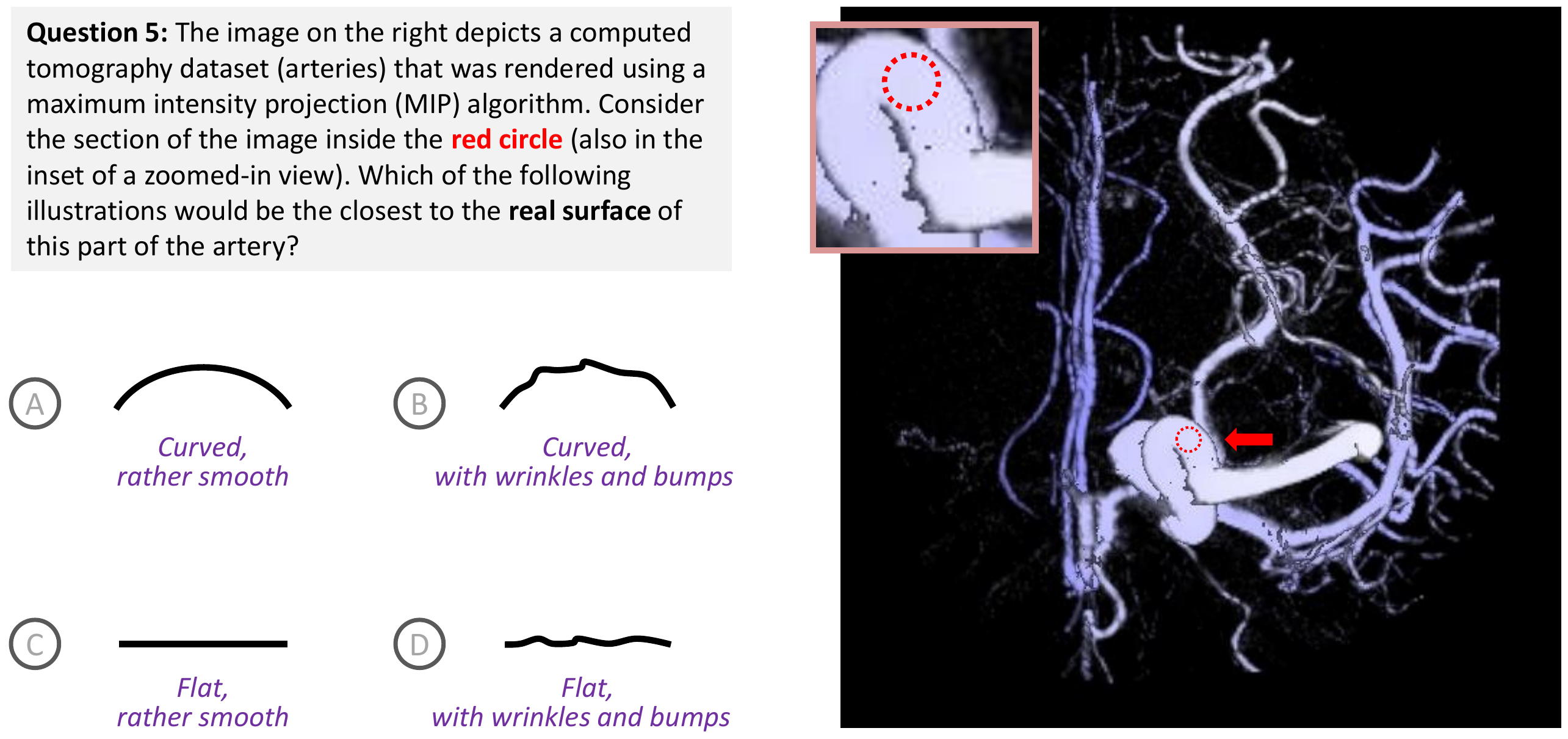}
\caption{A volume dataset was rendered using the MIP method. A question about a ``flat area'' in the image can be used to tease out a viewer's knowledge that is useful in a visualization process.}
\label{fig:Arteries}
\vspace{-4mm}
\end{figure}

Consider the visual representation of a network of arteries in Figure \ref{fig:Arteries}.
The image was generated from a volume dataset using the maximum intensity projection (MIP) method.
While it is known that MIP cannot convey depth information well, it has been widely used for observing some classes of medical imaging data, such as arteries.
The highlighted area in Figure \ref{fig:Arteries} shows an apparently flat area, which is a distortion from the actuality of a tubular surface likely with some small wrinkles and bumps.
The doctors who deal with such medical data are expected to have sufficient knowledge to reconstruct the reality adequately from the ``distorted'' visualization, while being able to focus on the more important task of making diagnostic decisions, e.g., about aneurysm.

As shown in some recent works, it is possible for visualization designers to estimate AC, PD, and Cost qualitatively \cite{Chen:2019:TVCG,Chen:2019:CGF} and quantitatively \cite{Tam:2017:TVCG,Kijmongkolchai:2017:CGF}.
It is highly desirable to advance the scientific methods for quantitative estimation, towards the eventual realization of computer-assisted analysis and optimization in designing visual representations.
This work focuses on one challenge of quantitative estimation, i.e., how to estimate the benefit of visualization to human users with different knowledge about the depicted data and visual encoding.

Building on the methods of observational estimation \cite{Tam:2017:TVCG} and controlled experiment \cite{Kijmongkolchai:2017:CGF}, one may reasonably anticipate a systematic method based on a short interview by asking potential viewers a few questions.
For example, one may use the question in Figure \ref{fig:Arteries} to estimate the knowledge of doctors, patients, and any other people who may view such a visualization.
The question is intended to tease out two pieces of knowledge that may help reduce the potential distortion due to the ``flat area'' depiction.
One piece is about the general knowledge that associates arteries with tube-like shapes.
Another, which is more advanced, is about the surface texture of arteries and the limitations of the MIP method.

In a separate paper (see the supplementary materials), the question in Figure \ref{fig:Arteries} is one of eight questions used in a survey for collecting empirical data for evaluating the bounded measures considered in this paper.
As this paper focuses on the theoretical discourse and conceptual evaluation, we use a highly abstracted version of this example to introduce the relevant information-theoretic notations and elaborate the problem statement addressed by this paper.

\begin{figure}[t]
    \centering
    \includegraphics[width=80mm]{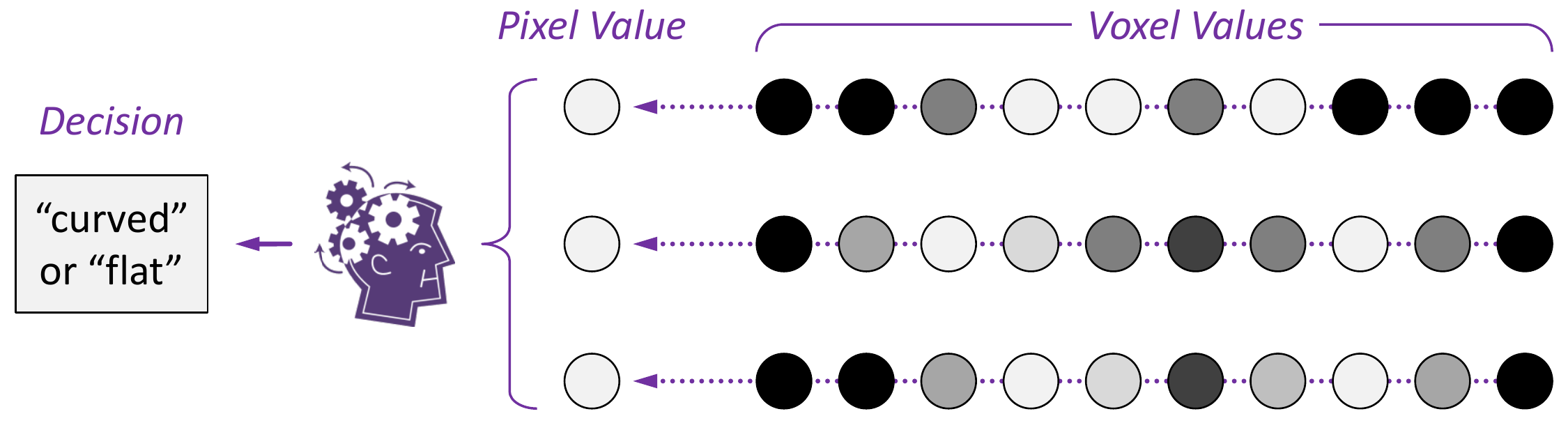}
    \caption{A simplified scenario of visualizing three sequences of voxels, which are rendered using the MIP method. A viewer needs to determine if the surface in the volume is curved or flat (or straight in the illustration).}
    \label{fig:MIP}
    \vspace{-4mm}
\end{figure}

% ====================
\section{Mathematical Notations and Problem Statement}
\label{sec:Problem}

\textbf{Mathematical Notation.} \quad
Consider a simplified scenario in Figure \ref{fig:MIP}, where three sequences of voxels are rendered using the MIT method, resulting in three pixel values on the left.
Here the three sequence voxels examplifies a volume with $N_x \times N_y \times N_Z$ voxels (illustrated as $1 \times 3 \times 10$).
Let each voxel value be an 8-bit unsigned integer.
In information theory, the possible 256 values are referred to as an \emph{alphabet}, denoted here as $\mathbb{D}_\text{vxl} = \{0, 1, 2, \ldots, 255\}$.
The 256 valid values $[0, 255]$ are its \emph{letters}.
The alphabet is associated with a \emph{probability mass function} (PMF) $P(\mathbb{D}_\text{vxl})$.
The Shannon entropy of this alphabet $\mathcal{H}(\mathbb{D}_\text{vxl})$ measures the average uncertainty and information of the voxel, and is defined as:
\[
    \SE(\mathbb{D}_\text{vxl}) = - \sum_0^{255} p_i \log_2 p_i \quad \text{where } p_i \in [0, 1], \sum_0^{255} p_i = 1
\]
\noindent where $p_i$ indicates the probability for the voxel to have its value equal to $i \in [0, 255]$.
When all 256 values are equally probable (i.e., $\forall i \in [0, 255], p_i = 1/256$), we have $\SE(\mathbb{D}_\text{vxl}) = 8$ bits.
In practice, an application usually deals with a specific type of volume data, the probability of different values may vary noticeably.
For example, in medical imaging, a voxel at the concern of a volume is more likely to have a value indicating an empty space.

The entire volume of $N_x \times N_y \times N_Z$ voxels be defined as a composite alphabet $\mathbb{D}_\text{vlm}$. Its letters are all valid combinations of voxel values. If the $N_x \times N_y \times N_Z$ voxels are modelled as independent and identically distributed random variables, we have:
\[
    \mathcal{H}(\mathbb{D}_\text{vml}) = \sum_{k=1}^M \mathcal{H}(\mathbb{D}_{\text{vxl}, k}) = 8 \times N_x \times N_y \times N_Z \text{ bits}
\]
\noindent where $M = N_x \times N_y \times N_z$. For the volume illustrated in Figure \ref{fig:MIP}, $\SE(\mathbb{D}_\text{vml})$ would be 240 bits.
However, this is the maximum entropy of such a volume.
In real world applications, it is very unlikely for the $N_x \times N_y \times N_Z$ voxels to be independent and identically distributed random variables.
Although domain experts may not have acquired the ground truth PMF, by measuring a very large corpus of volume data, they have intuitive knowledge as to what may be possible or not.
For example, doctors, who handle medical imaging data, do not expect to see a car, ship, aeroplane, or other ``weird'' objects in a volume dataset.
This intuitive and imprecise knowledge about the PMF can explain the humans' ability to decode visualization featuring some ``short comings'' such as various visual multiplexing phenomena (e.g., occlusion, displacement, and information omission) \cite{Chen:2014:CGF}. In the follow-on paper (see supplementary materials), we will explore means to measure such ability quantitatively. 

Similarly, we can define an alphabet for an pixel $\mathbb{R}_\text{pxl}$ and a composite alphabet for an image $\mathbb{R}_\text{img}$.
For the example in Figure \ref{fig:MIP}, we assume a simple MIP algorithm that selects the maximum voxel value along each ray, and assigns it as the corresponding pixel as an 8-bit monochromatic value.
It is obvious that the potential variations of a pixel is much less than the potential combined variations of all voxels along a ray.
Hence in terms of Shannon entropy, most likely
$\SE(\mathbb{D}_\text{vlm}) - \SE(\mathbb{D}_\text{img}) \ggg 0$, indicating significant information loss during the rendering process.

Given an analytical task to be performed through visualization, the analytical decision alphabet $\mathbb{A}$ usually contains a small number of letters, such as
$\{ \textit{contain artefact X}, \textit{no artefact X} \}$ or
$\{ \textit{big}, \textit{medium}, \textit{small}, \textit{tiny}, \textit{none} \}$.
The entropy of $\mathbb{A}$ is usually much lower than that of $\mathbb{D}_\text{img}$, i.e., 
$\SE(\mathbb{D}_\text{img}) - \SE(\mathbb{A}) \gg 0$,
indicating further information loss during perception and cognition.
As discussed in Section \ref{sec:Motivation}, this is referred to as \emph{alphabet compression} and is a general trend of all data intelligence workflows.
The question is thus about how much the analytical decision was disadvantaged by the loss of information.
This is referred to as \emph{potential distortion}.

In the original quantitative formula proposed in \cite{Chen:2016:TVCG}, the potential distortion is measured using Kullback–Leibler divergence (or KL-divergence) \cite{Kullback:1951:AMS}.
Given an alphabet $\mathbb{Z}$ with two PMFs $P$ and $Q$, KL-divergence measures how much $Q$ differs from the reference distribution $P$:
\begin{equation}\label{eq:DKL}
    \DKL(P(\mathbb{Z}) || Q(\mathbb{Z}))
    = \sum_{i=1}^n p_i \bigl( \log_2 p_i - \log_2 q_i \bigr)
    = \sum_{i=1}^n p_i \log_2 \frac{p_i}{q_i}    
\end{equation}
\noindent where $n = \| \mathbb{Z} \|$ is the number of letters in the alphabet $\mathbb{Z}$, and $p_i$ and $q_i$ are the probability values associated with letter $z_i \in \mathbb{Z}$.
$\DKL$ is also measured in $bit$.
Because $\DKL$ is an unbounded measure regardless the maximum entropy of $\mathbb{Z}$, it is easy to relate, quantitatively, the value of potential distortion and that of alphabet compression.
This leads the problem to be addressed in this paper and the follow-on paper in the supplementary materials.

\noindent Note: In this paper, to simplify the notations in different contexts, for an information-theoretic measure, we use an alphabet $\mathbb{Z}$ and its PMF $P$ interchangeably, e.g., $\SE(P(\mathbb{Z})) = \SE(P) = \SE(\mathbb{Z})$.
Appendix \ref{app:InfoTheory} provides more mathematical background about information theory, which may be helpful to some readers.

% ----------------------------------
\noindent\textbf{Problem Statement.} \quad
Recall our brief discussion about an analytical task that may be affected by the MIP image in Figure \ref{fig:Arteries} in Section \ref{sec:Motivation}.
Let us define the analytical task as binary options about whether the ``flat area'' is actually flat or curved.
In other words, it is an alphabet $\mathbb{A} = \{ \textit{curved}, \textit{flat} \}$.
The likelihood of the two options is represented by a probability distribution or probability mass function (PMF) $P(\mathbb{A}) = \{1-\epsilon, 0+\epsilon\}$, where $0<\epsilon<1$.
Since most arteries in the real world are of tubular shapes, one can imagine that a ground truth alphabet $\mathbb{A}_\text{G.T.}$ might have a PMF $P(\mathbb{A}_\text{G.T.})$ strongly in favor of the \textit{curved} option.
However, the visualization seems to suggest the opposite, implying a PMF $P(\mathbb{A}_\text{MIP})$ strongly in favor of the \textit{flat} option.
It is not difficult to interview some potential viewers, enquiring how they would answer the question.
One may estimate a PMF $P(\mathbb{A}_\text{doctors})$ from doctors' answers, and another $P(\mathbb{A}_\text{patients})$ from patients' answers.

\begin{table}[t]
\caption{Imaginary scenarios where probability data is collected for estimating knowledge related to alphabet
$\mathbb{A} = \{ \textit{curved}, \, \textit{flat} \}$.
The ground truth (G.T.) PMFs are defined with $\epsilon = 0.01$ and $0.0001$ respectively.
The potential distortion (as ``$\rightarrow$ value'') is computed using the KL-divergence.}
\label{tab:KL-ex1}
\centering
\begin{tabular}{@{}l@{\hspace{3mm}}l@{\hspace{4mm}}l@{}}
  & \textbf{Scenario 1} & \textbf{Scenario 2}\\
  \hline
  $Q(\mathbb{A}_\text{G.T.})$:
    & $\{0.99, 0.01\}$
    & $\{0.9999, 0.0001\}$ \\
  $P(\mathbb{A}_\text{MIP})$:
    & $\{0.01, 0.99\} \rightarrow 6.50$
    & $\{0.0001, 0.9999\} \rightarrow 13.28$\\
  $P(\mathbb{A}_\text{doctors})$:
    & $\{0.99, 0.01\} \rightarrow 0.00$
    & $\{0.99, 0.01\} \rightarrow 0.05$ \\
  $P(\mathbb{A}_\text{patients})$:
    & $\{0.7, 0.3\} \rightarrow 1.12$
    & $\{0.7, 0.3\} \rightarrow 3.11$\\
  \hline
\end{tabular}
\vspace{0mm}
\end{table}

\begin{table}[t]
\caption{Imaginary scenarios for estimating knowledge related to alphabet $\mathbb{B} = \{ \textit{wrinkles-and-bumps}, \, \textit{smooth} \}$.
The ground truth (G.T.) PMFs are defined with $\epsilon = 0.1$ and $0.001$ respectively.
The potential distortion (as ``$\rightarrow$ value'') is computed using the KL-divergence.}
\label{tab:KL-ex2}
\centering
\begin{tabular}{@{}l@{\hspace{7mm}}l@{\hspace{8mm}}l@{}}
  & \textbf{Scenario 3} & \textbf{Scenario 4}\\
  \hline
  $Q(\mathbb{B}_\text{G.T.})$:
    & $\{0.9, 0.1\}$
    & $\{0.001, 0.999\}$ \\
  $P(\mathbb{B}_\text{MIP})$:
    & $\{0.1, 0.9\} \rightarrow 2.54$
    & $\{0.001, 0.999\} \rightarrow 9.94$\\
  $P(\mathbb{B}_\text{doctors})$:
    & $\{0.8, 0.2\} \rightarrow 0.06$
    & $\{0.8, 0.2\} \rightarrow 1.27$ \\
  $P(\mathbb{B}_\text{patients})$:
    & $\{0.1, 0.9\} \rightarrow 2.54$
    & $\{0.1, 0.9\} \rightarrow 8.50$\\
  \hline
\end{tabular}
\vspace{-4mm}
\end{table}

Table \ref{tab:KL-ex1} shows two scenarios where different probability data is obtained. The values of PD are computed using the KL-divergence as proposed in \cite{Chen:2016:TVCG}.
In Scenario 1, without any knowledge, the visualization process would suffer 6.50 bits of potential distortion (PD).
As doctors are not fooled by the ``flat area'' shown in the MIP visualization, their knowledge is worth 6.50 bits.
Meanwhile, patients would suffer 1.12 bits of PD on average, their knowledge is worth $5.38 = 6.50 - 1.12$ bits.

In Scenario 2, the PMFs of $P(\mathbb{A}_\text{G.T.})$ and $P(\mathbb{A}_\text{MIP})$ depart further away, while $P(\mathbb{A}_\text{doctors})$ and $P(\mathbb{A}_\text{patients})$ remain the same.
Although doctors and patients would suffer more PD, their knowledge is worth more than that in Scenario 1 (i.e., $13.28 - 0.05 = 13.23$ bits and $13.28 - 3.11 = 10.17$ bits respectively).

Similarly, the binary options about whether the ``flat area'' is actually smooth or not can be defined by an alphabet
$\mathbb{A} = \{ \textit{wrinkles-and-bumps}, \, \textit{smooth} \}$.
Table \ref{tab:KL-ex2} shows two scenarios about collected probability data.
In these two scenarios, doctors exhibit much more knowledge than patients, indicating that the surface texture of arteries is a piece of specialized knowledge.

The above example demonstrates that using the KL-divergence to estimate PD can differentiate the knowledge variation between doctors and patients regarding the two pieces of knowledge that may reduce the distortion due to the ``flat area''.
When it is used in Eq.\, \ref{eq:CBM-1} in a relative or qualitative context (e.g., \cite{Chen:2019:TVCG,Chen:2019:CGF}), the unboundedness of the KL-divergence does not pose an issue.

However, this does become an issue when the KL-divergence is used to measure PD in an absolute and quantitative context.
From the two diverging PMFs $P(\mathbb{A}_\text{G.T.})$ and $P(\mathbb{A}_\text{MIP})$ in Table \ref{tab:KL-ex1}, or $P(\mathbb{B}_\text{G.T.})$ and $P(\mathbb{B}_\text{MIP})$ in Table \ref{tab:KL-ex2}, we can observe that the smaller $\epsilon$ is, the more divergent the two PMFs become and the higher value the PD has.
Indeed, consider an arbitrary alphabet $\mathbb{Z} = \{z_1, z_2 \}$, and two PMFs defined upon $\mathbb{Z}$: $P=[0+\epsilon, \, 1-\epsilon]$ and $Q=[1-\epsilon, \, 0+\epsilon]$.
When $\epsilon \rightarrow 0$, we have the KL-divergence $\DKL(Q||P) \rightarrow \infty$.

Meanwhile, the Shannon entropy of $\mathbb{Z}$, $\mathcal{H}(\mathbb{Z})$, has an upper bound of 1 bit.
It is thus not intuitive or practical to relate the value of $\DKL(Q||P)$ to that of $\mathcal{H}(\mathbb{Z})$.
Many applications of information theory do not relate these two types of values explicitly.
When reasoning such relations is required, the common approach is to impose a lower-bound threshold for $\epsilon$ (e.g., \cite{Kijmongkolchai:2017:CGF}).
However, there is yet a consistent method for defining such a threshold for various alphabets in different applications, while preventing a range of small or large values (i.e., $[0, \sigma)$ or $(1-\sigma, 1]$) in a PMF is often inconvenient in practice.
Indeed, for a binary alphabet with two arbitrary $P$ and $Q$, in order to restrict its $\DKL(P||Q) \leq 1$, one has to set $0.0658 \lesssim \sigma \lesssim 0.9342$, rendering some $13\%$ of the probability range [0, 1] unusable. 
In the following section, we discuss several approaches to defining a bounded measure for PD.

% =====================
\section{Bounded Measures for Potential Distortion (PD)} 
Let $\mathbf{P}_i$ be a process in a data intelligence workflow, $\mathbb{Z}_i$ be its input alphabet, and $\mathbb{Z}_{i+1}$ be its output alphabet.
$\mathbf{P}_i$ can be a human-centric process (e.g., visualization and interaction) or a machine-centric process (e.g., statistics and algorithms).
In the original proposal \cite{Chen:2016:TVCG}, the value of Benefit in Eq.\,\ref{eq:CBM-1} is measured using:
\begin{equation} \label{eq:CBM-2}
  \text{Benefit} = \text{AC} - \text{PD}
                 = \SE(\mathbb{Z}_i) - \SE(\mathbb{Z}_{i+1})
                 - \DKL(\mathbb{Z}'_i||\mathbb{Z}_i)
\end{equation}
\noindent where $\SE()$ is the Shannon entropy of an alphabet and $\DKL()$ is KL-divergence of an alphabet from a reference alphabet.
Because the Shannon entropy of an alphabet with a finite number of letters is bounded, AC, which is the entropic difference between the input and output alphabets, is also bounded.
On the other hand, as discussed in the previous section, PD is unbounded.
Although Eq.\,\ref{eq:CBM-2} can be used for relative comparison, it is not quite intuitive in an absolute context, and it is difficult to imagine that the amount of informative distortion can be more than the maximum amount of information available.

In this section, we present the unpublished work by Chen and Sbert \cite{Chen:2019:arXiv}, which shows mathematically that for alphabets of a finite size, the KL-divergence used in Eq.\,\ref{eq:CBM-2} should ideally be bounded.
In their arXiv report, they also outlined a new divergence measure and compare it with a few other bounded measures.
Building on the initial comparison in \cite{Chen:2019:arXiv}, we use visualization in Section \ref{sec:VisualAnalysis} to assist the multi-criteria analysis and selection of a bounded divergence measure to replace the KL-divergence used in Eq.\,\ref{eq:CBM-2}.
In the separate follow-on paper in the supplementary materials, we further examine the practical usability of a subset of bounded measures by evaluating them using synthetic and experimental data. 

% --------------------------------------------------
\subsection{A Conceptual Proof of Boundedness}
\label{sec:Proof}
According to the mathematical definition of $\DKL$ in Eq.\,\ref{eq:DKL}, $\DKL$ is of course unbounded.
We do not in anyway try to prove that this formula is bounded.
We are interested in a scenario where an alphabet $\mathbb{Z}$ is associated with two PMF, $P$ and $Q$, which is very much the scenario of measuring the penitential distortion in Eq.\,\ref{eq:CBM-1}. We ask a question: is \textbf{conceptually} necessary for $\DKL$ to yield a unbounded value to describe the divergence between $P$ and $Q$ in this scenario despite that $\SE(P)$ and $\SE(Q)$ are both bounded?  

We highlight the word ``conceptually'' because this relates to the \textbf{concept} about another information-theoretic measure, cross entropy, which is defined as:
\begin{equation} \label{eq:CE}
  \begin{split}
    \CE(P, Q) &= - \sum_{i=1}^n p_i \log_2 q_i = \sum_{i=1}^n p_i \log_2 \frac{1}{q_i}\\
    &=\sum_{i=1}^n p_i \log_2 \frac{p_i}{q_i} - \sum_{i=1}^n p_i \log_2 p_i\\
    &= \DKL(P||Q) + \SE(P)
  \end{split}
\end{equation}
Conceptually, cross entropy measures the cost of a coding scheme.
If a code (i.e., an alphabet $\mathbb{Z}$) has a true PMF $P$, the optimal coding scheme should require only $\SE(P)$ bits according to Shannon's source coding theorem \cite{Cover:2006:book}.
However, if the code designer mistakes the PMF as $Q$, the resulting coding scheme will have $\CE(P, Q)$ bits.
From Eq.\,\ref{eq:CE}, we can observe that the inefficiency is described by the term $\DKL(P||Q)$.
Naturally, we can translate our aforementioned question to: should such inefficiency be bounded if there is a finite number of codewords (letters) in the code (alphabet).

Coding theory has been applied to visualization, e.g., for explaining the efficiency of logarithmic plots in displaying data of a family of skewed PMFs and the usefulness of redundancy in visual design \cite{Chen:2010:TVCG}. Here we focus on proving that $\CE(P, Q)$ is conceptually bounded.    

Let $\mathbb{Z}$ be an alphabet with a finite number of letters, $\{ z_1, z_2, \ldots, z_n \}$, and $\mathbb{Z}$ is associated with a PMF, $Q$, such that: 
\begin{equation} \label{eq:CodeP}
\begin{split}
  q(z_n) &= \epsilon, \quad\text{(where $0 < \epsilon < 2^{-(n-1)}$}),\\
  q(z_{n-1}) &= (1-\epsilon)2^{-(n-1)},\\
  q(z_{n-2}) &= (1-\epsilon)2^{-(n-2)},\\
  &\cdots\\
  q(z_{2}) &= (1-\epsilon)2^{-2},\\
  q(z_{1}) &= (1-\epsilon)2^{-1} + (1-\epsilon)2^{-(n-1)}.
\end{split}
\end{equation}
When we encode this alphabet using an entropy binary coding scheme \cite{Moser:2012:book}, we can be assured to achieve an optimal code with the lowest average length for
codewords. One example of such a code for the above probability is:
\begin{equation} \label{eq:Code}
\begin{split}
  z_1&: 0, \qquad z_2: 10, \qquad z_3: 110\\
  &\cdots\\
  z_{n-1}&: 111\ldots10 \quad\text{(with $n-2$ ``1''s and one ``0'') }\\
  z_n&: 111\ldots11 \quad\text{(with $n-1$ ``1''s and no ``0'') }
\end{split}
\end{equation}
In this way, $z_n$, which has the smallest probability, will always be assigned a codeword with the maximal length of $n-1$.
Entropy coding is designed to minimize the average number of bits per letter when one transmits a ``very long'' sequence of letters in the alphabet over a communication channel.
Here the phrase ``very long'' implies that the string exhibits the above PMF $Q$ (Eq.\,\ref{eq:CodeP}).

Suppose that $\mathbb{Z}$ is actually of PMF $P$, but is encoded as Eq.\,\ref{eq:Code} based on $Q$.
The transmission of $\mathbb{Z}$ using this code will have inefficiency.
As mentioned above, the cost is measured by cross entropy $\CE(P, Q)$, and the inefficiency is measured by the term $\DKL(P||Q)$ in Eq.\,\ref{eq:CE}.
%
%\begin{equation} \label{eq:CE-2}
%   \CE(P, Q) = \SE(P) + \DKL(P||Q) = -\sum_{i=1}^n p_i \log_2 q_i
%\end{equation}

Clearly, the worst case is that the letter, $z_n$, which was encoded using $n-1$ bits, turns out to be the most frequently used letter in $P$ (instead of the least in $Q$).
It is so frequent that all letters in the long string are of $z_n$.
So the average codeword length per letter of this string is $n-1$.
The situation cannot be worse.
Therefore, $n-1$ is the upper bound of the cross entropy.
From Eq.\,\ref{eq:CE}, we can also observe that $\DKL(P||Q)$ must also be bounded since $\CE(P, Q)$ and $\SE(P)$ are both bounded as long as $\mathbb{Z}$ has a finite number of letters.
Let $\top_{\text{CE}}$ be the upper bound of $\CE(P, Q)$.
The upper bound for $\DKL(P||Q)$, $\top_{\text{KL}}$, is thus:
\begin{equation}\label{eq:CEandKL}
	\DKL(P||Q) = \CE(P, Q) - \SE(P) \le \top_{\text{\text{CE}}} - \min_{\forall P(\mathbb{Z})}\bigl( \SE(P) \bigr)
\end{equation}

There is a special case worth noting. In practice, it is common to
assume that $Q$ is a uniform distribution, i.e., $q_i = 1/n, \forall q_i \in Q$, typically because $Q$ is unknown or varies frequently.
Hence the assumption leads to a code with an average length equaling $\log_2 n$ (or in practice, the smallest integer $\ge \log_2 n$).
Under this special (but rather common) condition, all letters in a very long string have codewords of the same length.
The worst case is that all letters in the string turn out to the same letter. Since there is no informative variation in the PMF $P$ for this very long string, i.e., $\mathcal{H}(P) = 0$, in principle, the transmission of this string is unnecessary.
The maximal amount of inefficiency is thus $\log_2 n$.
This is indeed much lower than the upper bound $\top_{\text{CE}} = n-1$, justifying the assumption or use of a uniform $Q$ in many situations.

A more formal proof of the boundedness of $\CE(P, Q)$ and $\DKL(P||Q)$ for an alphabet with a finite number of letters can be found in Appendix \ref{app:Proof} with more detailed discussions.
It is necessary to note again that the discourse in this section and Appendix \ref{app:Proof} does not imply that the KL-divergence is incorrect.
Firstly, the KL-divergence applies to both discrete probability distributions (PMFs) and continuous distributions.
Secondly, the KL-divergence is one of the many divergence measures found in information theory, and a member of the huge collection of statistical distance or difference measures.
There is no simply answer as to which measure is correct and incorrect or which is better.
We therefore should not over-generalize the proof to undermine the general usefulness of the KL-divergence.

% --------------------------------------------------
\subsection{Existing Candidates of Bounded Measures}
\label{sec:OldMetric}
While numerical approximation may provide a bounded KL-divergence, it is not easy to determine the value of $\epsilon$ and it is difficult to ensure everyone to use the same $\epsilon$ for the same alphabet or comparable alphabets.
It is therefore desirable to consider bounded measures that may be used in place of $\DKL$.

Jensen-Shannon divergence is such a measure:
\begin{equation} \label{eq:DJS}
\begin{split}
    \DJS(P||Q) &= \frac{1}{2} \bigl( \DKL(P||M) + \DKL(Q||M) \bigr) = \DJS(Q||P)\\
    &= \frac{1}{2} \sum_{i=1}^n \biggl(p_i \log_2 \frac{2 p_i}{p_i + q_i} + q_i \log_2 \frac{2 q_i}{p_i + q_i} \biggr)
\end{split}
\end{equation}
\noindent where $P$ and $Q$ are two PMFs associated with the same alphabet $\mathbb{Z}$ and $M$ is the average distribution of $P$ and $Q$.
Each letter $z_i \in \mathbb{Z}$ is associated with a probability value $p_i \in P$ and another $q_i \in Q$. 
With the base 2 logarithm as in Eq.\,\ref{eq:DJS}, $\DJS(P||Q)$ is bounded by 0 and 1.

Another bounded measure is the conditional entropy $\SE(P|Q)$:
\begin{equation} \label{eq:CondSE}
\begin{split}
    \SE({P|Q}) &= \SE(P) - \MI(P;Q)\\
    &= \SE(P) - \sum_{i=1}^n\sum_{j=1}^n r_{i,j} \log_2 \frac{r_{i,j}}{p_i q_j}
\end{split}
\end{equation}
\noindent where $\MI(P;Q)$ is the mutual information between $P$ and $Q$ and $r_{i,j}$ is the joint probability of the two conditions of $z_i, z_j \in \mathbb{Z}$ that are associated with $P$ and $Q$.
$\SE(P|Q)$ is bounded by 0 and $\SE(P)$.
Because $\MI(P;Q)$ measures the amount of shared information between $P$ and $Q$ (and therefore a kind of similarity), $\SE({P|Q})$ thus increases if $P$ and $Q$ are less similar. Note that we use $\SE({P|Q})$ and $\MI(P;Q)$ here in the context that $P$ and $Q$ to be associated with the same alphabet $\mathbb{Z}$, though the general definitions of $\SE({P|Q})$ and $\MI(P;Q)$ are more flexible.

The above two measures in Eqs.\,\ref{eq:DJS} and \ref{eq:CondSE} consist of logarithmic scaling of probability values, in the same form of Shannon entropy.
They are entropic measures.
There are many other divergence measures in information theory, including many in the family of $f$-divergences \cite{Liese:2006:TIT}.
However, many are also unbounded.

Meanwhile, entropic divergence measures belong to the broader family of statistical distances or difference measures. In this work, we considered a set of non-entropic measures in the form of Minkowski distances, which have the following general form:
\begin{equation}
    D^k_{\text{M}}(P,Q) = \sqrt[\leftroot{2}\uproot{10} k]{\sum_{i=1}^n |p_i - q_i|^k}
    \quad (k > 0)
\end{equation}
\noindent where we use symbol $D$ instead of $\mathcal{D}$ because it is not entropic.

% ---------------
\begin{table*}[t]
  \centering
  \caption{A summary of multi-criteria decision analysis. Each measure is scored against a conceptual criterion using an integer in [0, 5] with 5 being the best.
  The symbol $\blacktriangleright$ indicates an interim conclusion after considering one or a few criteria.}
  \label{tab:MultiCriteria}
  \begin{tabular}{@{}l@{\hspace{3mm}}c@{\hspace{3mm}}c@{\hspace{3mm}}c@{\hspace{2mm}}c@{\hspace{2mm}}c@{\hspace{3mm}}c@{\hspace{3mm}}c@{\hspace{3mm}}c@{\hspace{3mm}}c@{\hspace{3mm}}c@{}}
  \textbf{Criteria} & \textbf{Importance}
  & $0.3\DKL$ & $\DJS$ & $\SE(P|Q)$
  & $\DnewA$ & $\DnewB$
  & $\DncmA$ & $\DncmB$
  & $D^{k=2}_{\text{M}}$ & $D^{k=200}_{\text{M}}$\\[0.5mm]
  \hline
  1. Boundedness & critical
        & 0 & 5 & 5 & 5 & 5 & 5 & 5 & 3 & 3 \\
        \multicolumn{11}{l}{$\blacktriangleright$
        \emph{$0.3\DKL$ is eliminated but used below only for comparison.
        The other scores are carried forward.}} \\
  \hline
  2. Number of PMFs & important
        & \textcolor{gray}{5} & 5 & 2 & 5 & 5 & 5 & 5 & 5 & 5\\ 
  3. Entropic measures & important
        & \textcolor{gray}{5} & 5 & 5 & 5 & 5 & 5 & 5 & 1 & 1 \\
  4. Curve shapes (Fig. \ref{fig:P1_0_1}) & helpful
        & \textcolor{gray}{5} & 5 & 1 & 2 & 4 & 2 & 4 & 3 & 3 \\
  5. Curve shapes (Fig. \ref{fig:NearZero}) & helpful
        & \textcolor{gray}{5} & 4 & 1 & 3 & 5 & 3 & 5 & 2 & 3\\[0.5mm]
  \multicolumn{2}{l}{$\blacktriangleright$
        \emph{Eliminate} \emph{$\SE(P|Q)$}, $\mathcal{D}^2_{\text{M}}$, $\mathcal{D}^{200}_{\text{M}}$ \emph{based on criteria 1-5}}
        & \textbf{sum:} & \textbf{24} & \textcolor{gray}{\textbf{14}}
        & \textbf{20} & \textbf{24} & \textbf{20} & \textbf{24}
        & \textcolor{gray}{\textbf{14}} & \textcolor{gray}{\textbf{15}}\\[0.5mm]
  \hline
%  6. Scenario: \emph{good} and \emph{bad} (Fig. \ref{fig:Compare-GoodBad}) & helpful
%        & $-$ & 3 & $-$ & 5 & 4 & 5 & 4 & $-$ & $-$\\
%  7. Scenario: A, B, C, D (Fig. \ref{fig:Compare-ABCD}) & helpful
%        & $-$ & 4 & $-$ & 5 & 3 & 2 & 1 & $-$ & $-$\\
%  8. Case Study 1 (Section \ref{sec:VolVis}) & important
%        & $-$ & 5 & $-$ & 1 & 5 & 5 & 5 & $-$ & $-$\\
%  9. Case Study 2: (Section \ref{sec:London}) & important
%        & $-$ & 3 & $-$ & 1 & 5 & 3 & 3 & $-$ & $-$\\[0.5mm]
%        \multicolumn{2}{l}{$\blacktriangleright$
%        \emph{$\DnewB$ has the highest score based on criteria 6-9 (1-9)}}
%        & \textbf{sum:} & \textcolor{gray}{\textbf{15}(39)} &
%        & \textcolor{gray}{\textbf{12}(32)} & \textbf{17}(41)
%        & \textcolor{gray}{\textbf{15}(35)} & \textcolor{gray}{\textbf{13}(37)}\\[0.5mm]
%   \hline
  \end{tabular}
  \vspace{-0mm}
\end{table*}
% -----------

% --------------------------------------------------
\subsection{New Candidates of Bounded Measures}
\label{sec:NewMetric}

For each letter $z_i \in \mathbb{Z}$, $\DKL(P||Q)$ measures the difference between its self-information $-\log_2 (p_i)$ and $-\log_2 (q_i)$ with respect to $P$ and $Q$. Similarly, $\DJS(P||Q)$ measures the difference of self-information with the involvement an average distribution $(P+Q)/2$.
Meanwhile, it will be interesting to consider the difference of two probability values, i.e., $|p_i - q_i|$, and the information content of the difference.
This would lead to measuring $\log_2 |p_i - q_i|$, which is unfortunately an unbounded term in $[-\infty, 0]$.

Let $u = |p_i - q_i|$, the function $\log_2 u^k + 1$ (where $k>0$) is an isomorphic transformation of $\log_2 u$. The former preserves all information of the latter, while offering a bounded measure in $[0, 1]$.
Although $\log_2 u^k+1$ and $\log_2 u$ are both monotonically increasing measures, they have different gradient functions, or visually, different shapes.
We thus introduce a power parameter $k$ to enable our investigation into different shapes.
The introduction of $k$ reflects the open-minded nature of this work.
It follows the same generalization approach as Minkowski distances and $\alpha$-divergences \cite{Erven:2014:TIT}, avoiding a fixation on their special cases such as the Euclidean distance or $\DKL$.

We first consider a commutative measure $\Dnew$:
\begin{equation} \label{eq:New}
    \Dnew(P||Q) = \frac{1}{2} \sum_{i=1}^n (p_i + q_i) \log_2 \bigl( |p_i - q_i|^k + 1 \bigr)
\end{equation}
\noindent where $k>0$. Because $0 \leq |p_i - q_i|^k \leq 1$, we have
\[
    \frac{1}{2} \sum_{i=1}^n (p_i + q_i) \log_2 (0+1)
    \leq \Dnew(P||Q) \leq \frac{1}{2} \sum_{i=1}^n (p_i + q_i) \log_2 (1+1)
\]
Since $\log_2 1=0$, $\log_2 2=1$, $\sum p_i = 1$, $\sum q_i = 1$, $\Dnew(P||Q)$ is thus bounded by 0 and 1.
The formulation of $\Dnew(P||Q)$ was derived from its non-commutative version: 
\begin{equation} \label{eq:NewA}
    \Dncm(P||Q) = \sum_{i=1}^n p_i \log_2 \bigl( |p_i - q_i|^k + 1 \bigr)
\end{equation}
which captures the non-commutative property of $\DKL$.
% Derivation of $\Dnew(P||Q)$ from $\Dncm(P||Q)$ is easier than that of $\DJS$ from $\DKL$ \cite{Lin:1991:TIT}.
In this work, we focus on two options of $\Dnew$ and $\Dncm(P||Q)$, i.e., when $k=1$ and $k=2$.

As $\DJS$, $\Dnew$, and $\Dncm$ are bounded by [0, 1], if any of them is selected to replace $\DKL$, Eq.\,\ref{eq:CBM-2} can be rewritten as 
\begin{equation} \label{eq:CBM-3}
  \text{Benefit} = \SE(\mathbb{Z}_i) - \SE(\mathbb{Z}_{i+1})
                 - \SE_{\text{max}}(\mathbb{Z}_i) \mathcal{D}(\mathbb{Z}'_i||\mathbb{Z}_i)
\end{equation}
\noindent where $\SE_{\text{max}}$ denotes maximum entropy, while $\mathcal{D}$ is a placeholder for $\DJS$, $\Dnew$, or $\Dncm$.

We have considered the option of using $\SE(\mathbb{Z}_i)$ instead of $\SE_{\text{max}}(\mathbb{Z}_i)$. However, this would lead to an undesirable paradox.
Consider an alphabet $\mathbb{Z}_i=\{z_a, z_b\}$ with a PMF $P_i = \{p_a, 1-p_a\}$.
Consider a simple visual mapping that is supposed to encode the probability value $p_a$ using the luminance of a monochrome shape with, $\text{luminance}(p_a) = p_a$, black = 0, and white = 1. Unfortunately, the accompanying legend displays incorrect labels as black for $p_a = 1$ and while for $p_a = 0$.
The visualization results thus feature a ``lie'' distribution $P_i = \{1-p_a, p_a\}$.
An obvious paradoxical scenario is when $P_i = \{1, 0\}$, which has an entropy value $\SE(\mathbb{Z}_i) = 0$.
Although $\DJS$, $\Dnew$, and $\Dncm$ would all return 1 as the maximum value of divergence for the visual mapping, the term $\SE(\mathbb{Z}_i) \mathcal{D}(\mathbb{Z}'_i||\mathbb{Z}_i)$ would indicate that there would be no divergence. Hence $\SE(\mathbb{Z}_i)$ cannot be used instead of $\SE_{\text{max}}(\mathbb{Z}_i)$.

% --------------------
\begin{figure*}[t]
\centering
\begin{tabular}{@{}c@{\hspace{2mm}}c@{\hspace{2mm}}c@{\hspace{2mm}}c@{}}
    \includegraphics[width=42.5mm]{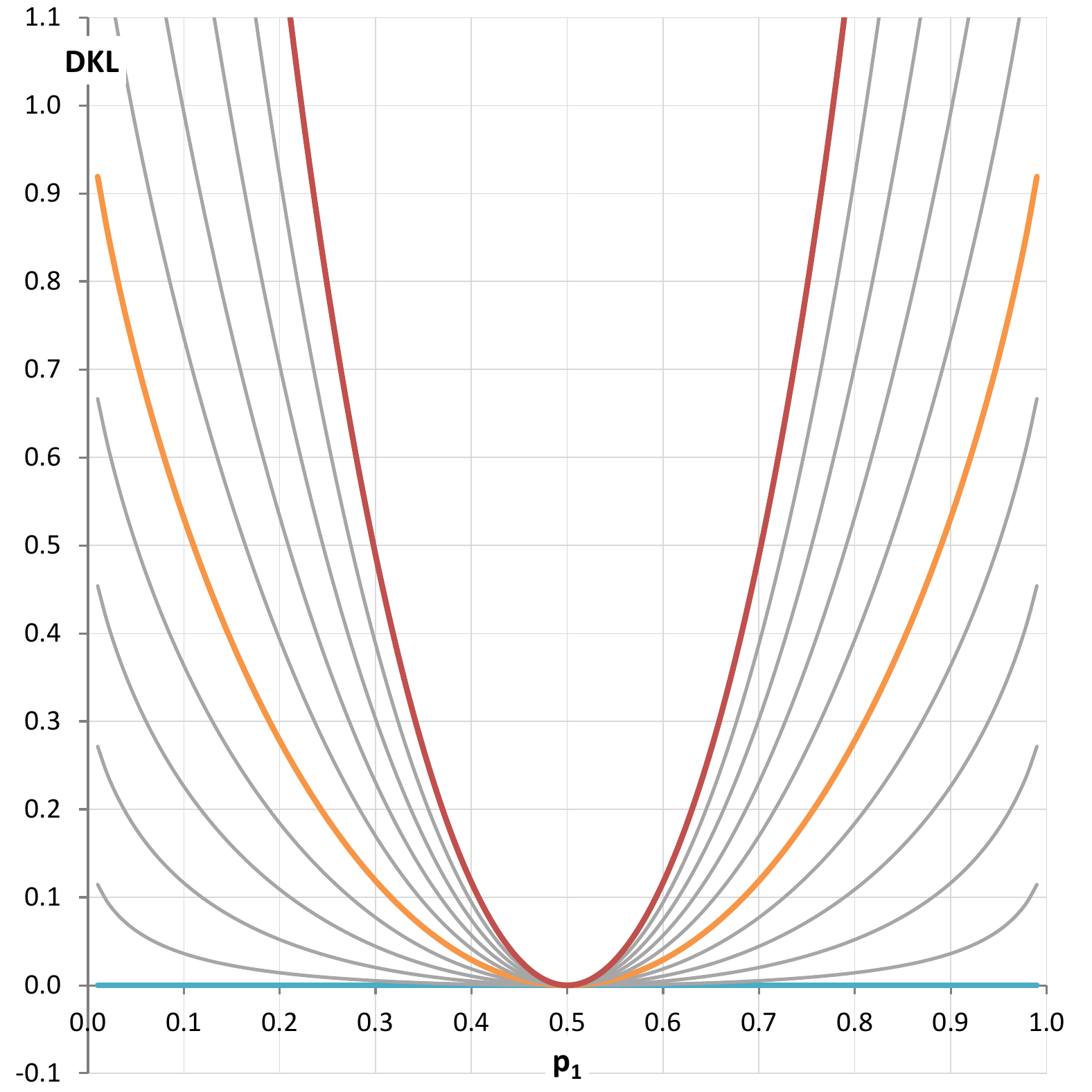} &
    \includegraphics[width=42.5mm]{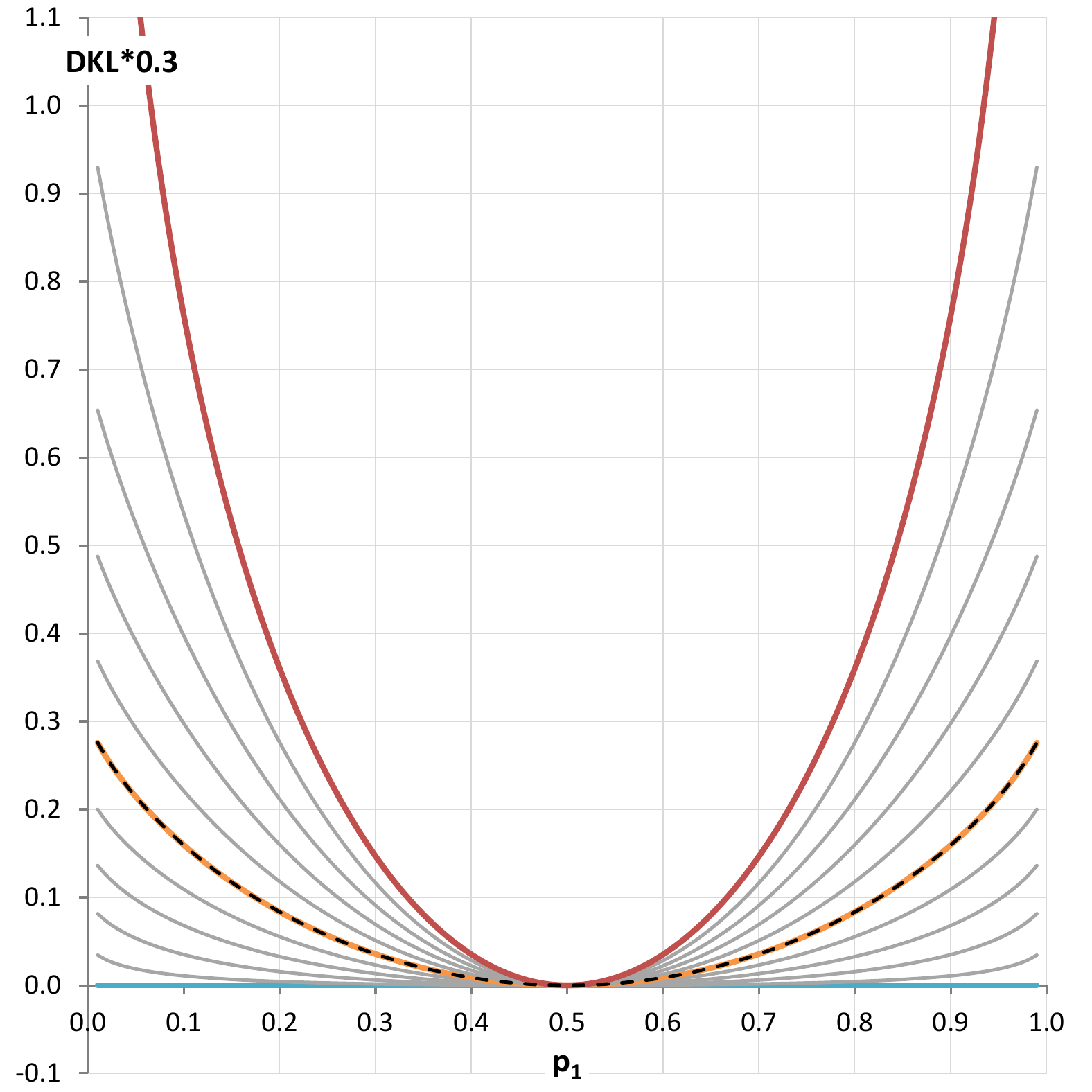} &
    \includegraphics[width=42.5mm]{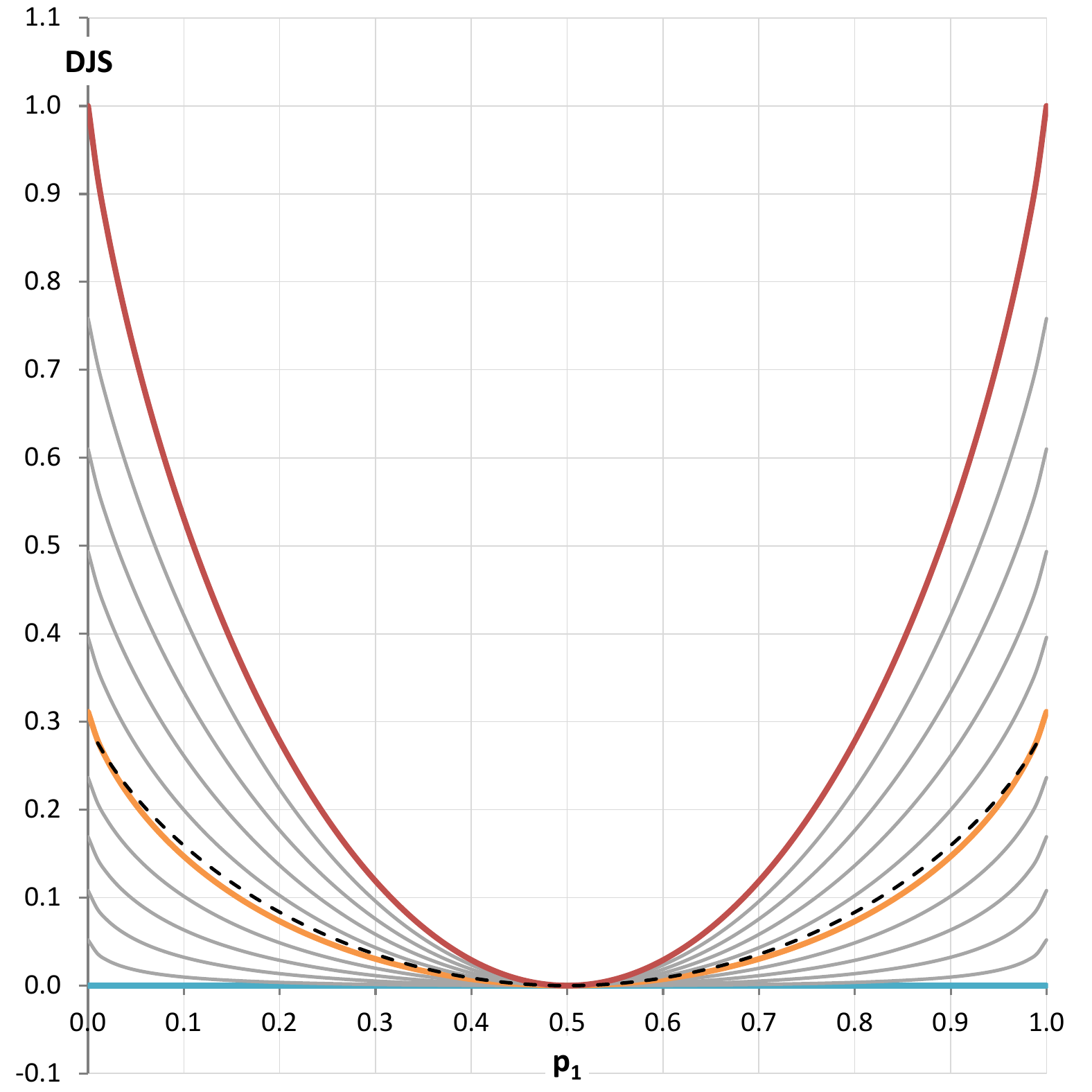} &
    \includegraphics[width=42.5mm]{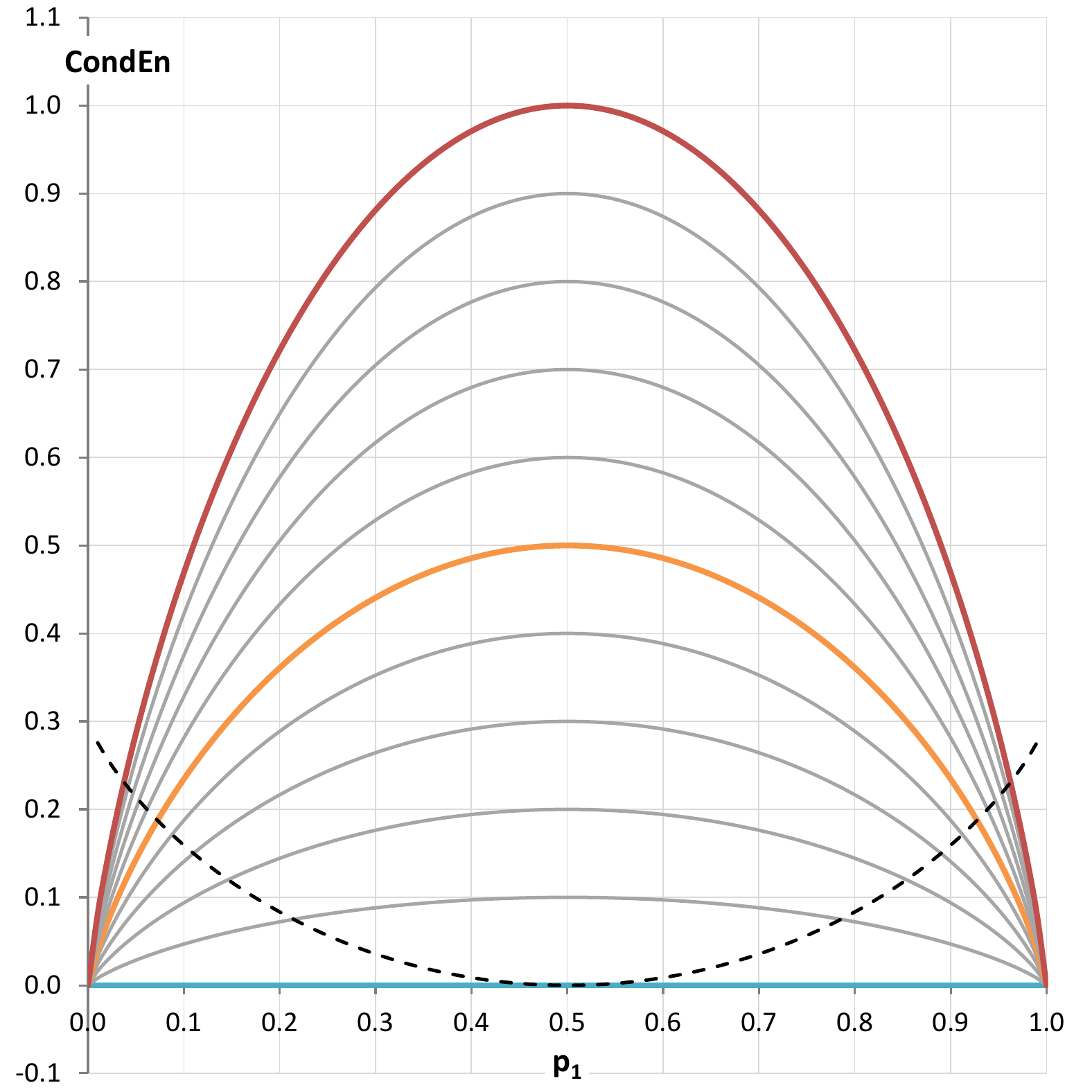} \\
    (a) $\DKL(P||Q)$ & (b) $0.3\DKL(P||Q)$ & (c) $\DJS(P||Q) $ & (d) $\SE(P|Q)$\\[2mm]
    \includegraphics[width=42.5mm]{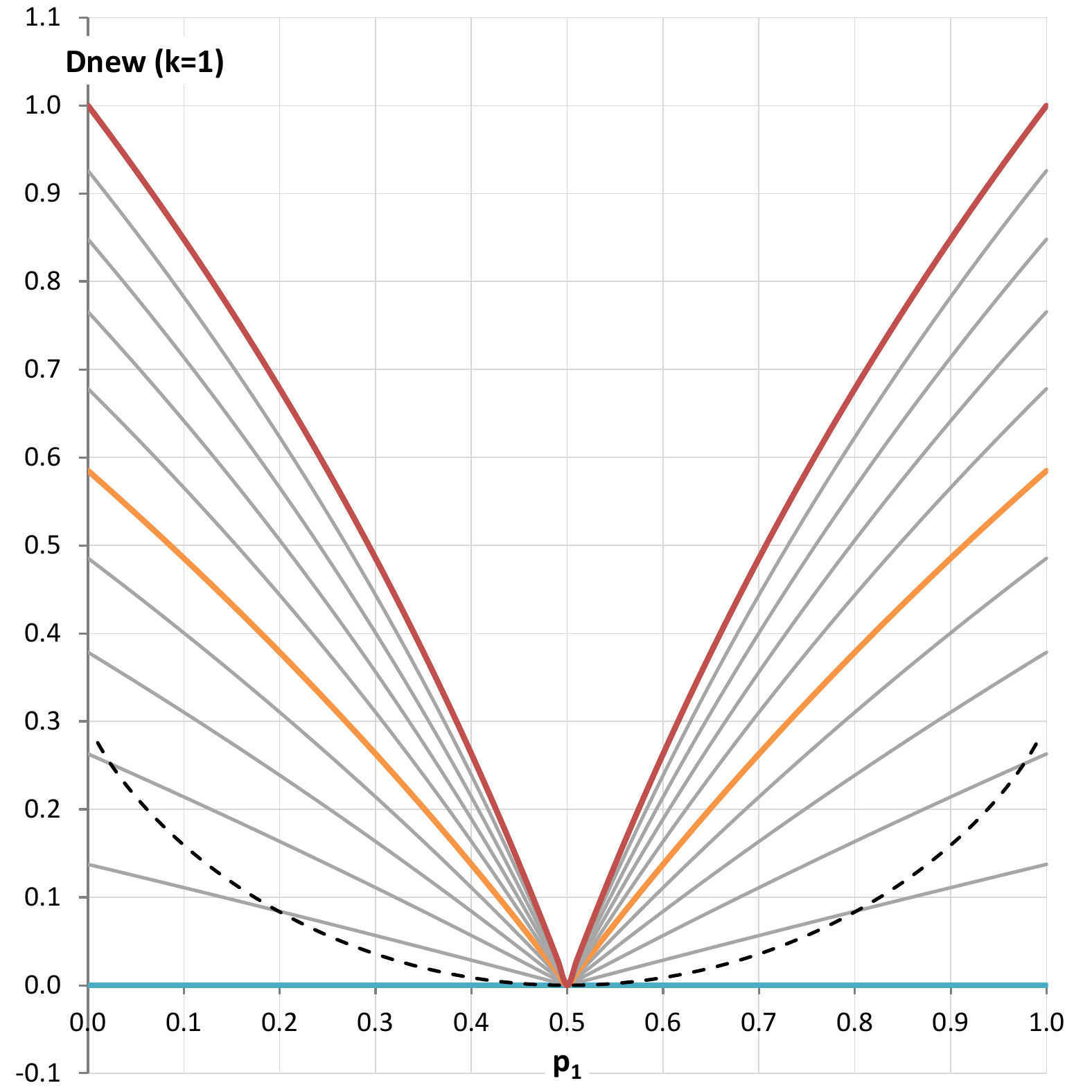} &
    \includegraphics[width=42.5mm]{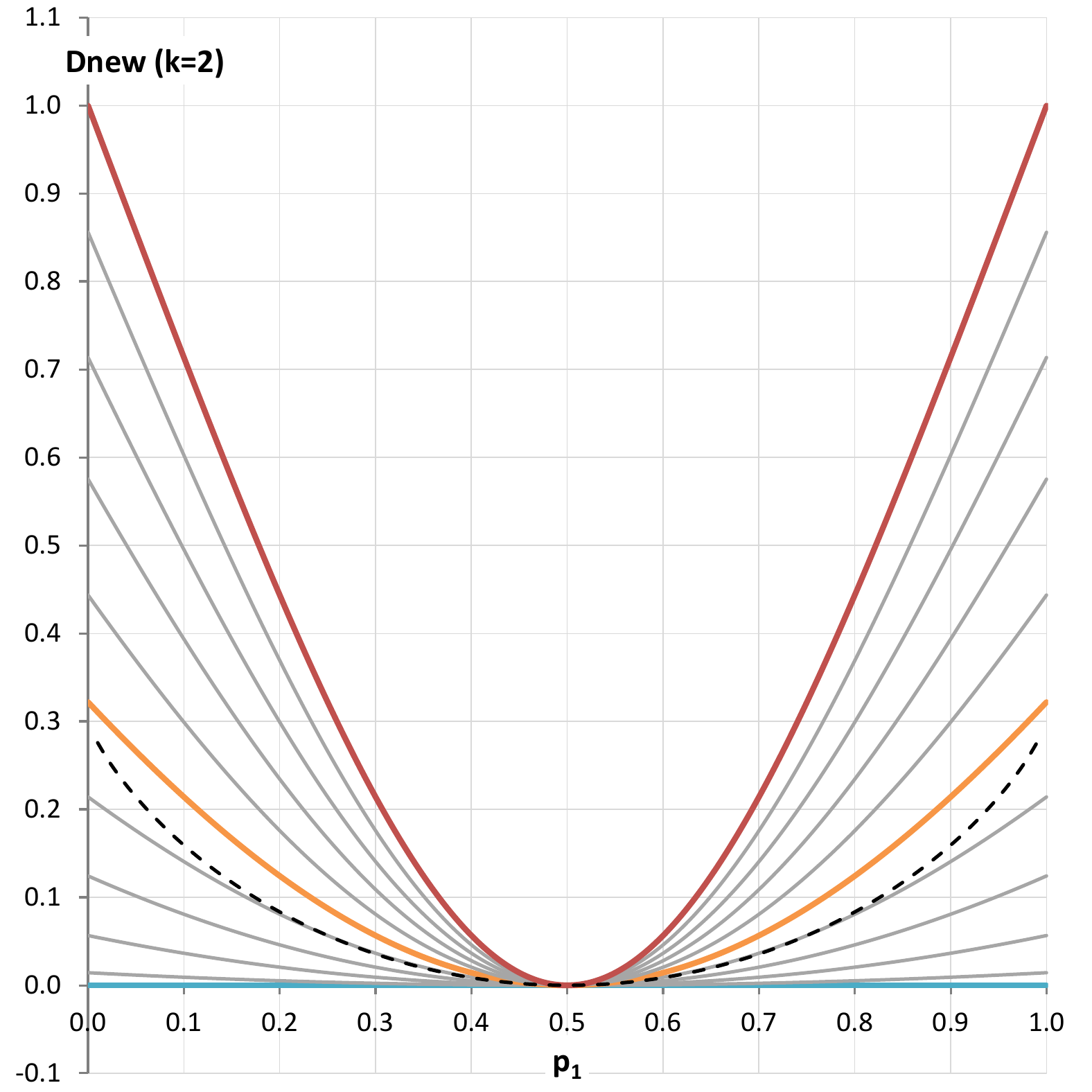} &
    \includegraphics[width=42.5mm]{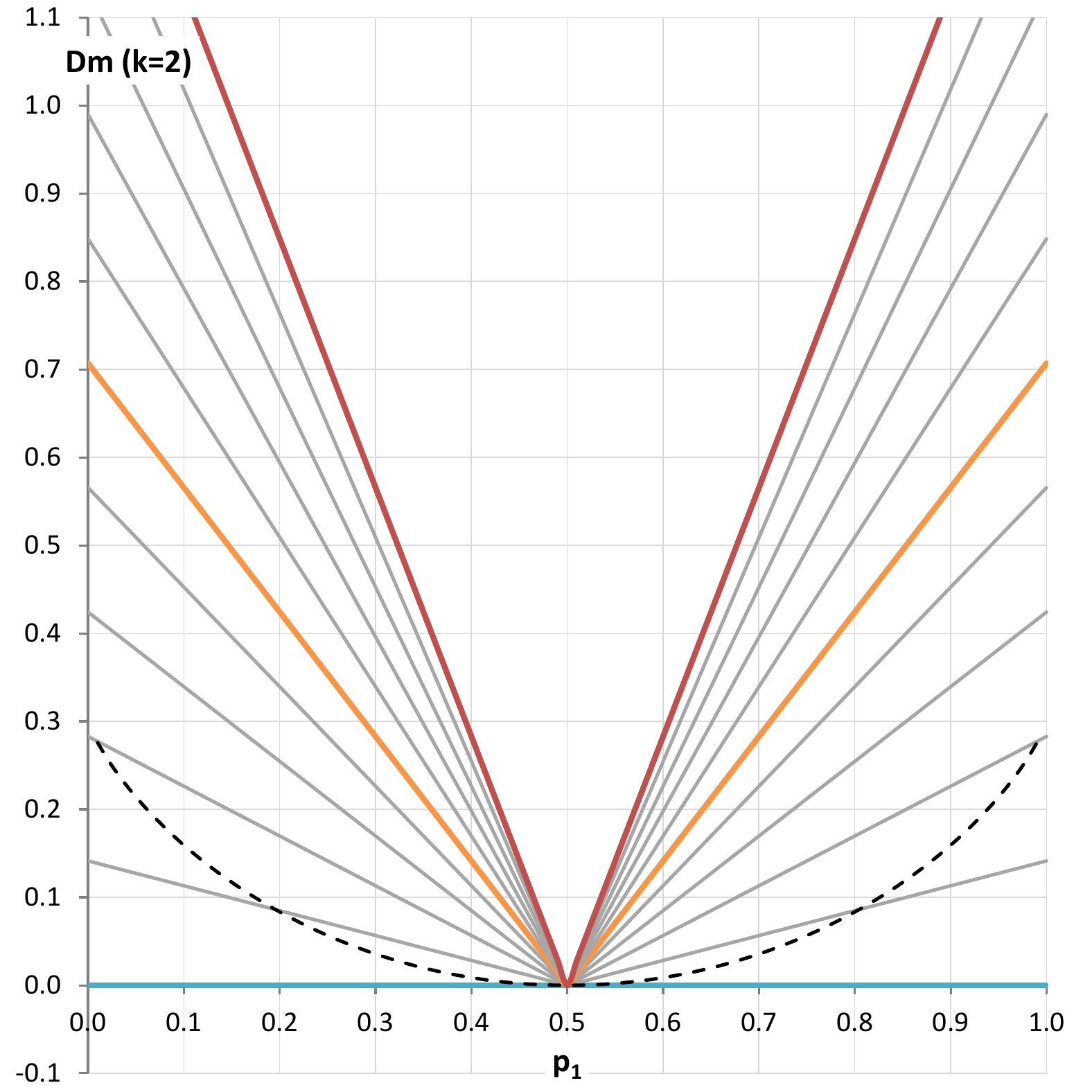} &
    \includegraphics[width=42.5mm]{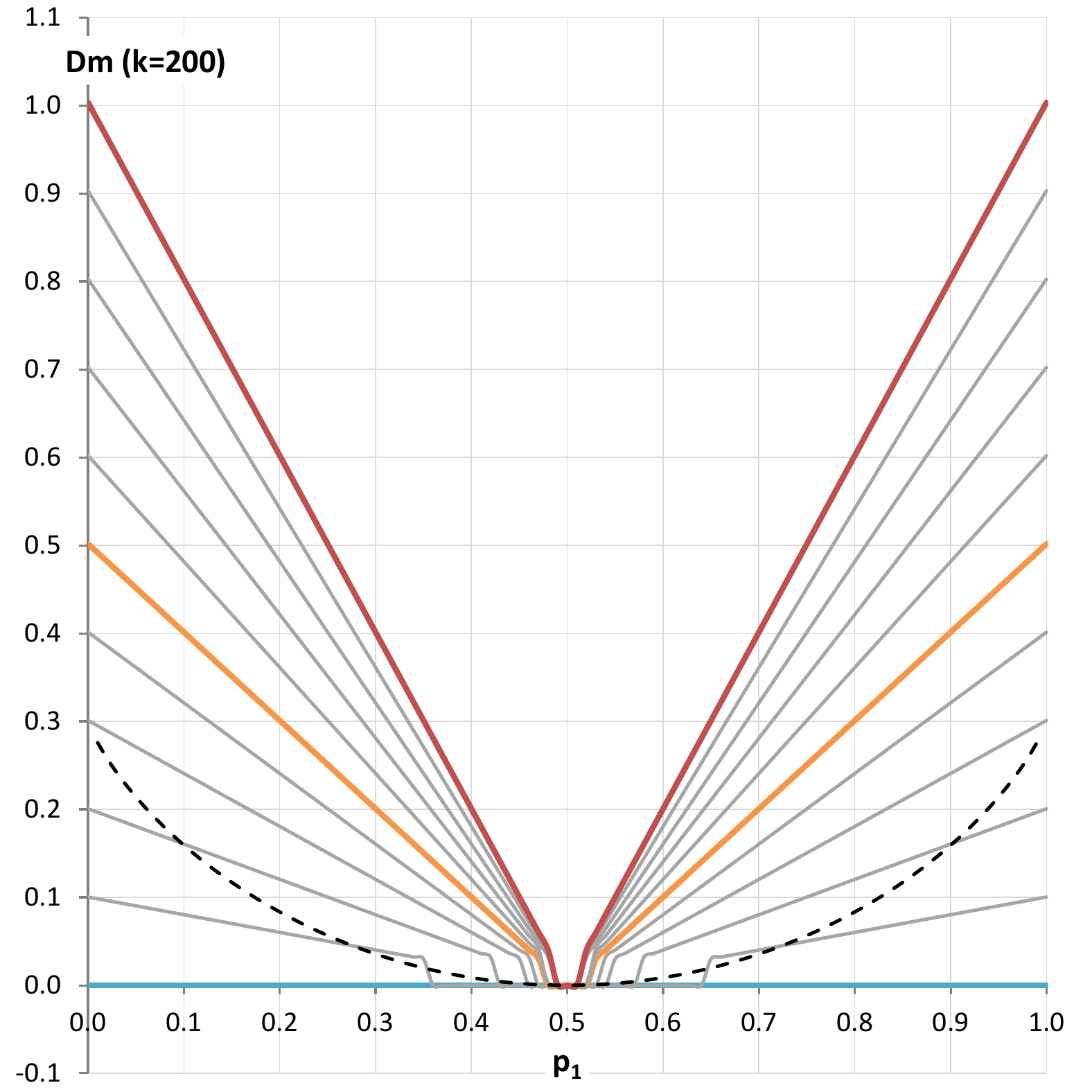} \\
    (e) $\Dnew(P||Q)$, $k=1$ & (f) $\Dnew(P||Q)$, $k=2$ &
    (g) $D^k_\text{M}(P,Q)$, $k=2$ & (h) $D^k_\text{M}(P,Q)$, $k=200$ \\[2mm]
\end{tabular}
\includegraphics[width=176mm]{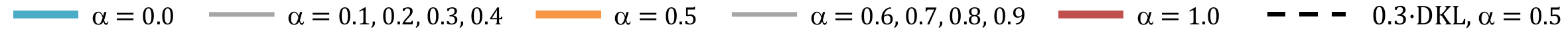}
\caption{The different measurements of the divergence of two PMFs, $P=\{ p_1, 1-p_1 \}$ and $Q=\{ q_1, 1-q_1 \}$. The $x$-axis shows $p_1$, varying from 0 to 1, while we set $q_1 = (1-\alpha)p_1 + \alpha(1-p_1), \alpha \in [0, 1]$. When $\alpha = 1$, $Q$ is most divergent away from $P$.}
\label{fig:P1_0_1}
\vspace{-4mm}
\end{figure*}

% ===================================================
\section{Comparing Bounded Measures: Conceptual Evaluation}
\label{sec:VisualAnalysis}
Given those bounded candidates in the previous section, we would like to select the most suitable measure to be used in Eq.\,\ref{eq:CBM-3}.
In the history of measurement science \cite{Klein:2012:book}, there have been an enormous amount of research effort devoted to inventing, evaluating, and selecting different candidate measures (e.g., metric vs. imperial measurement systems; temperatute scales: Celsius, Fahrenheit, kelvin, Rankine, and Reaumur; and Seismic magnitude scales: Richter, Mercalli, moment magnitude, and many others).
There is usually no ground truth as to which is correct, and the selection decision is rarely determined only by mathematical definitions or rules \cite{Pedhazur:1991:book}.
Similarly, there are numerous statistical distance and difference measures.
selecting a measure in a certain application is often an informed decision based on multiple factors. 

Measuring the benefit of visualization and the related informative divergence in visualization processes is a new topic in the field of visualization. It is not unreasonable to expect that more research effort will be made in the coming years, decades, or unsurprisingly, centuries.
The work presented in this paper and the follow-on paper in the supplementary materials represents the early thought and early effort in this endeavor.  
In this work, we devised a set of criteria and conducted multi-criteria decision analysis (MCDA) \cite{Ishizaka:2013:book} to evaluate the candidate measures described in the previous section.

Our criteria fall into two main categories. The first group of criteria reflect five desirable conceptual or mathematical properties, as shown in Table \ref{tab:MultiCriteria}.
The second criteria reflect the assessments based on numerical instances constructed synthetically or obtained from experiments.
This paper focuses conceptual evaluation based on the first group of criteria, while the follow-on paper in the supplementary materials focuses on empirical evaluation based on the second group criteria.

For criteria 1, 4, and 5 in the first group, we use visualization plots to aid our analysis of the mathematical properties.
Based on our analysis, we score each divergence measure against a criterion using ordinal values between 0 and 5 (0 unacceptable, 1 fall-short, 2 inadequate, 3 mediocre, 4 good, 5 best).
We intentionally do not assign weights to these criteria.
While we will offer our view as to the importance of different criteria, we encourage readers to apply their own judgement to weight these criteria.
We hope that readers will reach the same conclusion as ours.
We draw our conclusion about the conceptual evaluation in Section \ref{sec:Conclusions}, where we also outline the need for empirical evaluation.

% while for criteria 8 and 9, we use visualization applications to provide instances of using divergence measures.
% We detail our analysis of criteria 1-7 in this section, and that of criteria 8 and 9 in Section \ref{sec:CaseStudies}.

\textbf{Criterion 1.}
This is essential since the selected divergence measure is to be bounded.
Otherwise we could just use the KL-divergence.
Let us consider a simple alphabet $\mathbb{Z} = \{z_1, z_2 \}$, which is associated with two PMFs, $P=\{ p_1, 1-p_1 \}$ and $Q=\{ q_1, 1-q_1 \}$.
We set $q_1 = (1-\alpha)p_1 + \alpha(1-p_1), \alpha \in [0, 1]$, such that when $\alpha = 1$, $Q$ is most divergent away from $P$.
The entropy values of $P$ and $Q$ fall into the range of [0, 1].
Hence semantically, it is more intuitive to reason an unsigned value representing their divergence within the same range.

Figure \ref{fig:P1_0_1} shows several measures by varying the values of $p_1$ in the range of $[0, 1]$.
We can obverse that $\DKL$ raises its values quickly above 1 when $\alpha = 1, p_1 \leq 0.22$.
Its scaled version, $0.3\DKL$, does not rise up as quick as $\DKL$ but raises above 1 when $\alpha = 1, p_1 \leq 0.18$.
In fact $\DKL$ and $0.3\DKL$ are not only unbounded, they do not return valid values  when $p_1 = 0$ or $p_1 = 1$.
We therefore score them 0 for Criterion 1.

$\DJS$, $\SE(P|Q)$, $\Dnew$, and $\Dncm$ are all bounded by [0, 1], and
semantically intuitive.
We score them 5.
Although $\DM$ is a bounded measure, its semantic interpretation is not ideal, because its upper bound depends on $k$ and is always $> 1$.
We thus score it 3.
Although $0.3\DKL$ is eliminated based on criterion 1, it is kept in Table \ref{tab:MultiCriteria} as a benchmark in analyzing criteria 2-5.
Meanwhile, we carry all other scores forward to the next stage of analysis.

% We can visualize how different measures numerically convey the divergence between $P$ and $Q$ by observing their relationship with $0.3\DKL$.

\textbf{Criterion 2.}
For criteria 2-5, we follow the base-criterion method \cite{Haseli:2020:IJMSEM} by considering 
$\DKL$ and $0.3\DKL$ as the benchmark.
Criterion 2 concerns the number of PMFs as the input variables of each measure.
$\DKL$ and $0.3\DKL$ depend on two PMFs, $P$ and $Q$.
All candidates of the bounded measures depend on two PMFs, except the conditional entropy $\SE(P|Q)$ that depends on three.
Because it requires some effort to obtain a PMF, especially a joint probability distribution, this makes $\SE(P|Q)$ less favourable and it is scored 2.

\textbf{Criterion 3.}
In addition, we prefer to have an entropic measure as it is more compatible with the measure of alphabet compression as well as $\DKL$ that is to be replaced.
For this reason, $\DM$ is scored 1.

\textbf{Criterion 4.}
One may wish for a bounded measure to have a geometric behaviour similar to $\DKL$ since it is the most popular divergence measure.
Since $\DKL$ rises up far too quickly as shown in Figure \ref{fig:P1_0_1}, we use $0.3\DKL$ as a benchmark, though it is still unbounded.
As Figure \ref{fig:P1_0_1} plots the curves for $\alpha = 0.0, 0.1, \ldots, 1.0$, we can visualize the ``geometric shape'' of each bounded measure, and compare it with that of $0.3\DKL$.  

From Figure \ref{fig:P1_0_1}, we can observe that $\DJS$ has almost a perfect match when $\alpha = 0.5$, while $\Dnew (k=2)$ is also fairly close.
They thus score 5 and 4 respectively in Table \ref{tab:MultiCriteria}.
Meanwhile, the lines of $\SE(P|Q)$ curve in the opposite direction of $0.3\DKL$.
We score it 1.
$\Dnew (k=1)$ and $\DM (k=2, k=200)$ are of similar shapes, with $\DM$ correlating with $0.3\DKL$ slightly better.
We thus score $\Dnew (k=1)$ 2 and $\DM (k=2, k=200)$ 3.
For the PMFs $P$ and $Q$ concerned, $\Dncm$ has the same curves as $\Dnew$.
Hence $\Dncm$ has the same score as $\Dnew$ in Table \ref{tab:MultiCriteria}.

\begin{figure}[t]
  \centering
  \includegraphics[width=\linewidth]{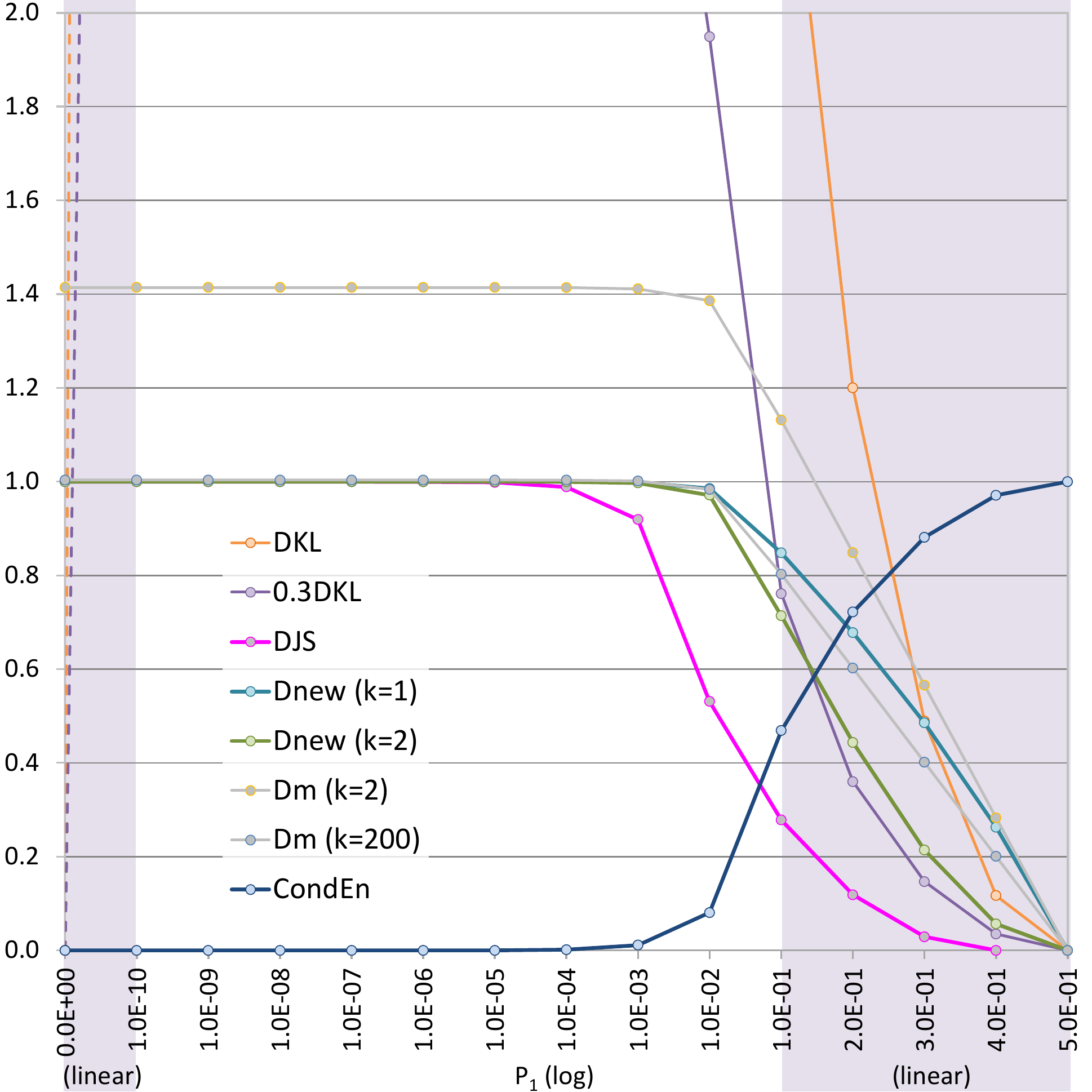}
  \caption{A visual comparison of the candidate measures in a range near zero. Similar to Figure \ref{fig:P1_0_1}, $P=\{ p_1, 1-p_1 \}$ and $Q=\{ q_1, 1-q_1 \}$, but only the curve $\alpha=1$ is shown, i.e., $q_1 = 1-p_1$. The line segments of $\DKL$ and $0.3\DKL$ in the range $[0, 0.1^{10}]$ do not represent the actual curves. The ranges $[0, 0.1^{10}]$ and $[0.1, 0.5]$ are only for references to the nearby contexts as they do not use the same logarithmic scale as in $[0.1^{10}, 0.1]$.}%
  \label{fig:NearZero}
  \vspace{-4mm}
\end{figure}

\textbf{Criterion 5.} We now consider Figure \ref{fig:NearZero}, where the candidate measures are visualized in comparison with $\DKL$ and $0.3\DKL$ in a range close to zero, i.e., $[0.1^{10}, 0.1]$.
The ranges $[0, 0.1^{10}]$ and $[0.1, 0.5]$ are there only for references to the nearby contexts as they do not have the same logarithmic scale as that in the range
$[0.1^{10}, 0.1]$.
We can observe that in $[0.1^{10}, 0.1]$ the curve of $0.3\DKL$ rises
as almost quickly as $\DKL$.
This confirms that simply scaling the KL-divergence is not an adequate solution.
The curves of $\DnewA$ and $\DnewB$ converge to their maximum value 1.0 earlier than that of $\DJS$.
If the curve of $0.3\DKL$ is used as a benchmark as in Figure \ref{fig:P1_0_1}, the curve of $\DnewB$ is closer to $0.3\DKL$ than that of $\DJS$.
We thus score $\DnewB$: 5, $\DJS$: 4, $\DnewA$: 3, $\DM (k=200)$: 3, $\DM (k=200)$: 2, and $\SE(P|Q)$: 1.
Same as Figure \ref{fig:P1_0_1}, $\Dncm$ has the same curves and thus the same score as $\Dnew$.

The sums of the scores for criteria 1-5 indicate that $\SE(P|Q)$ and $\DM$ are much less favourable than $\DJS$, $\Dnew$, and $\Dncm$.
Because these criteria have more holistic significance than the instance-based analysis for criteria 6-9, we can eliminate $\SE(P|Q)$ and $\DM$ for further consideration.
Ordinal scores in MCDA are typically subjective.
Nevertheless, in our analysis, $\pm 1$ in those scores would not affect the elimination. 

% ====================
\section{Discussions and Conclusions}
\label{sec:Conclusions}
%
% \textcolor{red}{To revise completely.}

In this paper, we have considered the need to improve the mathematical formulation of an information-theoretic measure for analyzing the cost-benefit of visualization as well as other processes in a data intelligence workflow \cite{Chen:2016:TVCG}.
The concern about the original measure is its unbounded term based on the KL-divergence.
As discussed in the early sections of this paper, although using the KL-divergence measure in \cite{Chen:2016:TVCG} as part of the cost-benefit measure is a conventional or orthodox choice, its unboundedness leads to several issues in the potential applications of the cost-benefit measure to practical problems:
\begin{itemize}
    \item It is not intuitive to interpret a set of values that would indicate that the amount of distortion in viewing a visualization that features some information loss, could be much more than the total amount of information contained in the visualization.
    
    \item It is difficult to specify some simple visualization phenomena. For example, before a viewer observes a variable $x$ using visualization, the viewer incorrectly assumes that the variable is a constant (e.g., $x \equiv 10$, and probability $p(10) = 1$). The KL-divergence cannot measure the potential distortion of this phenomenon of bias because this is a singularity condition, unless one changes $p(10)$ by subtracting a small value $0<\epsilon<1$.   
    
    \item If one tries to restrict the KL-divergence to return values within a bounded range, e.g., determined by the maximum entropy of the visualization space or the underlying data space, one could potentially lose a non-trivial portion of the probability range (e.g., $13\%$ in the case of a binary alphabet).
\end{itemize}

To address these problems, we have proposed to replace the KL-divergence in the cost-benefit measure with a bounded measure. 
We have obtained a proof that the divergence used in the cost-benefit formula is conceptually bounded, as long as the input and output alphabets of a process have a finite number of letters.

We have considered a number of bounded measures to replace the unbounded term, including a new divergence measure $\Dnew$ and its variant $\Dncm$.
We have conducted multi-criteria decision analysis to select the best measure among these candidates.
In particular, we have used visualization to aid the observation of the mathematical properties of the candidate measures, assisting in the analysis of three criteria in considered in this paper.

From Table \ref{tab:MultiCriteria}, we can observe the process of narrowing down from eight candidate measures to five measures.
In particular, three candidate measures $\DJS$, $\Dnew (k=2)$, and $\Dncm (k=2)$ received the same total scores. It is not easy to separate them.

In the history of measurement science \cite{Klein:2012:book}, scientists encountered many similar dilemma in choosing different measures. For example, temperature measures Celsius, Fahrenheit, R\'{e}aumur, R{\o}mer, and Delisle scales exhibit similar mathematical properties, their proposal and adoption were largely determined by practical instances:
\begin{itemize}
    \item Fahrenheit (original) --- 0 degree: the freezing point of brine (a high-concentration solution of salt in water), 32 degree: ice water, 96 degree: average human body temperature; 
    \item Fahrenheit (present) --- 32 degree: the freezing point of water, 212 degree: the boiling point of water;
    \item R\'{e}aumur --- 0 degree: the freezing point of water, 80 degree: the boiling point of water;
    \item R{\o}mer --- 7.5 degree: the freezing point of water, 60 degree: the boiling point of water;
    \item Celsius (original) --- 0 degree: the boiling point of water, 100 degree: the freezing point of water;
    \item Celsius (1744--1954) --- 0 degree: the freezing point of water, 100 degree: the boiling point of water;
    \item Celsius (1954--2019) --- redefined based on absolute zero and the triple point of VSMOW (specially prepared water);
    \item Celsius (2019--now) ---- redefined based on the Boltzmann constant.
    \item Delisle --- 0 degree: the boiling point of water, $-1$ degree; the contraction of the mercury in hundred-thousandths.
\end{itemize}

In addition, Newton proposed two temperatute systems based on his observation of some 18 instance values \cite{NewtonScale:2020:wiki}.

The effort of these scientific pioneers suggested that observing how candidate measures relate to practical instances was part of the scientific processes for selecting different candidate measures. Therefore, building on the work presented in this paper, we carried our further investigation into a group of criteria based on observed instances in synthetic and experimental data.
This empirical evaluation is reported in a separate follow-on paper in the supplementary materials, where we narrow the remaining five candidate measures to one measure, and propose to the original cost-benefit ratio in \cite{Chen:2016:TVCG} based on the combined conclusion derived from the conceptual evaluation and empirical evaluation.

% This cost-benefit measure was developed in the field of visualization, for optimizing visualization processes and visual analytics workflows.
% It is now being improved by using visual analysis and with the survey data collected in the context of visualization applications.
% We would like to continue our theoretical investigation into the mathematical properties of the new divergence measure.
% Meanwhile, having a bounded cost-benefit measure offers many new opportunities of developing tools for aiding the measurement and using such tools in practical applications, especially in visualization and visual analytics.

% The authors wish to thank A, B, and C. This work was supported
% in part by a grant from XYZ (\# 12345-67890).}

%\bibliographystyle{abbrv}
% \bibliographystyle{abbrv-doi}
% \bibliographystyle{abbrv-doi-narrow}
% \bibliographystyle{abbrv-doi-hyperref}
% \bibliographystyle{abbrv-doi-hyperref-narrow}

\bibliographystyle{eg-alpha-doi}  
\bibliography{EstimatePD}

\newcommand{\etalchar}[1]{$^{#1}$}
\begin{thebibliography}{\uppercase{PAJKW08}}

\bibitem[BBB{\etalchar{*}}12]{Bramon:2012:TVCG}
\textsc{Bramon R., Boada I., Bardera A., Rodr\'{i}guez Q., Feixas M., Puig J.,
  Sbert M.}:
\newblock Multimodal data fusion based on mutual information.
\newblock \emph{IEEE Transactions on Visualization and Computer Graphics 18}, 9
  (2012), 1574--1587.

\bibitem[BDSW13]{Biswas:2013:TVCG}
\textsc{Biswas A., Dutta S., Shen H.-W., Woodring J.}:
\newblock An information-aware framework for exploring multivariate data sets.
\newblock \emph{IEEE Transactions on Visualization and Computer Graphics 19},
  12 (2013), 2683--2692.

\bibitem[BM10]{Bruckner:2010:CGF}
\textsc{Bruckner S., M\"{o}ller T.}:
\newblock Isosurface similarity maps.
\newblock \emph{Computer Graphics Forum 29}, 3 (2010), 773--782.

\bibitem[BRB{\etalchar{*}}13a]{Bramon:2013:CGF}
\textsc{Bramon R., Ruiz M., Bardera A., Boada I., Feixas M., Sbert M.}:
\newblock An information-theoretic observation channel for volume
  visualization.
\newblock \emph{Computer Graphics Forum 32}, 3pt4 (2013), 411--420.

\bibitem[BRB{\etalchar{*}}13b]{Bramon:2013:JBHI}
\textsc{Bramon R., Ruiz M., Bardera A., Boada I., Feixas M., Sbert M.}:
\newblock Information theory-based automatic multimodal transfer function
  design.
\newblock \emph{IEEE Journal of Biomedical and Health Informatics 17}, 4
  (2013), 870--880.

\bibitem[BS05]{Bordoloi:2005:Vis}
\textsc{Bordoloi U., Shen H.-W.}:
\newblock View selection for volume rendering.
\newblock In \emph{Proc. IEEE Visualization} (2005), pp.~487--494.

\bibitem[CE19]{Chen:2019:CGF}
\textsc{Chen M., Ebert D.~S.}:
\newblock An ontological framework for supporting the design and evaluation of
  visual analytics systems.
\newblock \emph{Computer Graphics Forum 38}, 3 (2019), 131--144.

\bibitem[CE20]{Chen:2020:book}
\textsc{Chen M., Edwards D.~J.}:
\newblock `isms' in visualization.
\newblock In \emph{Foundations of Data Visualization}, Chen M., Hauser H.,
  Rheingans P., Scheuermann G., (Eds.). Springer, 2020.

\bibitem[CFV{\etalchar{*}}16]{Chen:2016:book}
\textsc{Chen M., Feixas M., Viola I., Bardera A., Shen H.-W., Sbert M.}:
\newblock \emph{Information Theory Tools for Visualization}.
\newblock A K Peters, 2016.

\bibitem[CG16]{Chen:2016:TVCG}
\textsc{Chen M., Golan A.}:
\newblock What may visualization processes optimize?
\newblock \emph{IEEE Transactions on Visualization and Computer Graphics 22},
  12 (2016), 2619--2632.

\bibitem[CGJ{\etalchar{*}}17]{Chen:2017:CGA}
\textsc{Chen M., Grinstein G., Johnson C.~R., Kennedy J., Tory M.}:
\newblock Pathways for theoretical advances in visualization.
\newblock \emph{IEEE Computer Graphics and Applications 37}, 4 (2017),
  103--112.

\bibitem[CGJM19]{Chen:2019:TVCG}
\textsc{Chen M., Gaither K., John N.~W., McCann B.}:
\newblock Cost-benefit analysis of visualization in virtual environments.
\newblock \emph{IEEE Transactions on Visualization and Computer Graphics 25}, 1
  (2019), 32--42.

\bibitem[CJ10]{Chen:2010:TVCG}
\textsc{Chen M., J\"anicke H.}:
\newblock An information-theoretic framework for visualization.
\newblock \emph{IEEE Transactions on Visualization and Computer Graphics 16}, 6
  (2010), 1206--1215.

\bibitem[CS19]{Chen:2019:arXiv}
\textsc{Chen M., Sbert M.}:
\newblock On the upper bound of the kullback-leibler divergence and cross
  entropy.
\newblock arXiv:1911.08334, 2019.

\bibitem[CT06]{Cover:2006:book}
\textsc{Cover T.~M., Thomas J.~A.}:
\newblock \emph{Elements of Information Theory}.
\newblock John Wiley \& Sons, 2006.

\bibitem[CWB{\etalchar{*}}14]{Chen:2014:CGF}
\textsc{Chen M., Walton S., Berger K., Thiyagalingam J., Duffy B., Fang H.,
  Holloway C., Trefethen A.~E.}:
\newblock Visual multiplexing.
\newblock \emph{Computer Graphics Forum 33}, 3 (2014), 241--250.

\bibitem[DCK12]{Dasgupta:2012:CGF}
\textsc{Dasgupta A., Chen M., Kosara R.}:
\newblock Conceptualizing visual uncertainty in parallel coordinates.
\newblock \emph{Computer Graphics Forum 31}, 3 (2012), 1015–1024.

\bibitem[FdBS99]{Feixas:2001:CGF}
\textsc{Feixas M., {del Acebo} E., Bekaert P., Sbert M.}:
\newblock An information theory framework for the analysis of scene complexity.
\newblock \emph{Computer Graphics Forum 18}, 3 (1999), 95--106.

\bibitem[FSG09]{Feixas:2009:AP}
\textsc{Feixas M., Sbert M., Gonz\'{a}lez F.}:
\newblock A unified information-theoretic framework for viewpoint selection and
  mesh saliency.
\newblock \emph{ACM Transactions on Applied Perception 6}, 1 (2009), 1--23.

\bibitem[Gol20]{Golin:2020:web}
\textsc{Golin M.~J.}:
\newblock Lecture 17: Huffman coding.
\newblock
  \url{http://home.cse.ust.hk/faculty/golin/COMP271Sp03/Notes/MyL17.pdf},
  accessed in March 2020.

\bibitem[Gum02]{Gumhold:2002:Vis}
\textsc{Gumhold S.}:
\newblock Maximum entropy light source placement.
\newblock In \emph{Proc. IEEE Visualization} (2002), pp.~275--282.

\bibitem[HSS20]{Haseli:2020:IJMSEM}
\textsc{Haseli G., Sheikh R., Sana S.~S.}:
\newblock Base-criterion on multi-criteria decision-making method and its
  applications.
\newblock \emph{International Journal of Management Science and Engineering
  Management 15}, 2 (2020), 79--88.

\bibitem[IN13]{Ishizaka:2013:book}
\textsc{Ishizaka A., Nemery P.}:
\newblock \emph{Multi-Criteria Decision Analysis: Methods and Software}.
\newblock John Wiley \& Sons, 2013.

\bibitem[JS10]{Jaenicke:2010:CGA}
\textsc{J\"anicke H., Scheuermann G.}:
\newblock Visual analysis of flow features using information theory.
\newblock \emph{IEEE Computer Graphics and Applications 30}, 1 (2010), 40--49.

\bibitem[JWSK07]{Jaenicke:2007:TVCG}
\textsc{J\"anicke H., Wiebel A., Scheuermann G., Kollmann W.}:
\newblock Multifield visualization using local statistical complexity.
\newblock \emph{IEEE Transactions on Visualization and Computer Graphics 13}, 6
  (2007), 1384--1391.

\bibitem[KARC17]{Kijmongkolchai:2017:CGF}
\textsc{Kijmongkolchai N., Abdul-Rahman A., Chen M.}:
\newblock Empirically measuring soft knowledge in visualization.
\newblock \emph{Computer Graphics Forum 36}, 3 (2017), 73--85.

\bibitem[KL51]{Kullback:1951:AMS}
\textsc{Kullback S., Leibler R.~A.}:
\newblock On information and sufficiency.
\newblock \emph{Annals of Mathematical Statistics 22}, 1 (1951), 79--86.

\bibitem[Kle12]{Klein:2012:book}
\textsc{Klein H.~A.}:
\newblock \emph{The Science of Measurement: A Historical Survey}.
\newblock Dover Publications, 2012.

\bibitem[Lin91]{Lin:1991:TIT}
\textsc{Lin J.}:
\newblock Divergence measures based on the shannon entropy.
\newblock \emph{IEEE Transactions on Information Theory 37} (1991), 145–151.

\bibitem[LV06]{Liese:2006:TIT}
\textsc{Liese F., Vajda I.}:
\newblock On divergences and informations in statistics and information theory.
\newblock \emph{IEEE Transactions on Information Theory 52}, 10 (2006),
  4394–4412.

\bibitem[Mos12]{Moser:2012:book}
\textsc{Moser S.~M.}:
\newblock \emph{A Student's Guide to Coding and Information Theory}.
\newblock Cambridge University Press, 2012.

\bibitem[NM04]{Ng:2004:IV}
\textsc{Ng C.~U., Martin G.}:
\newblock Automatic selection of attributes by importance in relevance feedback
  visualisation.
\newblock In \emph{Proc. Information Visualisation} (2004), pp.~588--595.

\bibitem[PAJKW08]{Purchase:2008:LNCS}
\textsc{Purchase H.~C., Andrienko N., Jankun-Kelly T.~J., Ward M.}:
\newblock Theoretical foundations of information visualization.
\newblock In \emph{Information Visualization: Human-Centered Issues and
  Perspectives}, Springer LNCS 4950. 2008, pp.~46--64.

\bibitem[PS91]{Pedhazur:1991:book}
\textsc{Pedhazur E.~J., Schmelkin L.~P.}:
\newblock \emph{Measurement, Design, and Analysis: An Integrated Approach}.
\newblock Lawrence Erlbaum Associates, 1991.

\bibitem[RBB{\etalchar{*}}11]{Ruiz:2011:TVCG}
\textsc{Ruiz M., Bardera A., Boada I., Viola I., Feixas M., Sbert M.}:
\newblock Automatic transfer functions based on informational divergence.
\newblock \emph{IEEE Transactions on Visualization and Computer Graphics 17},
  12 (2011), 1932--1941.

\bibitem[RFS05]{Rigau:2005:SMA}
\textsc{Rigau J., Feixas M., Sbert M.}:
\newblock Shape complexity based on mutual information.
\newblock In \emph{Proc. IEEE Shape Modeling and Applications} (2005).

\bibitem[SEAKC19]{Streeb:2019:TVCG}
\textsc{Streeb D., El-Assady M., Keim D., Chen M.}:
\newblock Why visualize? untangling a large network of arguments.
\newblock \emph{IEEE Transactions on Visualization and Computer Graphics early
  view} (2019).
\newblock 10.1109/TVCG.2019.2940026.

\bibitem[Sha48]{Shannon:1948:BSTJ}
\textsc{Shannon C.~E.}:
\newblock A mathematical theory of communication.
\newblock \emph{Bell System Technical Journal 27} (1948), 379--423.

\bibitem[TKC17]{Tam:2017:TVCG}
\textsc{Tam G. K.~L., Kothari V., Chen M.}:
\newblock An analysis of machine- and human-analytics in classification.
\newblock \emph{IEEE Transactions on Visualization and Computer Graphics 23}, 1
  (2017).

\bibitem[TT05]{Takahashi:2005:Vis}
\textsc{Takahashi S., Takeshima Y.}:
\newblock A feature-driven approach to locating optimal viewpoints for volume
  visualization.
\newblock In \emph{Proc. IEEE Visualization} (2005), pp.~495--502.

\bibitem[VCI20]{Viola:2019:book}
\textsc{Viola I., Chen M., Isenberg T.}:
\newblock Visual abstraction.
\newblock In \emph{Foundations of Data Visualization}, Chen M., Hauser H.,
  Rheingans P., Scheuermann G., (Eds.). Springer, 2020.
\newblock Preprint at arXiv:1910.03310, 2019.

\bibitem[VFSG06]{Viola:2006:TVCG}
\textsc{Viola I., Feixas M., Sbert M., Gr{\"o}ller M.~E.}:
\newblock Importance-driven focus of attention.
\newblock \emph{IEEE Transactions on Visualization and Computer Graphics 12}, 5
  (2006), 933--940.

\bibitem[VFSH04]{Vazquez:2004:CGF}
\textsc{V\'{a}zquez P.-P., Feixas M., Sbert M., Heidrich W.}:
\newblock Automatic view selection using viewpoint entropy and its application
  to image-based modelling.
\newblock \emph{Computer Graphics Forum 22}, 4 (2004), 689--700.

\bibitem[vH14]{Erven:2014:TIT}
\textsc{{van Erven} T., Harremos P.}:
\newblock R\'{e}nyi divergence and {Kullback-Leibler} divergence.
\newblock \emph{IEEE Transactions on Information Theory 60}, 7 (2014),
  3797--3820.

\bibitem[Wik20]{NewtonScale:2020:wiki}
\textsc{Wikipedia}:
\newblock Newton scale.
\newblock \url{https://en.wikipedia.org/wiki/Newton_scale}, accessed in Nov.
  2020.

\bibitem[WLS13]{Wei:2013:CGF}
\textsc{Wei T.-H., Lee T.-Y., Shen H.-W.}:
\newblock Evaluating isosurfaces with level-set-based information maps.
\newblock \emph{Computer Graphics Forum 32}, 3 (2013), 1--10.

\bibitem[WS05]{Wang:2006:TVCG}
\textsc{Wang C., Shen H.-W.}:
\newblock {LOD Map} - a visual interface for navigating multiresolution volume
  visualization.
\newblock \emph{IEEE Transactions on Visualization and Computer Graphics 12}, 5
  (2005), 1029--1036.

\bibitem[WS11]{Wang:2011:E}
\textsc{Wang C., Shen H.-W.}:
\newblock Information theory in scientific visualization.
\newblock \emph{Entropy 13} (2011), 254--273.

\bibitem[WYM08]{Wang:2008:TVCG}
\textsc{Wang C., Yu H., Ma K.-L.}:
\newblock Importance-driven time-varying data visualization.
\newblock \emph{IEEE Transactions on Visualization and Computer Graphics 14}, 6
  (2008), 1547--1554.

\bibitem[XLS10]{Xu:2010:TVCG}
\textsc{Xu L., Lee T.~Y., Shen H.~W.}:
\newblock An information-theoretic framework for flow visualization.
\newblock \emph{IEEE Transactions on Visualization and Computer Graphics 16}, 6
  (2010), 1216--1224.

\end{thebibliography}
% biblatex with biber
% \printbibliography 

\clearpage
\newpage
\noindent\huge%
\textsc{\textsf{\textbf{Appendices}}}

\noindent\LARGE%
\textbf{A Bounded Measure for Estimating\\the Benefit of Visualization:\\%
Theoretical Discourse and\\Conceptual Evaluation}

\noindent\large%
~\\
Min Chen, University of Oxford, UK\\
Mateu Sbert, University of Girona, Spain\\

\normalsize%

\appendix
% =================================================
\section{\textbf{Explanation of the Original Cost-Benefit Measure}}
\label{app:OriginalTheory}
This appendix contains an extraction from a previous publication \cite{Chen:2019:CGF}, which provides a relatively concise but informative description of the cost-benefit ratio proposed in \cite{Chen:2016:TVCG}. The inclusion of this is to minimize the readers' effort to locate such an explanation. The extraction has been slightly modified.
In addition, at the end of this appendix, we provide a relatively informal and somehow conversational discussion about using this measure to explain why visualization is useful.

Chen and Golan introduced an information-theoretic metric for measuring the cost-benefit ratio of a visual analytics (VA) workflow or any of its component processes \cite{Chen:2016:TVCG}.
The metric consists of three fundamental measures that are abstract representations of a variety of qualitative and quantitative criteria used in practice, including
operational requirements (e.g., accuracy, speed, errors, uncertainty, provenance, automation),
analytical capability (e.g., filtering, clustering, classification, summarization),
cognitive capabilities (e.g., memorization, learning, context-awareness, confidence), and so on.
The abstraction results in a metric with the desirable mathematical simplicity \cite{Chen:2016:TVCG}.
The qualitative form of the metric is as follows:
\begin{equation}
\label{eq:CBR}
\frac{\textit{Benefit}}{\textit{Cost}} = \frac{\textit{Alphabet Compression} - \textit{Potential Distortion}}{\textit{Cost}}
\end{equation}

The metric describes the trade-off among the three measures:

\begin{itemize}
\vspace{-1mm}
\item
\emph{Alphabet Compression} (AC) measures the amount of entropy reduction (or information loss) achieved by a process.
As it was noticed in \cite{Chen:2016:TVCG}, most visual analytics processes (e.g., statistical aggregation, sorting, clustering, visual mapping, and interaction), feature many-to-one mappings from input to output, hence losing information.
Although information loss is commonly regarded harmful, it cannot be all bad if it is a general trend of VA workflows.
Thus the cost-benefit metric makes AC a positive component.
\vspace{-1mm}
\item
\emph{Potential Distortion} (PD) balances the positive nature of AC by measuring the errors typically due to information loss. Instead of measuring mapping errors using some third party metrics, PD measures the potential distortion when one reconstructs inputs from outputs.
The measurement takes into account humans' knowledge that can be used to improve the reconstruction processes. For example, given an average mark of 62\%, the teacher who taught the class can normally guess the distribution of the marks among the students better than an arbitrary person.
\vspace{-1mm}
\item
\emph{Cost} (Ct) of the forward transformation from input to output and the inverse transformation of reconstruction provides a further balancing factor in the cost-benefit metric in addition to the trade-off between AC and PD. In practice, one may measure the cost using \emph{time} or a monetary measurement.
\end{itemize}

\vspace{2mm}
\noindent\textbf{Why is visualization useful?}
There have been many arguments about why visualization is useful. 
Streeb et al. collected a large number of arguments and found many arguments are in conflict with each other \cite{Streeb:2019:TVCG}.
Chen and Edwards presented an overview of schools of thought in the field of visualization, and showed that the ``why'' question is a bone of major contention \cite{Chen:2020:book}.

The most common argument about ``why'' question is because visualization offers insight or helps humans to gain insight. When this argument is used outside the visualization community, there are often counter-arguments that statistics and algorithms can offer insight automatically and often with better accuracy and efficiency. There are also concerns that visualization may mislead viewers, which cast further doubts about the usefulness of visualization, while leading to a related argument that ``visualization must be accurate'' in order for it to be useful.
The accuracy argument itself is not bullet-proof since there are many types of uncertainty in a visualization process, from uncertainty in data, to that caused by visual mapping, and to that during perception and cognition \cite{Dasgupta:2012:CGF}.
Nevertheless, it is easier to postulate that visualization must be accurate, as it seems to be counter-intuitive to condone the idea that ``visualization can be inaccurate,'' not mentioning the idea of ``visualization is normally inaccurate,'' or ``visualization should be inaccurate.''

The word ``inaccurate'' is itself an abstraction of many different types of inaccuracy.
Misrepresentation truth is a type of inaccuracy.
Such acts are mostly wrong, but some (such as wordplay and sarcasm) may cause less harm.
Converting a student's mark in the range of [0, 100] to the range of [A, B, C, D, E, F] is another type of inaccuracy.
This is a common practice, and must be useful.
From an information-theoretic perspective, these two types of inaccuracy are information loss.

In their paper \cite{Chen:2016:TVCG}, Chen and Golan observed that statistics and algorithms usually lose more information than visualization. Hence, this provides the first hint about the usefulness of visualization. They also noticed that like wordplay and sarcasm, the harm of information loss can be alleviated by knowledge. For someone who can understand a workplay (e.g., a pun) or can sense a sarcastic comment, the misrepresentation can be corrected by that person at the receiving end. This provides the second hint about the usefulness of visualization because any ``misrepresentation'' in visualization may be corrected by a viewer with appropriate knowledge.

On the other hand, statistics and algorithms are also useful, and sometimes more useful than visualization. Because statistics and algorithms usually cause more information loss, some aspects of information loss must be useful.
One important merit of losing information in one process is that the succeeding process has less information to handle, and thus incurs less cost.
This is why Chen and Golan divided information loss into two components, a positive component called alphabet compression and a negative component called potential distortion \cite{Chen:2016:TVCG}.

The positive component explains why statistics, algorithms, visualization, and interaction are useful because they all lose information.
The negative component explains why they are sometimes less useful because information loss may cause distortion during information reconstruction.
Both components are moderated by the cost of a process (i.e., statistics, algorithms, visualization, or interaction) in losing information and reconstructing the original information.
Hence, given a dataset, the best visualization is the one that loses most information while causing the least distortion.
This also explains why visual abstraction is effective when the viewers have adequate knowledge to reconstruct the lost information and may not be effective otherwise \cite{Viola:2019:book}.

The central thesis by Chen and Golan \cite{Chen:2016:TVCG} may appear to be counter-intuitive to many as it suggests ``inaccuracy is a good thing'', partly because the word ``inaccuracy'' is an abstraction of many meanings and itself features information loss. Perhaps the reason for the conventional wisdom is that it is relatively easy to think that ``visualization must be accurate''. To a very small extent, this is a bit like the easiness to think ``the earth is flat'' a few centuries ago, because the evidence for supporting that wisdom was available everywhere, right in front of everyone at that time.
Once we step outside the field of visualization, we can see the phenomena of inaccuracy everywhere, in statistics and algorithms as well as in visualization and interaction.
All these suggest that ``the earth may not be flat,'' or ``inaccuracy can be a good thing.''

In summary, the cost-benefit measure by Chen and Golan \cite{Chen:2016:TVCG} explains that when visualization is useful, it is because visualization has a better trade-off than simply reading the data, simply using statistics alone, or simply relying on algorithms alone.
The ways to achieve a better trade-off include: (i) visualization may lose some information to reduce the human cost in observing and analyzing the data, (ii) it may lose some information since the viewers have adequate knowledge to recover such information or can acquire such knowledge at a lower cost, (iii) it may preserve some information because it reduces the reconstruction distortion in the current and/or succeeding processes, and (iv) it may preserve some information because the viewers do not have adequate knowledge to reconstruct such information or it would cost too much to acquire such knowledge.  

% =================================================
\section{\textbf{Formulae of the Basic and Relevant Information-Theoretic Measures}}
\label{app:InfoTheory}
This section is included for self-containment. Some readers who have the essential knowledge of probability theory but are unfamiliar with information theory may find these formulas useful.

Let $\mathbb{Z} = \{ z_1, z_2, \ldots, z_n \}$ be an alphabet and $z_i$ be one of its letters.
$\mathbb{Z}$ is associated with a probability distribution or probability mass function (PMF) $P(\mathbb{Z}) = \{ p_1, p_2, \ldots, p_n \}$ such that
$p_i = p(z_i) \ge 0$ and $\sum_{1}^n p_i = 1$. The \textbf{Shannon Entropy} of $\mathbb{Z}$ is:

\[
  \mathcal{H}(\mathbb{Z}) = \mathcal{H}(P)= - \sum_{i=1}^n p_i \log_2 p_i \quad \text{(unit: bit)}
\]

Here we use base 2 logarithm as the unit of bit is more intuitive in context of computer science and data science.

An alphabet $\mathbb{Z}$ may have different PMFs in different conditions.
Let $P$ and $Q$ be such PMFs. The \textbf{KL-Divergence} $\mathcal{D}_{KL}(P||Q)$ describes the difference between the two PMFs in bits:
\[
  \mathcal{D}_{KL}(P||Q) = \sum_{i=1}^n p_i \log_2 \frac{p_i}{q_i} \quad \text{(unit: bit)}
\]
$\mathcal{D}_{KL}(P||Q)$ is referred as the divergence of $P$ from $Q$.
This is not a metric since $\mathcal{D}_{KL}(P||Q) \equiv \mathcal{D}_{KL}(Q||P)$ cannot be assured.

Related to the above two measures, \textbf{Cross Entropy} is defined as:
\[
  \mathcal{H}(P, Q) = \mathcal{H}(P) + \mathcal{D}_{KL}(P||Q) = - \sum_{i=1}^n p_i \log_2 q_i \quad \text{(unit: bit)}
\]
Sometimes, one may consider $\mathbb{Z}$ as two alphabets $\mathbb{Z}_a$ and $\mathbb{Z}_b$ with the same ordered set of letters but two different PMFs.
In such case, one may denote the KL-Divergence as $\mathcal{D}_{KL}(\mathbb{Z}_a||\mathbb{Z}_b)$, and the cross entropy as $\mathcal{H}(\mathbb{Z}_a, \mathbb{Z}_b)$.

% =================================================
\section{\textbf{Conceptual Boundedness of} $\CE(P, Q)$ \textbf{and} $\mathcal{D}_\text{KL}$}
\label{app:Proof}
For readers who are not familiar with information-theoretic notations and measures, it is helpful to read Appendix \ref{app:InfoTheory} first.
According to the mathematical definition of cross entropy:
\begin{equation} \label{eq:CrossEntropy}
    \mathcal{H}(P, Q) = - \sum_{i=1}^n p_i \log_2 q_i = \sum_{i=1}^n p_i \log_2 \frac{1}{q_i}
\end{equation}
\noindent $\SE(P,Q)$ is of course unbounded.
When $q_i \rightarrow 0$, we have $\log_2 \frac{1}{q_i} \rightarrow \infty$. As long as $p_i \neq 0$ and is independent of $q_i$, $\SE(P,Q) \rightarrow \infty$.
Hence the discussion in this appendix is not about a literal proof that $\SE(P,Q)$ is unbounded when this mathematical formula is applied without any change.
It is about that the concept of cross entropy implies that it should be bounded when $n$ is a finite number.

\vspace{2mm}\noindent\textbf{Definition 1.}
\emph{Given an alphabet $\mathbb{Z}$ with a true PMF $P$, cross-entropy $\SE(P,Q)$ is the average number of bits required when encoding $\mathbb{Z}$ with an alternative PMF $Q$.}
\vspace{2mm}

This is a widely-accepted and used definition of cross-entropy in the literature of information theory.
The concept can be observed from the formula of Eq.\,\ref{eq:CrossEntropy}, where $\log_2 (1/q_i)$ is considered as the mathematically-supposed length of a codeword that is used to encode letter $z_i \in \mathbb{Z}$ with a probability value $q_i \in Q$. Because $\sum_{i=1}^n p_i = 1$, $\SE(P,Q)$ is thus the weighted or probabilistic average length of the codewords for all letters in $\mathbb{Z}$.

Here a \emph{codeword} is the digital representation of a letter in an alphabet. A \emph{code} is a collection of the codewords for all letters in an alphabet. In communication and computer science, we usually use binary codes as digital representations for alphabets, such as ASCII code and variable-length codes.

However, when a letter $z_i \in \mathbb{Z}$ is given a probability value $q_i$, it is not necessary for $z_i$ to be encoded using a codeword of length $\log_2 (1/q_i)$ bits, or more precisely, the nearest integer above or equal to it, i.e., $\lceil log_2 (1/q_i) \rceil$ bits, since a binary codeword cannot have fractional bits digitally.
For example, consider a simple alphabet $\mathbb{Z} = \{z_1, z_2\}$.
Regardless what PMF is associated with $\mathbb{Z}$, $\mathbb{Z}$ can always be encoded with a 1-bit code, e.g., codeword 0 for $z_1$ and codeword 1 for $z_2$, as long as neither of the two probability values in $P$ is zero, i.e., $p_1 \neq 0$ and $p_2 \neq 0$.
However, if we had followed Eq.\,\ref{eq:CrossEntropy} literally, we would have created codes similar to the following examples:
\begin{itemize}
    \item if $P = \{ \frac{1}{2}, \frac{1}{2} \}$, codeword 0 for $z_1$ and codeword 1 for $z_2$;
    \item if $P = \{ \frac{3}{4}, \frac{1}{4} \}$, codeword 0 for $z_1$ and codeword 10 for $z_2$;
    \item $\cdots$
    \item if $P = \{ \frac{63}{64}, \frac{1}{64} \}$, codeword 0 for $z_1$ and codeword 111111 for $z_2$;
    \item $\cdots$
\end{itemize}

\begin{figure} [t]
    \centering
    \includegraphics[width=88mm]{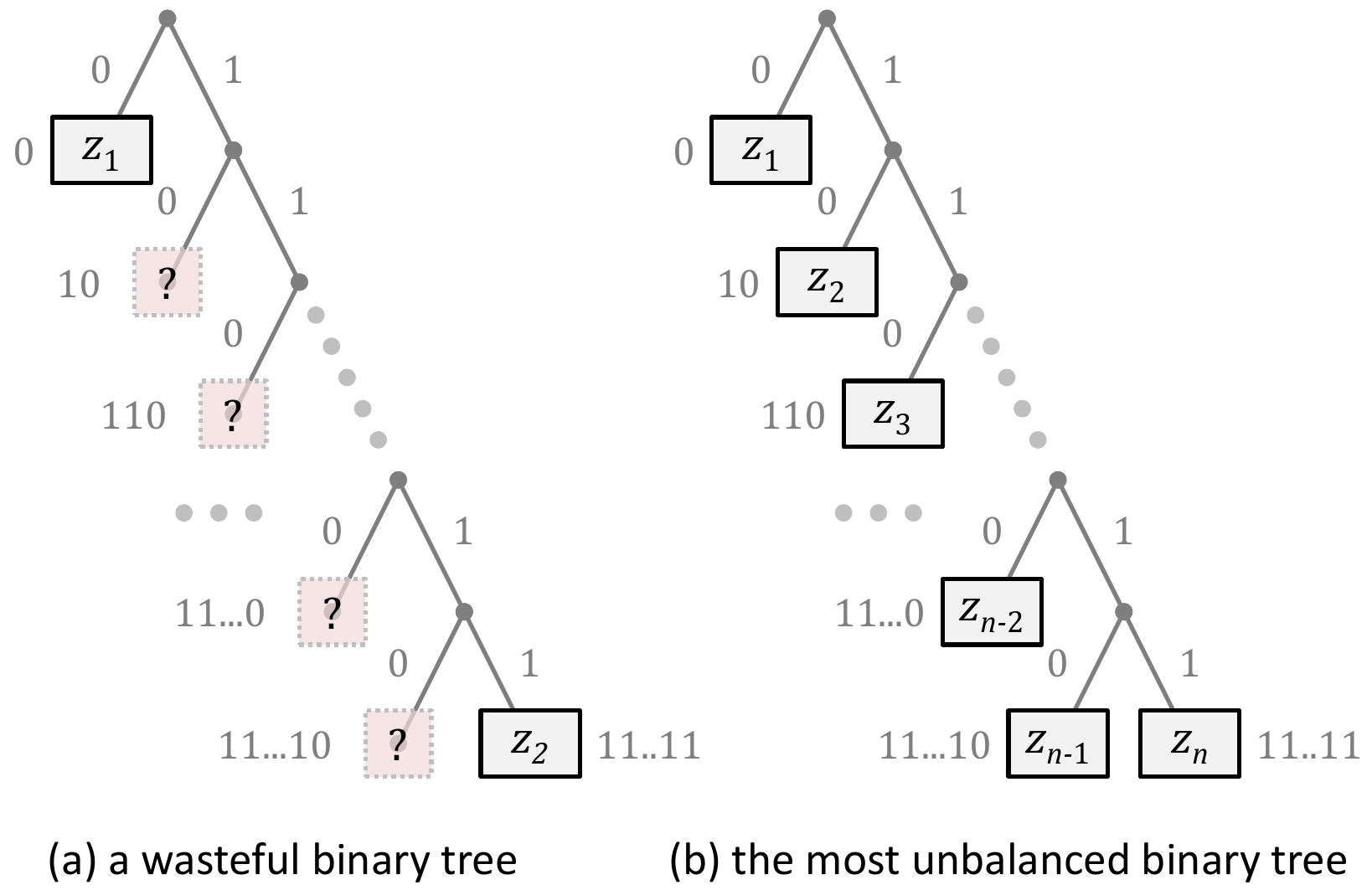}
    \caption{Two examples of binary codes illustrated as binary trees.}
    \label{fig:BinaryTree}
\end{figure}
As shown in Fig. \ref{fig:BinaryTree}(a), such a code is very wasteful.
Hence, in practice, encoding $\mathbb{Z}$ according to Eq.\,\ref{eq:CrossEntropy} literally is not desirable. Note that the discussion about encoding is normally conducted in conjunction with the Shannon entropy. Here we use the cross entropy formula for our discussion to avoid a deviation from the flow of reasoning.

Let $\mathbb{Z}$ be an alphabet with a finite number of letters, $\{ z_1, z_2, \ldots, z_n \}$, and $\mathbb{Z}$ is associated with a PMF, $Q$, such that: 
\begin{equation} \label{eq:CodeP-APP}
\begin{split}
  q(z_n) &= \epsilon, \quad\text{(where $0 < \epsilon < 2^{-(n-1)}$}),\\
  q(z_{n-1}) &= (1-\epsilon)2^{-(n-1)},\\
  q(z_{n-2}) &= (1-\epsilon)2^{-(n-2)},\\
  &\cdots\\
  q(z_{2}) &= (1-\epsilon)2^{-2},\\
  q(z_{1}) &= (1-\epsilon)2^{-1} + (1-\epsilon)2^{-(n-1)}.
\end{split}
\end{equation}
We can encode this alphabet using the Huffman encoding that is a practical binary coding scheme and adheres the principle to obtain a code with the Shannon entropy as the average length of codewords \cite{Moser:2012:book}.
Entropy coding is designed to minimize the average number of bits per letter when one transmits a ``very long'' sequence of letters in the alphabet over a communication channel.
Here the phrase ``very long'' implies that the string exhibits the above PMF $Q$ (Eq.\,\ref{eq:CodeP-APP}).
In other words, given an alphabet $\mathbb{Z}$ and a PMF $Q$, the Huffman encoding algorithm creates an optimal code with the lowest average length of codewords when the code is used to transmit a ``very long'' sequence of letters in $\mathbb{Z}$.
One example of such a code for the above PMF $Q$ is:
\begin{equation} \label{eq:Code-APP}
\begin{split}
  z_1&: 0, \qquad z_2: 10, \qquad z_3: 110\\
  &\cdots\\
  z_{n-1}&: 111\ldots10 \quad\text{(with $n-2$ ``1''s and one ``0'') }\\
  z_n&: 111\ldots11 \quad\text{(with $n-1$ ``1''s and no ``0'') }
\end{split}
\end{equation}
Fig. \ref{fig:BinaryTree}(b) shows illustrates such a code using a binary tree.
In this way, $z_n$, which has the smallest probability value, will always be assigned a codeword with the maximum length of $n-1$.

\vspace{2mm}\noindent\textbf{Lemma 1.}
\emph{Let $\mathbb{Z}$ be an alphabet with $n$ letters and $\mathbb{Z}$ is associated with a PMF $P$.
If $\mathbb{Z}$ is encoded using the aforementioned entropy coding, the maximum length of any codeword for $z_i \in \mathbb{Z}$ is always $\leq n-1$.}
\vspace{2mm}

We can prove this lemma by confirming that when one creates a binary code for an $n$-letter alphabet $\mathbb{Z}$, the binary tree shown in Fig. \ref{fig:BinaryTree}(b) is the worst unbalanced tree without any wasteful leaf nodes.
Visually, we can observe that the two letters with the lowest values always share the lowest internal node as their parent node. The remaining $n-2$ letters are to be hung on the rest binary subtree.
Because the subtree is not allowed to waste leaf space, the $n-2$ leaf nodes can be comfortably hung on the root and up to $n-3$ internal node.
A formal proof can be obtained using induction.
For details, readers may find Golin's lecture notes useful \cite{Golin:2020:web}.
See also \cite{Cover:2006:book} for related mathematical theorems.

\vspace{2mm}\noindent\textbf{Theorem 1.}
\emph{Let $\mathbb{Z}$ be an alphabet with a finite number of letters and $\mathbb{Z}$ is associated with two PMFs, $P$ and $Q$. With the Huffman encoding, conceptually the cross entropy $\SE(P,Q)$ should be bounded.}
\vspace{2mm}

Let $n$ be the number of letters in $\mathbb{Z}$. 
According to \textbf{Lemma 1}, when $\mathbb{Z}$ is encoded in conjunction with PMF $Q$ using the Huffman encoding, the maximum codeword length is $\leq n-1$.
In other words, in the worst case scenario, there is letter $z_k \in \mathbb{Z}$ that has the lowest probability value $q_k$, i.e., $q_k \leq q_j \forall j=1, 2, \ldots n \text{ and } j \neq k$.
With the Huffman encoding, $z_k$ will be encoded with the longest codeword of up to $n-1$ bits.

According to \textbf{Definition 1}, there is a true PMF $P$.
Let $L(z_i, q_i)$ be the codeword length of $z_i \in \mathbb{Z}$ determined by the Huffman encoding.
We can write a conceptual cross entropy formula as:
\[
  \SE(P, Q) = \sum_{i=1}^n p_i \cdot L(z_i, q_i) \leq \sum_{i=1}^n p_i \cdot L(z_k, q_k) \leq n-1
\]
\noindent where $q_k$ is the lowest probability value in $Q$ and $z_k$ is encoded with a codeword of up to $n-1$ bits (i.e., $L(z_k, q_k) \leq n-1$).
Hence conceptually $\SE(P, Q)$ is bounded by $n-1$ if the Huffman encoding is used.
Since we can find a bounded solution for any $n$-letter alphabet with any PMF, the claim of unboundedness has been falsified.
$\blacksquare$

\vspace{2mm}\noindent\textbf{Corollary 1.}
\emph{Let $\mathbb{Z}$ be an alphabet with a finite number of letters and $\mathbb{Z}$ is associated with two PMFs, $P$ and $Q$. With the Huffman encoding, conceptually the KL-divergence $\DKL(P\|Q)$ should be bounded.}
\vspace{2mm}

For an alphabet $\mathbb{Z}$ with a finite number of letters, the Shannon entropy $\mathcal{H}(P)$ is bounded regardless any PMF $P$.
The upper bound of $\mathcal{H}(P)$ is $\log_2 n$, where $n$ is the number of letters in $\mathbb{Z}$. 
Since we have
\begin{align*}
    &\mathcal{H}(P, Q) = \mathcal{H}(P) + \mathcal{D}_{KL}(P||Q)\\
    &\mathcal{D}_{KL}(P||Q) = \mathcal{H}(P, Q) - \mathcal{H}(P)
\end{align*}
\noindent using \textbf{Theorem 1}, we can infer that with the Huffman encoding, conceptually $\mathcal{D}_{KL}(P||Q)$ is also bounded.
$\blacksquare$

\vspace{2mm}
\noindent\textbf{Further discussion.}
The code created using Huffman encoding is also considered to be optimal for source coding (i.e., assuming without the need for error correction and detection). A formal proof can be found in \cite{Golin:2020:web}.

Let $\mathbb{Z}$ be an $n$-letter alphabet, and $Q$ be its associated PMF. When we use the Shannon entropy to determine the length of each codeword mathematically, we have:
\[
    L(z_i, q_i) = \lceil \log_2 \frac{1}{q_i} \rceil, \quad z_i \in \mathbb{Z}, q_i \in Q
\]
As we showed before, the length of a codeword can be infinitely long if $q_i \rightarrow 0$. Huffman encoding makes the length finite as long as $n$ is finite. This difference between the mathematically-literal entropy encoding and Huffman encoding is important to our proof that conceptually $\SE(P, Q)$ and $\DKL(P\|Q)$ are bounded.

However, we should not draw a conclusion that there is much difference between the communication efficiency gained based on the mathematically-literal entropy encoding and that gained using the Huffman encoding.
In fact, in terms of  the average length of codewords, they differ by less than one bit since both lie between $\SE(Q)$ and $\SE(Q)+1$ \cite{Cover:2006:book}, although their difference in terms of the maximum length of individual letters can be very different.

For example, if $\mathbb{Z}$ is a two-letter alphabet, and its PMF $Q$ is $\{0.999, 0.001\}$, the Huffman encoding results in a code with one bit for each letter, while the mathematically-literal entropy encoding results in 1 bit for $z_1 \in \mathbb{Z}$ and 10 bits for $z_2 \in \mathbb{Z}$. The probabilistic average length of the two codewords, which indicate the communication efficiency, is 1 bit for the Huffman encoding, and 1.009 bits for the mathematically-literal entropy encoding, while the entropy $\SE(Q)$ is 0.0114 bits. As predicted, $0.0114 < 1 < 1.009 < 1.0114$.

Consider another example with a five-letter alphabet and $Q = \{0.45, 0.20, 0.15, 0.15, 0.05\}$.
The mathematically-literal entropy encoding assigns five codewords with lengths of $\{2, 3, 3, 3, 5\}$, while the Huffman encoding assigns codewords with lengths of $\{1, 3, 3, 3, 3\}$.
The probabilistic average length of the former is 2.65, while that of the Huffman encoding is 2.1, while the entropy $\SE(Q)$ is 2.0999.
As predicted, $2.0999 < 2.1 < 2.65 < 3.0999$.

\section{\textbf{Authors' Revision Report following the SciVis 2020 Reviews}}
\label{app:Feedback}
2 December 2020

\noindent Dear EuroVis 2021 co-chairs, IPC members, and Reviewers,

\noindent \textbf{Bounded Measure for Estimating the Benefit of Visualization:\\
Theoretical Discourse and Conceptual Evaluation\\
Min Chen and Mateu Sbert}

The original version of this paper, ``1051: A Bounded Measure for Estimating the Benefit of Visualization, by Chen, Sbert, Abdul-Rahman, and Silver'', was submitted to EuroVis 2020, the paper received scores (5, 4.5, 2.5, 2). The co-chairs recommended for the paper to be resubmitted to CGF after a major revision (i.e., fast track). Unfortunately, the first author was an editor-in-chief of CGF until January 2020 and still has administrative access to the CGF review system at that time. Hence it will not be possible for CGF to arrange a confidential review process for this paper.

The paper was revised according to the reviewers' comments and was subsequently submitted to SciVis 2020. The SciVis submission, the revision report, and the EuroVis 2020 reviews can be found as an arXiv report https://arxiv.org/pdf/2002.05282.pdf.

The SciVis 2020 submission improved upon the EuroVis 2020 with additional clarifications in the main text and detailed explanations in appendices including a mathematical proof. The reviewers were not impressed and scored the submission as [3.5 2, 2, 1.5], which is much lower than the scores for the EuroVis 2020 submission. We appreciate that all four SciVis 2020 reviewers stated transparently that they had (and likely still have) doubts about the theoretic measure proposed by Chen and Golan, TVCG, 2016, which this paper tries to improve. We also appreciate that the reviewers stated that they tried not to let these doubts to prejudice this work.

We notice that almost all comments that have influenced the rejection decision are in the form of requiring further explanations and clarifications. As the summary review states ``The paper demands a lot from the reader; without detailed knowledge of information theory concepts, as well as details of the cost-benefit ratio, ...'', clearly simply addressing this issue through adding appendices has not met the reviewers' concerns. When we converted the SciVis 2020 submission into the CGF format, it resulted in 11.5 pages, 1.5 pages above the limit of EuroVis 2021 guideline. Considering the needs (i) to add a fair amount of additional explanation and clarification, and (ii) to reduce the paper length to 10 pages, we decided to split our paper into two papers. We are aware that the EuroVis and CGF rarely see two-part publications. We hence have rewritten the two submissions substantially, making them self-contained and relatively independent, while avoiding unnecessary duplications. This is the first paper on ``Theoretical Discourse and Conceptual Evaluation''. The second (follow-on) paper, entitled

\noindent\emph{Bounded Measure for Estimating the Benefit of Visualization: Case Studies and Empirical Evaluation\\
Min Chen, Alfie Abdul-Rahman, Deborah Silver, and Mateu Sbert}

\noindent is included in the supplementary materials. Similarly, this submission is also included the supplementary materials of the second paper. The both submissions are available as arXiv reports, together with the corresponding revision reports.

In this revision report, we focus on the reviewers' concerns and queries related to the theoretical discourses in the original EuroVis and SciVis submissions. Those comments are highlighted in \textcolor{orange}{orange}. We structure our feedback based on the summary review, and incorporate individual reviewers' comments when they are mentioned in the summary. The comments highlighted in \textcolor{violet}{purple} are for the follow-on paper to address.

\noindent Yours sincerely,\\
Min Chen and Mateu Sbert\\
\rule{4cm}{0.4pt}

\noindent\textbf{Summary Review:}

\noindent\textcolor{orange}{1.	The need for a bounded measure for PD is not sufficiently justified; there are several arguments (see in particular R2 and R3) that this requirement is not necessary. This, however, makes the premise of the paper doubtful. (R1, R3, R4)}

	\textbf{Clarification and Action:} The previous text accompanying Tables 1 and 2 has explained the difficulties in interpreting the unbounded PD measure. We have now added a new section, Section 4 Mathematical Notations and Problem Statement, which introduces more basic information theory concepts and hopefully helps readers to appreciate the original problem statement better. We have also added new text in Section 7 Discussions and Conclusions to summarize the needs to study new measures.

	Of course, one can argue that the original DKL can be used. Imagine that we had in the history a temperature measurement system that had been defined based humans' normal temperature (0 degree) and the maximum temperature when they were ill (100 degree), or a weight measurement system that had been a power function of the current metric system, e.g., $w' = w^{100}$ (which would show rapid growth as DKL). One might argue that they are adequate enough, and there is no need to consider other new measuring systems. The history of measurement science shows the opposite. As scientists, we would not discourage a serious exploration of alternative measures.

\noindent\textcolor{orange}{2.	Some of the authors' trains of thought cannot be fully comprehended, as there are discrepancies with earlier work that are not being clarified. (R1)}

	\textbf{Action:} We hope that the addition of Section 4 Mathematical Notations and Problem Statement helps explain the connection between cost-benefit ratio and volume rendering via information theory. We also hope that the addition of Section 5.3. New Candidates of Bounded Measures helps explain the thought process behind the proposed measures, as well as some technical considerations such as why using maximum entropy to scale the potential distortion instead of Shannon entropy itself.

\noindent\textcolor{orange}{3.	The alleged proofs are unsustainable. (R1, R3, R4)}

	\textbf{Clarification and Action:} We guess that the reviewers may misinterpret that the proof was for proving the mathematical formula of DKL is bounded. Obviously this is not the case. The proof is about a key concept that DKL and Cross Entropy are supposed to model. The proof showed that this key concept is bounded if the alphabet has a finite number of letters, while DKL cannot assure a bounded solution. This was explained in Appendix C, but was not in the main text due to the previous space limitation. Because we have a bit more space now after splitting the paper, we have added further clarification in Section 5.1 A Conceptual Proof of Boundedness.

\noindent\textcolor{orange}{4.	The proposed alternative measures are sometimes very ad hoc (R1, R3, R4), some are dimensionally doubtful (R4), they are (probably) not additive (R3) and there is no natural language interpretation (R1); other reasonable choices would have been possible (R3).}

\textcolor{orange}{R4:  I am also skeptical about the quantity $D^k_new$ in Eq. (9), except for the case  k=1. If one does a dimensional consideration by (for the moment) leaving out the summand 1 in the logarithm in Eq. (9), ...}

	\textbf{Clarification and Action:} Due to the page limits associated with the previous submissions, we did not explain the semantic meaning of the proposed measures. We have added a new subsection Section 5.3 New Candidates of Bounded Measures, where we described the semantic meaning of the proposed measures, addressing the ``\textcolor{orange}{natural language interpretation issue}'' raised by R1.

	We also explain the reason for considering the power factor $k$. This is a common approach in defining statistical measures and information-theoretic measures. This is just part of the research investigation to study different options while avoiding a potentially-biased assumption by fixing $k$ to 1. In fact, we started this investigation with $k=1$. It was our curiosity led us to consider other $k$ values. We believe such an approach in research should be encouraged.

	We would also like to point out the fact that Shannon entropy itself is a parameterized formula as different logarithmic scales have been used in different applications, such as natural log for thermodynamics, $\log_2$ in computer science and communication theory, and $\log_{10}$ and natural log in psychology. It is important for us scientists not to dismiss such diversity as ``undesirable''.

	On the additive nature of the measure, this depends on the definition of addition operator. Divergence measures are often not distance metrics. Addition of two divergence values is not always semantically meaningful. We encourage all researchers to continue the investigation into different mathematical properties of the measures considered in this paper as well as any measures that may be proposed in the future for visualization. There is no reason that visualization researchers should not be involved in developing mathematical concepts and formulae, especially when they are for visualization.

\noindent\textcolor{orange}{5.	The evaluation of measures has a strong ad hoc character (R1, R2, R3, R4); with the combination of Likert scales and MCDA almost any connection to information theory is lost (R3).}

	\textbf{Clarification and Action:} We disagree with the reviewers' critique about the ``strong ad hoc character''. If this critique is valid, the same could be applied most research efforts in the history of measurement science. For many measure systems, such as time, temperature, length, weight, earthquake magnitude, etc. there are usually only partial constraints defined by physical principles, such as starting point of the universe and absolutely zero temperature, which are also sometimes  associated with some uncertainty. Most parameters such as critical point, interval, and linearity are defined based on different criteria. Otherwise, we would not have debates about imperial vs. metric systems, or Celsius vs. Fahrenheit vs. kelvin vs. Rankine vs Reaumur for temperature measurement. Of course, we should all aspire to discover more mathematically-defined principles for evaluating different measurement systems designed to measure the benefit of visualization. However, it is not reasonable to reject the MCDA approach based on a speculation of existence of a complete set of principles. We have added new text at the beginning of Section 6 Comparing Bounded Measure: Conceptual Evaluation, and in Section 7 Discussions and Conclusions, drawing some historical facts in Measurement Science to support our approach.

	In terms of R3's comment about the connection with information theory, part of the problem studied in this paper is that there are several optional information-theoretical measures and statistical measures. They exhibit different properties, in ways similar to the options of Celsius, Fahrenheit, kelvin, Rankine, Reaumur, and other scales for temperature measurement. Our decisions on which temperature measurement system to use are based on our multi-criteria judgement. For now, choosing which measure for measuring the benefit of visualization is perhaps more complex. Using MCDA helps to make the process more transparent and rigorous than without.

\noindent\textcolor{orange}{6.	The paper demands a lot from the reader; without detailed knowledge of information theory concepts, as well as details of the cost-benefit ratio, the paper is difficult to understand. It would be better if the basic terms and concepts were briefly presented again in the paper, with reference to the appendix) (R1). It would be nice if some of the two case studies were interwoven into the mathematical presentation to better anchor the concepts (R2). The notation needs to be improved (R1).}

	\textbf{Action:} We have now added a new Section 4 Mathematical Notations and Problem Statement, which include 3/4 pages of new text and a new figure (Figure 4) for introducing the mathematical notions and basic concepts of information to help readers who are not familiar with information theory. We have adopted the R2's suggestion for making links to the case studies. In this paper, we have made links with the volume rendering case study, while in the follow-on paper, we made links with the metro map case study.

\noindent\textcolor{orange}{7.	The validation of the PD measure is questionable (R1, R3, R4); for details see in particular R3.}

\textcolor{orange}{R3:	Hence, the chosen measure was not, it turns out, actually determined by information theory, and its actual value for *quantitative* analysis was never demonstrated.}

	\textbf{Clarification and Action:} We guess that the reviewers were concerned about the semantic meaning of the proposed measure, rather than its ``validation''. We have added a new subsection Section 5.3 New Candidates of Bounded Measures, where we described the semantic meaning of the proposed measures.

	There are numerous entropy and divergence measures in information theory, exhibiting different semantic meanings and mathematical properties. One approach is to test the candidates of bounded measures in different circumstances to see how they affect the semantic interpretations, in a way similar to interpreting what a negative temperature measure means, or how earthquake magnitude 8 relates to magnitude 5. This is why we conducted two experiments to collect real data. Hence we do not agree R3's comment that ``\textcolor{orange}{actual value for *quantitative* analysis was never demonstrated.}'' The case studies of volume rendering and metro-map are exactly for that. R1 and R4 did consider the two case studies as part of validation process, while requesting for more analysis of the collected data (R1) and critiquing the experimental design (R4).

	Note that the two experiments have now been moved to a separate follow-on paper. They are no longer ``demonstrated'' in this paper as in the previous EuroVis 2020 and SciVis 2020 submissions.  R1 and R4's comments will thus be addressed in the separate follow-on paper. Nevertheless, we added a discussion in Section 7 Discussions and Conclusions to state that the two case studies are part of the evaluation process.

% =================================================
\section{\textbf{SciVis 2020 Reviews}}
\label{app:SciVis2020}
\setlength{\parindent}{0mm}
\setlength{\parskip}{3pt}
\small
\begin{narrowfont}

We regret to inform you that we are unable to accept your IEEE VIS 2020 SciVis Papers submission:
 
  1055 - A Bounded Measure for Estimating the Benefit of Visualization
 
The reviews are included below. IEEE VIS 2020 SciVis Papers had 125 submissions and we conditionally accepted 32, for a provisional acceptance rate of about 25\%.
 
......
 
----------------------------------------------------------------

Reviewer 1 review

  Paper type

    Theory \& Model

  Expertise

    Expert

  Overall Rating

    <b>2 - Reject</b>\\
    The paper is not ready for publication in SciVis / TVCG.\\
    The work may have some value but the paper requires major revisions or
    additional work that are beyond the scope of the conference review cycle to meet
    the quality standard. Without this I am not going to be able to return a score of
    '4 - Accept'.

  Supplemental Materials

    Acceptable with minor revisions (specify revisions in The Review section)

  Justification

    The paper presents a new measure to assess the benefits of using visualization.
    The measure revises Chen and Golan's cost-benefit ratio by proposing to measure
    distortion differently. The previous measure was unbounded and this issue is
    addressed in the revised measures. The work is evaluated using analysis of two
    case studies where subjects' distortion was assessed through questionnaires and
    compared with the distortion as indicated by the proposed measures.

    Overall, the paper presents a novel idea, but it is not ready for publication. It
    lacks detail on fundamental concepts that were presented in earlier work, but are
    not widely known in the community. There are notable discrepancies to prior work,
    that are never addressed. The premise of the work is dubious, since the unbounded
    measure of distortion indicates that, according to the measure, visualizations can
    be arbitrarily bad, but in practice, where the aim is to maximize benefit, this is
    not a problem, since benefit can be bounded from above in the measure. The rating-
    based (pre-)selection of the proposed alternatives needs more details on the
    rating process to be objective. The suggested alternatives are rather ad-hoc and
    only compared to each other, but never to the previous distortion measure in
    whether they capture distortion and thus benefit more accurately. The appropriate
    validation of a measure is to show that it does measure what it was intended to
    measure. The paper is currently still lacking this and it is questionable, whether
    the authors can provide this validation this within the reviewing cycle.

  The Review

    I will elaborate on the issues mentioned above.

    1. The paper lacks detail on fundamental concepts. This includes both information-
    theoretic concepts as well as details concerning the cost-benefit ratio. Although
    the authors provided a brief introduction in the appendix, this should really be
    in the paper in a brief form. Elaborations can remain in the appendix. Information
    that is crucial to understand Equation (2) is lacking: there is no discussion of
    what Z'$\_$i denotes, not how Z$\_$i, Z$\_$i+1, and Z'$\_$i are related. Note how much the
    paper by Chen and Jänicke 10 years ago elaborated on the basic concepts in
    contrast to the submitted work. The authors may have spent 10 years thinking
    deeply about information theory for visualization and are thus deeply familiar
    with it, but they should respect the needs of the readers, who mostly studied
    something else. Especially, if the authors want a large part of the field of
    visualization to understand and apply the methods they propose.

    2. There are also notable differences to prior work that are glossed over. Chen
    and Golan describe visualization as a sequence of transformations from data to a
    decision and say that the benefit for the whole sequences is H(Z$\_$0) - H(Z$\_$n+1) -
    sum$\_$i=0..n D$\_$KL(Z'$\_$i || Z$\_$i); i.e. the distortion accumulates. In contrast, the
    formula in the current submission comprises just one transition from the "ground
    truth" (which presumably aren't the data) to the decision variable about the
    ground truth. Why do the intermediate distortions no longer show up? And how does
    ground-truth relate to data?

    3. The premise of the work is dubious. The introduction notes that the measure of
    distortion used in the cost-benefit ratio is unbounded. It shows both that it is
    unbounded, as well as that even for simple cases, the distortion can be
    arbitrarily large, even though the information is low, and that this is counter-
    intuitive. However, I will hold, that the unboundedness is not a problem in
    practice. Since H(Z$\_$i) is bounded from above by log |Z$\_$i|, H(Z$\_$i+1) and D$\_$KL(Z'$\_$i
    || Z$\_$i) are bounded from below by 0, the benefit, defined as H(Z$\_$i) - H(Z$\_$i+1) -
    D$\_$KL(Z'$\_$i || Z$\_$i), is bounded from above by log |Z$\_$i| as well. It may not be
    bounded by below, but since the aim is to maximize the cost-benefit ratio, this is
    of no concern. Only maximizing functions that cannot be bounded by above are
    problematic. Note that log-likelihood scores also cannot be bounded from below,
    yet statistics has no reservations in using them for maximum-likelihood
    estimation. A different motivation would be Kulback-Leibler divergence
    overestimates distortion and therefore non-confusing visualizations are
    incorrectly claimed to be confusing by the cost-benefit ratio. Chen an Golan wrote
    that negative benefit indicates confusing visualizations. How well do different
    measures for distortion align with this assessment? For which measures does the
    sign of the benefit measure indicate confusing visualizations? For which measures
    do benefit values correlate better with visualization quality?

    4. It doesn't help the discussion that the paper claims that Kullback-Leibler
    divergence is unbounded, and then provide a "proof" in Section 4.1 that it is
    bounded for particular choices of Q. Since the measure of distortion is applied in
    the paper to relate the ground truth and the decision variable, both of which have
    probability distributions not under our control, the premises of the proof are not
    met and its result is therefore irrelevant. The section should be removed.

    5. The suggested alternatives are rather ad-hoc. While Kullback-Leibler
    divergence, cross entropy have interpretations in data transmission contexts
    (which are even recounted in the paper: the expected number of (excess) bits when
    using code lengths for symbols that would be optimal for a distribution P, when
    the actual symbol properties are given by distribution Q). There is no
    interpretation of the new measures given (in the sense of how to translate values
    for these measures to meaningful natural-language sentences), yet these previous
    measures are declared unintuitive and the new one's intuitive without any
    indication of why. The assessment of the different measures uses a catalogue of
    criteria and values are assigned to each measures, again, rather ad-hoc. There are
    no clear rules, under which conditions a certain number (ranging from 0-5) is
    assigned, nor a discussion of inter-coder consistency (Would different people
    assign the same numbers?). Overall, I do not see that such qualitative assessments
    are helping. See next point. [Note that Shannon put forth axioms that a measure
    for uncertainty should have and showed that entropy is the only formula meeting
    all the properties specified axiomatically. This however is quite different to the
    approach taken by the authors of the present submission.]

    6. \textcolor{violet}{The proper validation of a measure is to show that it measures what it is
    intended to measure.} Although there are numerous studies in visualization that
    establish one design as working or working better with respect to some task, the
    cost-benefit ratio was not applied to any of them. Instead the authors performed
    two studies. However, they do not fully model these studies in their framework.
    There is no mention of the sample space underlying the experiment, or the random
    variables. How can we estimate the probability distribution for the decision
    variable from the responses? The authors instead use one probability distribution
    for each answer. Is this meaningful? It removes all the probabilistic effects that
    underlie the foundation of the measures in the first place. Only some outcomes of
    the studies are discussed. Why? The distributions for ground truth are guessed,
    while there is arguably a ground truth present from which the distribution can be
    read off (e.g. the number of subway stations between two given stations). In the
    discussion, the images do not seem to play a role, only the distortion between
    ground truth and response. Furthermore it is confusing that both Q and Q' are
    described as probability distributions for ground truth, that F remains
    mysterious, and that the value for alphabet compression cannot be traced. Do the
    topological and the geographical subway maps have the same alphabet compression?
    Nor is it clear, how Q, Q', and F relate to Z$\_$i, Z$\_$i+1, and Z'$\_$i in Equation (2).
    The distribution function for Q in 5.2 is also ill-justified.

    7. Lastly: notation. The paper uses notation rather freely, e.g. to specify
    probability distributions, which are functions, using set or interval notation,
    which is non-standard, inconsistent, and confusing for people well-versed in
    probability theory. Similar concepts get different symbols, and the same thing is
    referred to by different expressions (e.g. H(Z), H(P(Z)), H(P)). It is also very
    confusing when probability distributions are denoted by their alphabets, in
    particular when different symbols for alphabets refer to probability distributions
    over the same alphabet. The authors should make notation consistent both within
    and between paper and appendix, and use established conventions in probability
    theory and information theory.

----------------------------------------------------------------

Reviewer 2 review

  Paper type

    Theory \& Model

  Expertise

    Knowledgeable

  Overall Rating

    <b>3.5 - Between Possible Accept and Accept</b>

  Supplemental Materials

    Acceptable

  Justification

    The manuscript is an interesting mathematical foray into how loss of information
    can actually benefit a visualization user, and what the bounds of the distortion-
    loss might be. The idea, the analysis, and the proof are valuable contributions to
    the visualization field, in particular since the conclusion is counterintuitive to
    many audiences, and a hangup of some domain experts (those who decline to use
    scientific visualization altogether because of inherent distortions). On the for
    improvement side, the manuscript is not an easy read, and could also benefit from
    certain clarifications. The two real world examples are intriguing, although the
    domain motivation (why these questions) could be further clarified and discussed.

  The Review

    The manuscript contributes a fascinating mathematical foray into how loss of
    information can actually benefit a visualization user. This is a valuable
    contribution to the field, in particular since the conclusion is counterintuitive.
    A major drawback is that the manuscript is not an easy read, in particular for a
    potential computer science or design graduate student, which is a pity, and that
    it could benefit from certain clarifications.

    The manuscript makes several significant contributions, including:
    
    ** a theoretical discussion of information loss, based on a theoretical basis in
    information theory.
    The information theory premise is valid, and builds on peer-reviewed prior
    publications. The same approach gave us, years ago, the first semblance of
    formalism in analyzing Shneiderman's "Overview First" mantra, and putting some
    bounds to it (good for novices, bad for experts). Albeit not perfect, that
    formalism opened up the possibility of a discussion in the field. Information
    Theory is one possible theoretic framework, leading to discussion in the field,
    and so it has merit.

    ** an interesting theoretical discussion of potential distortion (PD), as a term
    that could be calibrated and bounded.
    There seems to be disagreement in the field about whether PD should be absolutely
    minimized, or it could be calibrated as part of a benefit ratio. That is good: it
    means PD is something we need to discuss as a field. This manuscript is the
    beginning of that conversation.

    The PD subway example in the manuscript is brilliant, although somewhat cryptic,
    in that it illustrates well these different views of PD, and how we may be at
    cross-purposes when discussing it: In NYC, the shortest path involves exiting the
    subway system, walking for a few minutes on the street (hence the length judgment,
    and the importance of PD), and re-entering the subway system at a different line.
    In contrast, in the Berlin or Chicago subway, all the lines are well connected,
    and all that matters is the number of stops (PD almost unimportant). However, had
    I not lived in NYC for a few years, I wouldn't even know these things. It's
    possible the authors themselves are unaware of an alternative interpretation of
    subway maps and PD---than the one they used in the paper.

    This may be the case about the manuscript medical vis example, as well---I do not
    have experience educating novice medical students, but the authors do. This fact,
    that the manuscript looks at PD from a different point of view than me, has
    tremendous value to me. It would be helpful to clarify that this manuscript views
    things differently from several of us, and that overall, sometimes PD is just a
    quantity to be minimized, whereas other times PD is a quantity where bounding
    would be useful.

    ** an ad hoc construction of a bound, which the manuscript evaluates via a user
    study.
    Whereas the construction is ad hoc, and there must be other ways of approaching
    the problem, I do not think it detracts significantly from the manuscript. It
    works reasonably well as a proof of concept.

    Overall, this manuscript seems to me to present great opportunities for the sci
    vis field to grow.

    Suggestions for improvement:
    
    * It would be great if some part of the two case studies were weaved into the
    mathematical exposition, to help anchor the concepts. Humans are great at
    generalizing from examples. For example, I really can't parse this statement:
    ``Suppose that Z is actually of PMF P, but is encoded as Eq. 4 based on Q.''
    What does ``based on Q'' mean here?
    The easier to follow the proof and function construction, the better. In general,
    it might be worth moving to the appendix, or compacting the proof and ad hoc
    construction.

    * Instead of reiterating what was done, step by step, in the conclusion, it would
    be good to see some discussion/summary of assumptions, limitations, and where the
    work agrees or disagrees with other published observations.

    * The information theory approach and the PD discussion are interesting, and a
    valid way of approaching this problem. The user study is also a reasonable way to
    evaluate this theory. However, here and in previous manuscripts in this vein, it
    seems to me the supporting evidence is subject to how we interpret the term
    "information".  For example, in the two case studies, the questions asked as not
    typical of the two domains in some settings, but they may be typical in other
    settings. E.g., a clinician would be interested in where an anomaly is located and
    in its shape, and how much uncertainty there is; whereas a nursing student might
    be interested in general literacy of the type described in the manuscript
    evaluation. In the second case study, a Chicago subway user would often be
    interested in the number of stops and line changes, and not so much in walking
    distance, because the lines are well connected; whereas a NYC subway user often
    needs, for the fastest route, to come out of the subway system, walk for 5
    minutes, and enter the subway system at a different point--I see how the type of
    judgments used in the second case study would be useful there. These
    considerations should appear both in the introduction, to help motivate the
    manuscript, and in the discussion, to help anchor it.

    * A reworking of most figure captions would be good (see Minor comments below).

    Minor:
    I am afraid that we are becoming so mathematically illiterate as a field, that the
    manuscript would benefit from reminding the readers what ``unbounded'' means in math
    terms. Also, a gentle brief definition of terms unfamiliar to a graduate student
    (e.g., ``MCDA, a sub-discipline of operations research that explicitly evaluates
    multiple conflicting criteria in decision making'', `` data intelligence workflow,
    *def here*'', ``cross entropy'') might encourage students to keep reading the paper.

    When discussing the relationship between AC and PD, I'd be interested in any
    practical examples supporting the increase/decrease illustrated in Fig.2.

    Citations are grammatically invisible. Not good: ``observational estimation in
    [43]''. Good: ``observational estimation in Smith et al. [43]'' or ``observational
    estimation [43]''.

    Fig 6 and Fig 7
    * Some reorganization of the caption would be helpful (e.g., in Fig 7, I'd start
    with ``Divergence values for six users using five candidate measures, on an example
    scenario with four data values, A, B, C, D. ''
    * ``them'' -> ``the data values'' (?)

    Fig 7 
    * typo: ``A, B, C, are D.'' (and D). 
    * clarify ``pr.'' stands for ``process'', or remove

----------------------------------------------------------------

Reviewer 3 review

  Paper type

    Theory \& Model

  Expertise

    Expert

  Overall Rating

    <b>1.5 - Between Strong Reject and Reject</b>

  Supplemental Materials

    Not acceptable

  Justification

    The basic story of the paper is: information theory is a promising way to
    characterize the process of data vis because it gives ways to quantify information
    transmission. From some of the measures it provides (Shannon entropy, and KL
    divergence), the authors have previously [11] developed a measure of "benefit" (eq
    2).  But now, it is apparently a big problem that the negative term in benefit
    ("potential distortion" or PD) is unbounded, so a bounded measure is needed, and
    the paper aims to make a well-informed pick from among many possible choices.

    However, the authors do not make a case that, insofar as this is about data
    visualization, a bounded measure is actually a problem that merits a whole paper,
    and, the evaluation of the new measure, such as it is, has a strained relationship
    to both information theory and to visualization practice.

    This submission is on a topic of interest to visualization research, and it may be
    enthusiastically received by the adherents of [11] and the related work in the
    same vein. The submission should stand on its own as a piece of scholarship, but
    does not.

  The Review

    The authors are pursuing information theory as a means of putting data vis on
    mathematical footing.  The current submission relies heavily on a previous paper
    [11]. In response to a previous round of reviews (from a rejected Eurovis 2020
    submission), the authors were defensive about critiques that they felt were of
    [11], rather than of the new work, and have written a new summary of [11] work as
    Appendix A.  However, every submission presents a new opportunity for scrutiny,
    from a new set of reviewers, which ideally generates new opportunities for authors
    to strengthen or reconsider a line of work, even if initial steps have passed peer
    review.  Nonetheless, this review will strive to consider the new submission on
    its own terms.

    My high-level considerations are:

    ((1)) the motivation of needing a bounded information theoretic measure of PD, and
    the mathematical strength of the connection to information theory, and

    ((2)) "intuitive" and "practical" (as noted at the end of paragraph 4)
    applicability to vis applications

    First ((1)):

    We accomplish theoretical research when we can reframe new things in terms of pre-
    existing and trusted theory. The authors trust information theory, and advertise,
    in the Abstract and Intro, how they are drawing on information theory.  However,
    the authors seem bothered (top of 2nd column, 1st page) by the fact that the KL
    divergence when communicating a binary variable can be unbounded, because "the
    amount of distortion measured by the KL-divergence often has much more bits than
    the entropy of the information space itself. This is not intuitive to interpret
    .."  Later they say (middle 2nd column page page 3) "it is difficult to imagine
    that the amount of informative [sic?] distortion can be more than the maximum
    amount of information available."

    Actually, it is not difficult to imagine, if you accept information theory (as the
    authors surely want us to do), hence this is a worrisome start to the math
    foundations of this work.

    If you have an extremely biased coin Q (say, p(head)=0.99, p(tail)=0.01), and want
    to use bits to communicate a sequence of flips of Q, the optimal scheme will
    involving something like using "0" to represent some number (greater than 1) of
    heads in a row, i.e., some kind of compression, which can be very efficient for Q
    given how low the entropy is: ~0.081. But then, if you are tasked with encoding a
    sequence of flips of an *unbiased* coin P (p(head) = p(tail) = 0.5), you are going
    to be very disadvantaged with your Q-optimized encoding, so much so that it could
    easily more than double the expected length of the encoding, as compared to a
    P-optimized encoding (i.e. "0"=head "1"=tail). As the authors surely know, the KL
    divergence D$\_$KL(P|Q) tells you the expected number of extra bits needed to
    represent a sample from P, when using a code optimized for Q, rather than one
    optimized for P.  For this specific example, D$\_$KL(P|Q) = 2.32918, which motivates
    how D$\_$KL can naturally exceed 1 even for a binary event.  But again, the authors
    probably know this (in fact their section 4.1 is closely related to this
    situation), so it is unclear what they mean by saying this is "unintuitive".

    Also, if the worry is that KL divergence is unbounded because of the places where
    probabilities can be become very small, then why not try some smoothing?  That is
    a known thing in ML, for example this paper:

    https://papers.nips.cc/paper/8717-when-does-label-smoothing-help.pdf

    talks about "label smoothing" in the context of minimizing cross entropy (which is
    the same as minimizing KL divergence if the original or data distribution P is
    fixed).  So the connection to stats or information theory, as it is commonly used
    in current research, appears weak, which makes the proposed new measure appear
    contrived.

    Another good property of KL divergence is its additivity: P=P1*P2 and Q=Q1*Q2 ==>
    D$\_$KL(P|Q) = D$\_$KL(P1|Q1) + D$\_$KL(P2|Q2), which is relevant for considering how
    models and phenomena can have both coarse-grained and fine-grained components.  KL
    divergence nicely analyzes this in terms of simple addition, which would also seem
    to be valuable for the authors, given how in their driving equation (1) KL
    divergence appears in the numerator of a fraction.

    Instead, the authors invent some new measures (9) and (10), which raises various
    red flags.  Inventing new measures for information divergence seems like an
    interesting research direction, but (red flag \#1) the most informed peer reviewers
    for that will be found in statistics, so it is a little scary for the value of
    these measures to be evaluated by vis researchers, when fundamentally, the
    underlying math actually has nothing to do with visualization (any more than it
    has to do with any other application of statistics).  Red flag \#2 is that the new
    measures involve a free parameter "k", unlike the standard information-theoretic
    measures, but maybe this is okay in the same way that smoothing is okay.  Still,
    it doesn't seem like the new measures will be additive, or this isn't shown (red
    flag \#3).  Red flag \#4 is that section 4.1 ("A Mathematical Proof of Boundedness")
    is not a proof at all - it simply describes one specific scenario where the
    optimal encoding for Q will require more, but a bounded amount more, bit to encode
    a sample of P.  That does not prove anything, nor does it illuminate anything
    about how intuitive or reasonable information divergence measures should behave in
    general. The many red flags undermine the desired rigorous connection to
    information theory.

    If, at the end of the day, there is only a tenuous connection to information
    theory, and yet there is a perceived need to have a bounded information divergence
    measure, why not redirect the creative thought that went into formulating the new
    measures (9) and (10), into some bounded monotonic reparameterization of D$\_$KL,
    like arctan(D$\_$KL)?  The authors have not explained why this much simpler and
    obvious strategy is not acceptable, and yet, based on their motivation from Figure
    2, it should suffice.  Figure 2 nicely documents the desired *qualitative* effects
    of varying AC, PD, and Cost. None of the later quantitative arguments were
    compelling, so it was too bad the authors did not pursue a simpler way of meeting
    their goals that maintains at least some connection to D$\_$KL.

    We finally turn to how the different possible measures are actually compared and
    evaluated.  Criterion 1 of Section 4.3 essentially amounts to looking at lots of
    plots, and making qualitative judgments about what behaves in the desired way.  A
    suitably formulated monotonic reparameterization of D$\_$KL could have done well
    here, and the qualitative nature of this evaluation undermines the author's
    mission to find quantitative measures for visualization.  Criteria 2 through 5 in
    Table 3 involve a 1--5 Likert scale (!) for judging the various alternative
    measures.  Subsequent criteria (described below) are used in a "multi-criteria
    decision analysis (MCDA)" analysis, a fancy way of saying: a ranking based on some
    heuristically selected weighted averages of various desiderata.  The combination
    of Likert scales and MCDA is where we lose any remaining connection to information
    theory.  How we know: a different set of equally plausible candidate measures,
    under a different set of criteria, with a different Likert scales, and different
    weightings, could have produced a different winner.

    \textcolor{orange}{Hence, the chosen measure was not, it turns out, actually determined by
    information theory, and its actual value for *quantitative* analysis was never
    demonstrated.}

    Now for consideration ((2)): "intuitive" and "practical" (as noted at the end of
    paragraph 4) applicability to visualizations

    The exposition starts with the claim that "inaccuracy" is ubiquitous, even
    necessary, in visualization, and that information theory promises to quantify the
    inaccuracy.  Even if this claim was made in [11], it bears scrutiny as the
    foundation of this paper.  As Alfred Korzybski tells us, "the map is not the
    territory" and "the word is not the thing".  OF COURSE there is information
    missing in a visualization; it wouldn't be visualization, and it wouldn't be
    useful for visual analysis and communication, if it merely re-iterated all the
    data.  Visualizations are useful insofar as they make task-specific design
    decisions about WHAT to show (and what not to show), and audience-specific design
    decisions about HOW to show it.  The editing task of choosing what to show is
    where the authors would point out "inaccuracy", but that is not how visualization
    research, or scientific visualization research in particular, understands that
    term, regardless of [11].

    The later evaluation criteria in Sec 4.3 do not convey a realistic consideration
    of visualization applications.  In Criterion 6 the authors embrace an encompassing
    view of an analysis workflow that has to end in some binary decision, in which the
    mapping from data to decision actually has to transmit less than a single bit
    (since p(good)=0.8 so entropy is $\sim$0.72).  If some decision support system were so
    refined and mature that the human user need only made a binary decision, and the
    accuracy of that decision is the entirety of the story, does that seem like a
    compelling setting for visualization research?  There is no room in the analysis
    for using visualization to justify or contextualize the decision, and no down-
    stream stake-holders have any role in the analysis, so why even consider
    visualization, instead of a well-trained machine learning method?  The story for
    Criterion 7 is nearly as contrived, and again not actually admitting any
    visualizations per se, so it is not compelling as a setting for evaluation.

    Vis researchers use case studies with the belief that, even though they assess
    some limited and contrived scenario, they nonetheless reveal something informative
    and representative about broader and more general situations.  The two case
    studies were unconvincing individually, but there is also general problem: where
    are the encodings of the many alphabets Z$\_$i, Z$\_${i+1}, etc, one for each step of
    some workflow?  I thought the authors' theory is based on analyzing each of those
    steps, but the case studies collapse everything from some simplistic reduction of
    the data, to some simplistic reduction of a viewer's response. Because the case
    studies are so structurally disconnected from all the previous mathematical
    exposition, it's harder to see how the case study results support the goals of the
    paper.

    The Volume Visualization (Criterion 8) case study just doesn't work.  The
    questions asked about volume visualization in the questionnaire are completely
    unrecognizable to me, as someone with experience with real-world biomedical
    applications of volume rendering.  The questionnaire involves variations on the
    same false premise- that a good visualization is one that allows you to infer the
    data from the picture. No: in volume visualization the whole point is that there
    are huge families of transformations on the data that are purposely invisible (a
    task will typically require scrutinizing some materials or interfaces and ignoring
    others), and, the questions to answer for the task are not about recovering the
    original data values at any one location in the rendered image, but rather about
    inferring higher-order properties about structures and their geometric inter-
    relationships (e.g. what is the shape of the bone fracture, or, what is the size
    and extent of a lesion relative to other anatomy).  That all the questions here
    involve pre-existing renderings, of *old* test datasets (wholly detached from any
    real-world application), was economical on the part of the authors but deficient
    as scholarship. I am unconvinced that this questionnaire can shed light on how to
    evaluate volume visualizations, or how to  evaluate measures for evaluating volume
    visualizations.

    The London Tube map (Criterion 9) case study is more interesting, but describing
    Beck's map as "deformed" with a "significant loss of information" risks missing
    the point.  Beck recognized that for the commuter's task of navigating the subway,
    absolute geography is irrelevant (because you can't see it), but topology and
    counting stops matter (because that's what you do experience). Beck made a smart
    choice about what to show. Both kinds of maps used in the study are faithful: one
    is faithful to geography (while distorting the simplest depiction of the
    topology), and one is faithful to the topology (while distorting geography). There
    is something interesting to learn from a case study about measuring the mismatch
    between geographic vs topological maps, and geographic vs topological tasks, but
    this paper doesn't convince me that information theory is at all relevant to that.
    Instead, the case study is used here to help choose a bounded information
    distortion measure, but again the reliance on a Likert scale ("spot on", "close",
    "wild guess") is a sign that the ultimate choice of measure is much less connected
    to information theory than the Abstract and Intro imply.

    The more interesting component about the London Tube map case study was that cost
    was considered.  But hang on: why is the choice of a bounded information
    divergence measure -- for the numerator in (1) -- so crucially important, when the
    choice of cost -- for the denominator in (1) -- is basically left completely
    unspecified?  The authors say that cost can be different things "(e.g., in terms
    of energy, time, or money)", but if the goal is meaningful quantitation, how can
    cost be so under-specified?  It could have a huge effect on the relative
    cost/benefit ratios of comparable vis methods, which further undermines the idea
    that the choice of bounded divergence measure alone warrants a paper.

----------------------------------------------------------------

Reviewer 4 review

  Paper type

    Theory \& Model

  Expertise

    Knowledgeable

  Overall Rating

    <b>2 - Reject</b>\\
    The paper is not ready for publication in SciVis / TVCG.\\
    The work may have some value but the paper requires major revisions or
    additional work that are beyond the scope of the conference review cycle to meet
    the quality standard. Without this I am not going to be able to return a score of
    '4 - Accept'.

  Supplemental Materials

    Acceptable

  Justification

    The paper deals with the improvement of a quantitative measure for the cost-
    benefit ratio of visualizations. In previous work (Ref. [11]) a proposal was made
    for this ratio.
    In the counter there are 2 terms, one bounded, the other unbounded. The authors
    argue that both terms should be bounded. The argumentation seems to me not quite
    conclusive, at least it is not clear to me why it is so important.

    Furthermore, the authors are looking for a more suitable mathematical quantity
    that is bounded. They propose some candidates for this; the selection of the most
    suitable quantity seems to me to be rather ad hoc and arbitrary. The main reason
    is that the selection process is not based on realistic examples.

    Correcting these shortcomings seems to me to require a major revision, which is
    not possible within the short revision cycle.

  The Review

    Preliminary remark: 

    I'm convinced that information theory plays an important role in the foundation of
    visualization. However, I believe that information theory based on Shannon's
    mathematical theory of communication (which operates largely on a syntactic level)
    is not sufficient for this. Semantics and task-dependent weighting play too big a
    role in visualization. Although semantics can be represented by suitable
    alphabets, these have to be defined against the background of complex knowledge
    frameworks into which the user sorts the perceived information, utilizing
    perceptual and content-related prior knowledge. Therefore, I think that Shannon's
    theory represents a rigorous constraint to any further theory formation about all
    semantic and pragmatic aspects of information as well as about visualization. The
    question is how strong this constraint is and which dimensions need to be added to
    reflect the importance of semantics in data visualization.

    Furthermore, I think that the benefit of a visualization depends very much on how
    cognitive peculiarities (e.g. the particular strength of processing spatial
    information or the limited capacity of short-term memory) are taken into account
    in the visual design. I do not see (maybe I know too little about) how this is
    taken into account in the information-theoretic framework.

    As I am not an expert in these matters, I do not wish to include my fundamental
    doubts in my assessment. I try to evaluate the paper by asking what it brings to
    the table, *assuming* that the described information theory approach is adequate
    for "estimating the benefit of visualization".

    -----
    Review:

    Why should we be interested in estimating the benefits of visualization
    quantitatively? Of course, because it would help to advance the visualization. So
    the overall goal seems to make sense.

    The paper is based on Ref. [11], in which Chen and Golan proposed an information-
    theoretic measure for the cost-benefit of data visualization workflows. They
    propose to measure the cost-benefit ratio by (alphabet compression - potential
    distortion) / cost. Here, "alphabet compression" is the information loss due to
    visual abstraction and the "potential distortion" is the informative divergence
    between viewing the data through visualization with information loss and reading
    *all* the data.

    The "alphabet compression" is computed as the entropic difference between the
    input and output alphabets. This is a bounded quantity. The "potential distortion"
    was measured in Ref. [11] as the Kullback-Leibler (KL) divergence of an alphabet
    from some reference alphabet. This quantity is unbound. The only goal of the paper
    is to find an appropriate information-theoretical quantity for the "potential
    distortion", which is also *bounded*.

    As a reason for this goal, the authors say that the benefit calculated using the
    KL divergence "is not quite intuitive in an absolute context, and it is difficult
    to imagine that the amount of informative distortion can be more than the maximum
    amount of information available."

    From a practical point of view, I do not really see the problem: as I mentioned
    before, the goal is to improve visualization, which includes trying to keep the
    "potential distortion" small. In other words, you're practically working at the
    other end of the scale and you wouldn't even feel that the quantity is not
    bounded. For example, in the visualization of high-dimensional data, there is an
    entire branch of research that is mainly concerned with finding projections with
    the least possible distortion.

    The fact that the amount of information distortion can be greater than the maximum
    amount of available information could be seen as a theoretical disadvantage. To
    better understand the motivation for the paper, I would like to see this point
    worked out in more detail.

    Some of the statements in the paper puzzled me, especially the repeated statement
    that the following fact would seem counter-intuitive: that high alphabetical
    compression is a feature of a benefit of visualization. This surprises me, as so
    many efforts of visualization are directed in exactly this direction: Information
    reduction, visual abstraction, feature extraction, visual summaries, first
    overview then (user controlled) details... The intuition of visualization experts
    should rather tell them that information reduction is a quality feature - up to
    the point where essential information is no longer visually represented. The last
    point is missing in the model, unless the omission or loss of essential
    information is subsumed under "information distortion".

    I find the search for bounded measures for the Potential Distortion (PD)
    unconvincing. I am aware of the size of the problem: We try to find suitable
    abstractions to create a theoretical framework that covers a huge number of very
    different use cases. This is a mammoth task, which cannot be accomplished in one
    step. Rather, one has to approach it in many attempts with subsequent
    modifications and iterative improvements. Here we want even more, namely a useful
    quantitative formulation that captures essential aspects of information processing
    in the visualization pipeline. One basic problem is the vague imprecise
    terminology, another is the high dimensionality of the problem. I respect any
    attempt to address this huge problem. Nevertheless, the described selection of an
    appropriate mathematical quantity that measures PD I did not find convincing.

    Section 4 starts with the "mathematical proof" (Sect. 4.1). I don't think this is
    a mathematical proof, since a special PMF Q is used here (or is this distribution
    mathematically somehow distinguished?).

    In Sect. 4.2, again a reason is given (the arbitrariness of specifying epsilon)
    why "it is [...] desirable to consider bounded measures that may be used in place
    of D$\_$KL". I don't understand this argumentation.
    In equation (9), it should probably read k $\ge$ 1, so that slope of |...|$\land$k is not
    negative.

    In equation (11), H$\_$max should be explained: over which set is the maximum
    searched for?

    In Sect. 4.3, a search for the "most suitable measure" is compared to the search
    for a suitable unit (e.g., metric vs. imperial). This is a misleading comparison.
    While the mentioned example is about different units of measurement for the same
    operationally defined quantity (units of measurement that can be converted by
    multiplying by a constant factor), this paper is about selecting a suitable
    (operationally defined) quantity. The problem here corresponds more to the
    question which metric is the best one to solve a task, e.g. the dissimilarity of
    objects to. This has nothing to do with the choice of units of measurement (which
    in our case are [log$\_$2( p)] = bits).

    But the inappropriately chosen example reminds us of another point, which is
    extremely important and yet very elementary! If you add or subtract two
    quantities, they must have the same unit of measurement; otherwise you get
    complete nonsense. In our case we calculate AC - PD. AC has, because of log$\_$2 (p
    ), the unit bit; if we would calculate with log$\_$256 (p ) instead, we would have
    the result in the unit byte. So we have to calculate AC also in the unit "bit".
    Otherwise we pretend to be quantitative and yet we are only doing pseudo-science.

    This means that PD must also be an entropic quantity calculated with logarithms to
    base 2. This condition is fulfilled by the KL-divergence, Jensen-Shannon
    divergence and the conditional entropy. The condition is certainly not fulfilled
    for the Minkowski distances. The latter should therefore not be brought into play
    at all !

    \textcolor{orange}{I am also skeptical about the quantity D$\land$k$\_$new in Eq. (9), except for the case
    k=1.
    If one does a dimensional consideration by (for the moment) leaving out the
    summand 1 in the logarithm in Eq. (9),} then one can consider the exponent k as a
    multiplicative factor before the logarithm and thus before the sum; i.e. k is a
    factor that changes the unit of measurement. Although this is not a proof
    (especially since you cannot ignore the presence of the summand 1 in the
    logarithm), it is an indication that k unequal to 1 could be problematic. It could
    mean, so to say, to subtract bytes from bits.

    Aren't in information theory only terms of the type p * log (p ) considered, where
    p are always probabilities with values from the interval [0,1] and the logarithms
    are always taken to the same basis. I think there is a deep reason for this.

    I acknowledge the effort to make the selection of the "best" quantity halfway
    rational. But even the use of multi-criteria decision analysis (MCDA) cannot hide
    the fact that there is a great deal of arbitrariness in the analysis, as most
    numerical values were simply assumed. That these are somehow plausible is not
    enough to convince me. I think Sect. 4 is a rendition of the path the authors have
    taken, but not the way it should be presented. My recommendation would be to
    shorten this drastically.

    However, this does not dispel my fundamental reservations about the approach to
    finding an appropriate measure of PD. The examples shown with the tiny alphabets
    are, compared to realistic tasks, extremely simple and in my opinion not
    convincing. I would find it more convincing if, for example, 2 really realistic
    examples with realistic questions/tasks and suitable visualizations were used to
    find out which PD measure is most suitable.

    In doing so, I would leave out all candidates for whom it is actually clear from
    the outset that they will not perform well (e.g. for dimensionality reasons).

    In Section 5, the examples chosen again are too artificial and academic. The
    subway maps were primarily designed to quickly convey the number of stops between
    two locations (the most important information, e.g. to get off at the right
    station), and only secondly to estimate travel times (at least within the city,
    the travel time is approximately defined by the number of stops) and only thirdly
    to estimate actual distances.

    I also found the tomography example unworldly. Better would be e.g. the artery
    data set, but with reasonable questions. For example, a doctor would look for
    constrictions in the arteries, the spatial location of the constrictions in the
    arterial network; furthermore, she/he would estimate how narrow the constrictions
    really are (taking into account possible measurement and reconstruction errors)
    and what the medical risk is -- taking into account the overall blood flow in the
    arterial network and considering, which body regions are supplied.

    Overall, it is not quite clear to me why the motivating factor for this work,
    namely the boundedness of PD, is so important. Furthermore, I find the choice of a
    more suitable mathematical quantity not convincing. This selection should be made
    on the basis of real, realistic problems with appropriately designed
    visualizations. A summarizing discussion of the assumptions, limitations and scope
    of the theory is missing.

  Summary Rating

    <b>Reject</b>\\
    The paper is not ready for publication in SciVis / TVCG.\\
    The work may have some value but the paper requires major revisions or
    additional work that are beyond the scope of the conference review cycle to meet
    the quality standard. Without this I am not going to be able to return a score of
    '4 - Accept'.

  The Summary Review

    The authors aim to contribute to the foundation of data visualization by
    mathematically capturing essential characteristics of the visualization process.
    The main goal is a general mathematical expression that can be used to calculate
    the cost-benefit ratio of a visualization on an information-theoretical basis.
    While the costs are given by e.g. mean response times of the users, the benefit is
    more difficult to define. In Ref. [11], to which the first author of the current
    paper already contributed, a measure for the "benefit" was developed (Eq. 2). It
    consists of the difference "Alphabet Compression (AD)" - "Potential Distortion
    (PD)". The subject of the present work is PD, which measures the informative
    divergence between viewing the data through visualization with information loss
    and reading the data without any information loss. In [11] an expression based on
    Kullback-Leibler (KL) divergence was used for PD. Since the KL divergence is
    unbounded, PD can be arbitrarily large. In the eyes of the authors this is a
    problem which they - and this is the actual aim of the paper - want to solve by a
    more suitable mathematical expression.

    To this end, the authors first present a set of candidates for limited measures,
    oriented on common measures of information theory; some of these are
    parameterized, resulting in greater diversity. To select the most suitable
    measure, they define a set of criteria, apply them, using also multi-criteria
    decision analysis (MCDA). Then they validate the results with 2 case studies and
    conclude that a newly introduced measure, D$\land 2\_$new, is the most suitable one.

    All 4 reviewers assume that information theory will be part of the a foundational
    theory of visualization and will at least provide constraints for further
    approaches. \textcolor{teal}{However, the majority of the reviewers have doubts about the approach
    [11]. Since [11] and the follow-up work were positively assessed in peer reviews,
    these reviewers tried NOT to include their reservations in the evaluation, but to
    assess the manuscript on the basis of the assumption that the approach [11] is
    meaningful.}

    The main results of the review are:

    + The effort to make a contribution to the theoretical basis of the visualization
    is to be evaluated positively, especially since fundamental concepts are not
    sufficiently clarified and it is therefore a very difficult undertaking. (R1, R2,
    R3, R4)

    \textcolor{orange}{- The need for a bounded measure for PD is not sufficiently justified; there are
    several arguments (see in particular R2 and R3) that this requirement is not
    necessary. This, however, makes the premise of the paper doubtful. (R1, R3, R4)}

    \textcolor{orange}{- Some of the authors' trains of thought cannot be fully comprehended, as there
    are discrepancies with earlier work that are not being clarified. (R1)}

    \textcolor{orange}{- The alleged proofs are unsustainable. (R1, R3, R4)}

    \textcolor{orange}{- The proposed alternative measures are sometimes very ad hoc (R1, R3, R4), some
    are dimensionally doubtful (R4), they are (probably) not additive (R3) and there
    is no natural language interpretation (R1); other reasonable choices would have
    been possible (R3).}

    \textcolor{orange}{- The evaluation of measures has a strong ad hoc character (R1, R2, R3, R4); with
    the combination of Likert scales and MCDA almost any connection to information
    theory is lost (R3).}

    \textcolor{orange}{- The paper demands a lot from the reader; without detailed knowledge of
    information theory concepts, as well as details of the cost-benefit ratio, the
    paper is difficult to understand. It would be better if the basic terms and
    concepts were briefly presented again in the paper, with reference to the
    appendix) (R1). It would be nice if some of the two case studies were interwoven
    into the mathematical presentation to better anchor the concepts (R2). The
    notation needs to be improved (R1),}

    \textcolor{orange}{- The validation of the PD measure is questionable (R1, R3, R4); for details see
    in particular R3.}

   \textcolor{violet}{- When trying to apply the presented theory to real applications, huge gaps open
    up (R1, R3, R4); it starts with the fact that it is the goal of almost all
    visualizations to reduce the amount of information and to transmit only the
    information that is essential for the respective application situation; and it
    ends with the fact that the case studies in the paper, on closer inspection,
    hardly support the goals of the work (R1, R3, R4).}

    \textcolor{violet}{- A summarizing discussion of the assumptions, limitations and scope of the theory
    presented is missing (R1, R4).}

----------------------------------------------------------------

\end{narrowfont}
\normalfont
\normalsize
\setlength{\parindent}{5.1mm}
\setlength{\parskip}{0pt}

\end{document}